\documentclass[submission,Phys]{SciPost}

\usepackage{amssymb}
\usepackage{bm}
\usepackage{physics}
\usepackage{tikz}
\usetikzlibrary{calc}

\title{\Large
	Fermionization and Floquet dynamics of a voltage-biased\\
	Josephson junction coupled to a high-impedance transmission line
}

\author{
	J. Est\`eve\\
	\small Laboratoire de Physique des Solides,
	CNRS, Universit\'e Paris-Saclay,
	91405 Orsay, France
}

\date{\today}

\begin{document}
	
\maketitle

\begin{center}
\begin{abstract}
We investigate inelastic Cooper-pair tunneling through a dc-biased
Josephson junction coupled to a finite multimode transmission line of
characteristic impedance \(Z\). At the two special impedance points
\(Z=R_Q/2\) and \(Z=R_Q\), where \(R_Q=h/(2e)^2\), refermionization
reduces the coherent dynamics to a quadratic fermionic problem.
Although the resulting one-particle problem is exactly solvable,
the driven, dissipative many-body steady state remains nontrivial.
We characterize this state and derive predictions for the inelastic
tunneling current and the fluorescence spectrum of the emitted microwave
photons. In particular, we predict zero-delay antibunching
of forward Cooper-pair transfers at \(Z=R_Q/2\) and bunching at \(Z=R_Q\). These statistics
reflect the fermionic and bosonic algebras of the underlying refermionized models.
\end{abstract}
\end{center}

Quasiparticles or Cooper pairs tunneling across a junction can exchange
energy with environmental degrees of freedom. Such inelastic tunneling
processes have a long history and have been studied in a variety of settings,
including coupling to electromagnetic and vibronic modes~\cite{Devoret1990,JaklevicLambe1966}. 
In this work, we
focus on inelastic Cooper-pair tunneling in a high-impedance electromagnetic
environment, where the relevant degrees of freedom coupled to the tunneling
process are the microwave modes of the circuit connected to the Josephson
junction~\cite{Holst1994,Hofheinz2011,Rolland2019,Menard2022,Leppakangas2013}. 
By high impedance, we mean that the characteristic impedance of the
modes becomes comparable to the superconducting resistance quantum for
Cooper pairs, \(R_Q=h/(4e^2)\).

Several distinct dynamical regimes can then arise. In the incoherent regime,
the physics is well described by \(P(E)\) theory~\cite{IngoldNazarov1992,IngoldGrabertEberhardt1994,Joyez2013}. 
Individual tunneling
events may be regarded as statistically independent, and both the tunneling
rate and the associated emission or absorption of photons from an equilibrium
environment can be calculated. For a junction subject to a dc voltage bias
\(V\), the voltage source performs an amount of work \(2eV\) during each
Cooper-pair tunneling event. This energy is transferred to the electromagnetic
environment through the emission and absorption of one or many photons, with a net energy
transfer equal to \(2eV\). High impedance makes multiphoton processes appreciable
even without a large pre-existing photon population.

\begin{figure}
	\centering
	\includegraphics{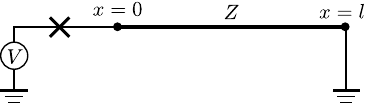}
	\caption{Schematic of the circuit considered in this work. A voltage-biased
		Josephson junction is coupled to a finite transmission line of characteristic
		impedance $Z$. The junction is located at $x=0$, while the remote end at
		$x=l$ is connected to a low-impedance measurement line (not shown), which is
		approximated here as an ideal ground.}
	\label{fig:circuit}
\end{figure}

This description ceases to apply when the lifetime of the microwave
excitations becomes comparable to the interval between tunneling events:
the environment need not relax before the next event, and successive
transfers are no longer independent~\cite{Leppakangas2016}. The dynamics must then be treated
explicitly, for example within a master-equation framework. This regime
is by now well understood when the junction is coupled to one or a few
microwave modes~\cite{Gramich2013,Armour2013,Armour2015,KubalaGramichAnkerhold2015}. 
Here, we extend the problem to the multimode electromagnetic environment that
naturally arises when a Josephson junction is connected to a finite section of
transmission line (see Fig.~\ref{fig:circuit}). The mode quality factors in such circuits typically exceed a few thousand, and the driven-dissipative dynamics is essential to understanding the properties of the tunneling current and the mode occupations. A self-consistent extension of \(P(E)\) theory,
with photon occupations determined by a kinetic equation, has shown
that photon accumulation can substantially modify the tunneling
current~\cite{Cailleaux2025}.

More specifically, we consider the modes of a finite transmission-line strip
with characteristic impedance \(Z\), along which microwave excitations
propagate between two boundaries: a voltage-biased Josephson junction at one
end and a low-impedance measurement line at the other. The strip supports a
ladder of commensurate modes with frequencies given by odd multiples of the
fundamental quarter-wave frequency \(\omega_0\). The superconducting phase at
the junction can be expanded over these free strip modes, while the Josephson
coupling mixes them through the phase rotation imposed by the dc voltage bias.
At the two particular impedance values \(Z=R_Q\) and \(Z=R_Q/2\), the
boundary phase field admits an exact
fermionic representation in which the Josephson coupling becomes quadratic
in fermionic fields~\cite{AmeduriKonikLeClair1995,PolchinskiThorlacius1994}. 
This exact algebra provides a tractable description of
the physical circuit below the scale at which junction capacitance and other
high-frequency corrections modify the harmonic mode ladder.
This boundary sine-Gordon description connects our finite, voltage-driven
circuit to work on the equilibrium Schmid transition, for which \(Z=R_Q\)
marks the critical impedance in the ideal Ohmic, infinite-line limit~\cite{Schmid1983,Bulgadaev1984,LukyanovWerner2007,Murani2020,GiacomelliCiuti2024,Kuzmin2025,Paris2025}.
It is also complementary to theoretical studies of inelastic microwave photon scattering
from quantum impurities~\cite{Goldstein2013} and to experiments that probe, by microwave spectroscopy and photon scattering, the spectrum of an unbiased junction coupled to the strip~\cite{Leger2019,PuertasMartinez2019,Kuzmin2019Superstrong,Kuzmin2021,Fraudet2025}. 

We formulate the problem in a Floquet framework, with the dc voltage \(V\) providing a periodic drive at the Josephson frequency \(\omega_J=2eV/\hbar\). The drive is resonant with the strip whenever $\omega_J$ is a multiple of $\omega_0$. These resonances produce a comb of current peaks in the \(I\)--\(V\) characteristic, separated by \(\hbar\omega_0/(2e)\), as observed experimentally~\cite{Aiello2020Thesis,Cailleaux2025Thesis}. When the renormalized Josephson coupling is small compared with the mode spacing, a secular approximation is justified. This provides a direct description of the Floquet states formed by Cooper-pair tunneling and connects their structure to the multimode photonic excitations that absorb the energy supplied by the voltage source. Two distinct physical pictures emerge at the solvable impedances:
\begin{itemize}
	\item At \(Z=R_Q/2\), the Toulouse, or Majorana, point, the bosonic Fock space can be reorganized in terms of fermionic orbitals~\cite{AmeduriKonikLeClair1995,Konik1995,LeClairLesageSaleur1996}.
	Near a given resonance, only one fermionic orbital participates in the resonant tunneling process. A forward tunneling event fills this orbital, and Pauli exclusion blocks an immediate second forward event, resulting in antibunching of the Cooper-pair current.
	\item At \(Z=R_Q\), a forward tunneling event creates a coherent superposition of particle--hole excitations of a filled Fermi sea, each carrying energy \(2eV\)~\cite{PolchinskiThorlacius1994,CallanKlebanovLudwigMaldacena1994,vonDelftSchoeller1998}.
	This collective excitation is bosonic, analogous to a charge-density wave, or plasmon, in a one-dimensional Luttinger liquid. Bosonic stimulation favors successive forward transfers, resulting in Cooper-pair current bunching.
\end{itemize}
A steady state arises when the electrical power \(IV\) supplied by Cooper-pair tunneling balances the power lost through the measurement port and through intrinsic dissipation in the strip. The Josephson drive takes a simple form in the refermionized basis, whereas photon loss is most naturally described in the bosonic basis. This makes the driven-dissipative many-body problem difficult to solve.

At the Toulouse point, the drive couples to a single fermionic orbital. Within some assumptions on the loss rates, its occupation obeys the optical Bloch equation of a driven two level atom. At the bosonic point, the dynamics is more involved, and we use a Gaussian approximation to obtain a numerically tractable model. In both cases, we characterize the steady state through the Cooper-pair current, the mean photon occupation of each strip mode, and the fluorescence spectrum at the measurement port. We also include the dephasing of superpositions of different numbers of transferred Cooper pairs, caused by low-frequency noise of the bias voltage, which makes Cooper-pair transfer incoherent.

The paper is organized as follows. Section~\ref{sec:floquet} derives the
Floquet Hamiltonian of the junction coupled to the finite strip, reviews the
refermionization of the Josephson coupling at $Z=R_Q/2$ and $Z=R_Q$, and
introduces the secular approximation. Section~\ref{sec:master} introduces the
master equation for photon loss and briefly summarizes the weak-drive steady
state, which reproduces $P(E)$ theory. Sections~\ref{sec:secular_toulouse}
and~\ref{sec:secular_RQ} use the refermionized models to obtain the steady
state beyond this perturbative regime.
From these steady states, we derive the microwave fluorescence spectrum, the
mean Cooper-pair current, and its second-order correlation function.
Section~\ref{sec:RQ_dephased} includes the dephasing of coherences between
states that differ in the number of transferred Cooper pairs.
Section~\ref{sec:beyond} goes beyond the secular
approximation and determines the single-particle Floquet quasienergy spectra
at both impedances, using the quadratic fermionic structure of the full
Floquet Hamiltonian.

\section{Refermionized Floquet Hamiltonian}
\label{sec:floquet}

We consider a transmission line of length \(l\), wave velocity \(v\), and
characteristic impedance \(Z\). At \(x=l\), the line is connected to a
low-impedance measurement line, which we approximate at this stage as an ideal
ground termination. At \(x=0\), it is terminated by a voltage-biased Josephson
junction, as shown in Fig.~\ref{fig:circuit}. In the absence of Josephson
tunneling, the transmission-line modes therefore obey a Neumann boundary
condition at the junction and a Dirichlet boundary condition at the grounded
end. Their resonance frequencies are
\begin{equation}
	\begin{aligned}
		\omega_r &= r\omega_0, \\
		\omega_0 &= \frac{\pi v}{2l},
		\qquad r=1,3,5,\ldots .
	\end{aligned}
\end{equation}

\begin{figure}
	\centering
	\includegraphics{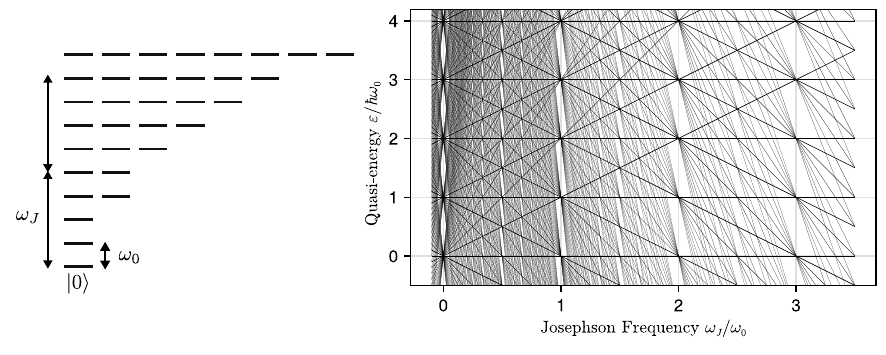}
	\caption{Free-state structure of the finite strip and its Floquet
		replicas. Left: the strip modes have frequencies
		\(\omega_0,3\omega_0,5\omega_0,\ldots\), generating degenerate
		multimode Fock-state manifolds separated by \(\hbar\omega_0\).
		A voltage-biased Josephson junction couples states whose energies differ
		by the energy \(\hbar\omega_J=2eV\) released per transferred Cooper pair,
		leading to resonances at
		\(\omega_J=m\omega_0\), with \(m=1,2,3,\ldots\).
		Right: free Floquet spectrum obtained by replicating the Fock spectrum
		in sectors shifted by \(-N\hbar\omega_J\), where the integer \(N\)
		counts the net number of Cooper pairs transferred across the junction.
		At the commensurate resonances, states from different Floquet sectors
		become degenerate. The Josephson coupling hybridizes these states and
		generates the dressed Floquet structure studied in this work.}
	\label{fig:fockstateladder}
\end{figure}

Because all mode frequencies are odd integer multiples of \(\omega_0\), the
multimode Fock states form degenerate manifolds at energies
\(n\hbar\omega_0\), with \(n=0,1,2,\ldots\). Their degeneracy grows rapidly
with \(n\). Remarkably, it is exactly equal to the degeneracy obtained by
filling free fermionic orbitals with energies \(n\hbar\omega_0\),
\(n=1,2,\ldots\), each orbital being occupied at most once. This correspondence
between the Fock space of the odd bosonic modes and a fermionic Fock space is
at the heart of the refermionization (see Fig.~\ref{fig:fockstateladder})~\cite{Anguelova2013}.

Including the Josephson coupling, the circuit Hamiltonian reads
\begin{equation}
	\begin{aligned}
		H(t) &= H_0 + H_J(t), \\
		H_0 &= \sum_{\substack{r>0\\ r\,{\rm odd}}}
		\hbar\omega_r a_r^\dagger a_r, \\
		H_J(t) &=
		-E_J\cos\!\left[\varphi(0)-\omega_J t\right],
	\end{aligned}
	\label{eq:Hlab}
\end{equation}
where the Josephson frequency is fixed by the dc voltage bias through
\(\hbar\omega_J=2eV\). The Josephson coupling is localized at the boundary and
couples the transmission-line modes through the superconducting phase
\(\varphi(0)\) across the junction. Expanding this boundary phase over the
normal modes gives
\begin{equation}
	\begin{aligned}
		\varphi(0)
		&= i\sum_{\substack{r>0\\ r\,{\rm odd}}} \beta_r
		\left(a_r-a_r^\dagger\right), \\
		\beta_r^2
		&= \frac{4\rho}{r},
		\qquad
		\rho=\frac{Z}{R_Q}.
	\end{aligned}
	\label{eq:cr}
\end{equation}
The dimensionless impedance \(\rho=Z/R_Q\) therefore controls the vacuum
phase fluctuations of the strip modes at the junction. Since the coupling to
mode \(r\) scales as \(\beta_r\propto\sqrt{Z/R_Q}\), appreciable multiphoton and
multimode processes naturally emerge when the line impedance becomes
comparable to \(R_Q\).

The voltage bias renders the Hamiltonian periodic,
\(H(t+T_J)=H(t)\), with \(T_J=2\pi/\omega_J\), so that the problem is
naturally formulated within Floquet theory. Introducing an auxiliary Floquet space spanned
by the states \(|N\rangle\), with
\begin{equation}
	\begin{aligned}
		\hat{N}|N\rangle &= N|N\rangle, \\
		T|N\rangle &= |N+1\rangle,
	\end{aligned}
	\label{eq:floquetShiftOperators}
\end{equation}
the quasienergies are obtained from the eigenvalues of the time-independent
Floquet Hamiltonian
\begin{align}
	\mathcal{H}_{F}
	&= H_0-\hbar\omega_J\hat{N}+\mathcal{H}_J, \\
	\mathcal{H}_J
	&= -\frac{E_J}{2}
	\left[
	T e^{i\varphi(0)}
	+T^\dagger e^{-i\varphi(0)}
	\right].
	\label{eq:circuitFloquetHamiltonian}
\end{align}
The Floquet index \(N\) has a direct physical interpretation as the net number
of Cooper pairs transferred across the junction. A forward tunneling event
increases \(N\) by one and releases an energy
\(\hbar\omega_J=2eV\) into the electromagnetic modes. Resonant tunneling
therefore occurs whenever
\(\omega_J\simeq m\omega_0\), with \(m=1,2,\ldots\).

Figure~\ref{fig:fockstateladder} illustrates the corresponding free Floquet
spectrum of \(H_0-\hbar\omega_J\hat{N}\). Each copy of the strip Fock spectrum
is shifted in energy by an integer multiple of \(\hbar\omega_J\). At the
commensurate points \(\omega_J=m\omega_0\), states belonging to different
Floquet sectors become degenerate, producing highly degenerate resonant
manifolds. The Josephson coupling hybridizes these states and lifts their
degeneracies, giving rise to dressed Floquet eigenstates. 

\subsection{Vertex operator}

Normal ordering of the operator \(e^{i\varphi(0)}\) gives
\begin{align}
	e^{i\varphi(0)}
	&= e^{-S/2}V, \\
	V = :e^{i\varphi(0)}:
	&=
	\exp\!\left(
	\sum_r \beta_r a_r^\dagger
	\right)
	\exp\!\left(
	-\sum_r \beta_r a_r
	\right),
\end{align}
where \(S=\sum_r \beta_r^2\). This sum diverges logarithmically for an ideal
transmission line and must be regularized by a physical high-frequency cutoff,
set by the junction capacitance, the superconducting gap, and other
high-frequency corrections to the circuit. Here, we focus on the regime in
which the normal modes of the linearized circuit, including all effects except
the Josephson inductance, remain well approximated by the odd-harmonic ladder
introduced above. This requires, in particular, a sufficiently small junction
capacitance. The cutoff-dependent vacuum renormalization is absorbed into
\(E_J^\ast=E_J e^{-S/2}\). The exact vertex identities used below refer to
the ideal-ladder scaling limit.

The operator \(V\) is the normal-ordered form of a multimode displacement
operator and is naturally identified as a vertex operator in the
refermionization language~\cite{vonDelftSchoeller1998,Anguelova2013}. In the interaction picture with respect to \(H_0\),
its time dependence can be written as
\begin{equation}
	V(z)=
	\exp\!\left[
	\sum_r \beta_r a_r^\dagger z^r
	\right]
	\exp\!\left[
	-\sum_r \beta_r a_r z^{-r}
	\right],
\end{equation}
where \(z=e^{i\omega_0 t}\). Using the standard algebra of displacement
operators, one obtains
\begin{align}
	V(u)V(v)
	&=
	F_\rho\!\left(\frac{v}{u}\right)
	:V(u)V(v): ,
	\\
	V(u)V^\dagger(v)
	&=
	F_\rho^{-1}\!\left(\frac{v}{u}\right)
	:V(u)V^\dagger(v): ,
	\\
	F_\rho(x)
	&=
	\exp\!\left[
	-4\rho
	\sum_{\substack{r>0\\ r\ {\rm odd}}}
	\frac{x^r}{r}
	\right]
	=
	\left(\frac{1-x}{1+x}\right)^{2\rho}.
\end{align}
The restriction to odd modes gives these relations a structure that differs
slightly from the more familiar vertex-operator formulas obtained when all
positive harmonics are present. In particular, as \(v\to u\), the vacuum
correlator has the short-distance singularity
\(\langle V(u)V^\dagger(v)\rangle\propto (u-v)^{-2\rho}\), showing that the
vertex operator has scaling dimension \(\rho\). The function \(F_\rho\) is closely related to the phase correlator entering \(P(E)\) theory: its Taylor expansion coefficients determine the weights
for transferring different amounts of energy to the electromagnetic
environment in a single Cooper-pair tunneling event
(see Appendix~\ref{app:spectral})~\cite{IngoldNazarov1992,vonDelftSchoeller1998}.

The free evolution of the vertex operator contains Fourier components at all
integer multiples of the fundamental frequency \(\omega_0\), and we may write
the Laurent expansion
\begin{equation}
	V(z)=\sum_{n\in\mathbb Z}V_n z^n .
\end{equation}
We define the energy-resolved coupling operators by
\begin{equation}
	A_n=-\frac{(-1)^n}{2}V_{-n},
	\qquad
	A_n^\dagger=-\frac{1}{2}V_n .
\end{equation}
These expressions are Hermitian conjugates because the strip contains
only odd harmonics, implying \(V_n^\dagger=(-1)^nV_{-n}\).
The Josephson term then takes the form
\begin{equation}
	\mathcal{H}_J
	=
	E_J^\ast
	\sum_{n\in\mathbb Z}
	\left(
	T A_n^\dagger + T^\dagger A_n
	\right).
	\label{eq:circuitFloquetHamiltonianMode}
\end{equation}

\subsection{Secular approximation}
The operators \(A_n\) are nonlinear multimode combinations of the original
bosonic creation and annihilation operators. Acting on a given Fock state,
\(A_n^\dagger\) generates a coherent superposition of states corresponding to
different redistributions of excitations among the strip modes, constrained
by a total increase in excitation energy of \(n\hbar\omega_0\). They therefore
provide a natural basis for describing resonant Cooper-pair tunneling in the
multimode system. 

For a bias close to the resonance \(\omega_J\simeq m\omega_0\), with \(m=1,2,\ldots\), the secular approximation
retains only the resonant \(n=m\) contribution, and the Josephson coupling in \eqref{eq:circuitFloquetHamiltonianMode} simplifies to
\begin{equation}
	T A_m^\dagger+T^\dagger A_m.
\end{equation}
An equivalent description is to consider that the secular approximation retains the Josephson
coupling within each resonant manifold that appears at exact resonance, $\omega_J=m\omega_0$, and neglects couplings between different manifolds (see Fig.~\ref{fig:fockstateladder}).
The approximation requires \(E_J^\ast\ll\hbar\omega_0\).

Within a given resonant manifold, the quantity
\(\ell=\sum_r r n_r-mN\) is fixed, so the photon energy uniquely
determines the Floquet index \(N\). The shift \(T\) accompanying
\(A_m^\dagger\) therefore only records the corresponding change
in \(N\) and can be omitted from the reduced description of the
strip dynamics. Keeping it explicit remains useful for counting
Cooper-pair transfers and evaluating the tunneling current.
Transforming to the frame where each mode \(r\) rotates at
\(r\omega_J/m\) gives the time-independent secular Hamiltonian
\begin{equation}
	H_{\rm sec}
	=
	\frac{-\delta}{m\omega_0}H_0
	+
	E_J^\ast\left(A_m+A_m^\dagger\right),
	\label{eq:Hrot}
\end{equation}
where \(\delta=\omega_J-m\omega_0\) is a small detuning,
\(|\delta|\ll\omega_0\). Equivalently, this Hamiltonian is obtained by eliminating \(N\)
from the diagonal Floquet term \(H_0-\hbar\omega_JN\) and
subtracting the constant offset \(\hbar\omega_J\ell/m\).
This offset fixes the position of each manifold in the Floquet
spectrum. Laboratory emission frequencies are recovered either
by retaining these offsets in the Floquet transition energies
or by restoring the phase \(e^{-ir\omega_Jt/m}\) of each photon
operator in a rotating-frame calculation.

\begin{figure}
	\centering
	\includegraphics[width=0.9\textwidth]{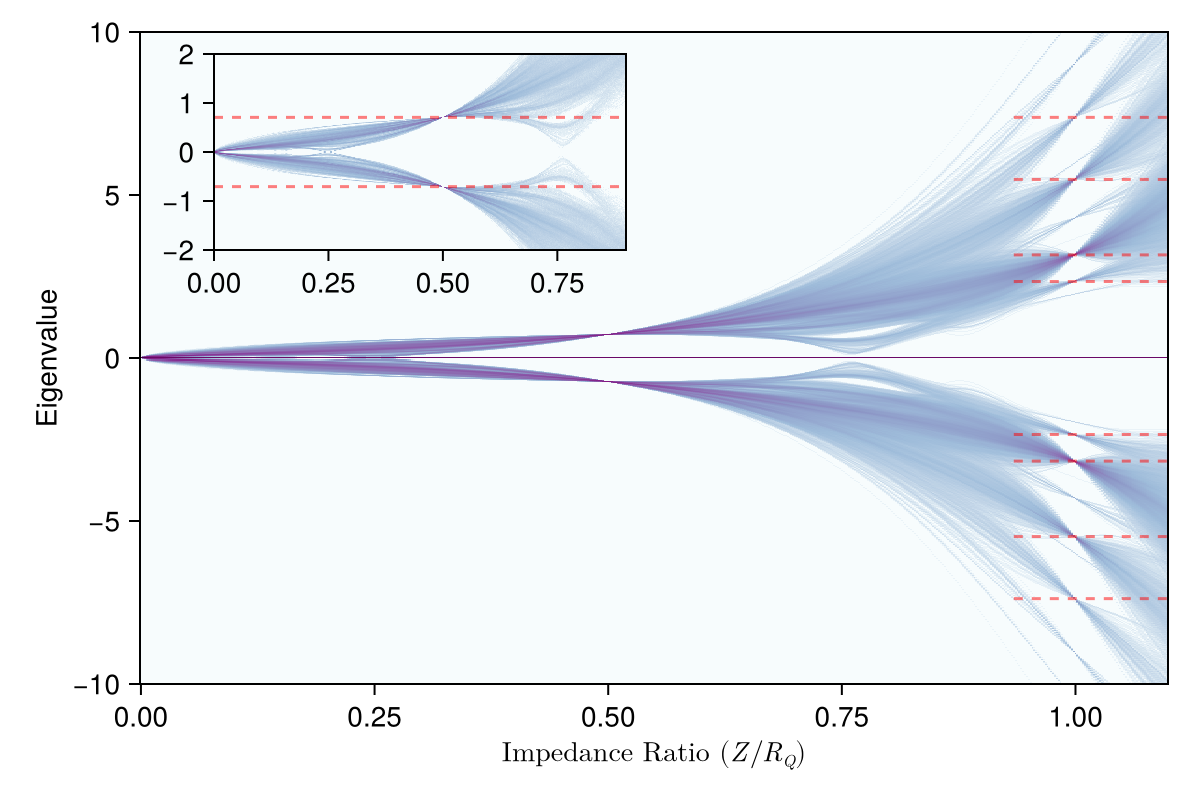}
	\caption{Eigenvalues of \(A_m^\dagger+A_m\) for \(m=10\) as a
		function of \(Z/R_Q\). The operator is projected onto the
		resonant subspace spanned by photon Fock states
		\(\ket{n_1,n_3,n_5,\ldots}\) with
		\(E_0=mN\hbar\omega_0\), where \(N=0,1,\ldots\) and
		\(E_0/(\hbar\omega_0)=n_1+3n_3+5n_5+\cdots\).
		The Fock space is truncated at \(E_0\leq60\hbar\omega_0\),
		and the spectrum is symmetric about zero.
		At \(Z=R_Q/2\), the nonzero eigenvalues collapse onto
		\(\pm1/\sqrt{2}\), reflecting the coupling of Cooper-pair
		tunneling to a single fermionic orbital (inset).
		At \(Z=R_Q\), the spectrum reduces to that of a harmonic-oscillator
		quadrature in a truncated space; the eigenvalues are
		\(\sqrt{2m}\) times the zeros of the Hermite polynomial
		\(H_d\), with \(d=1,\ldots,7\) for the
		chosen cutoff. Dashed lines correspond to selected roots with large degeneracies.}
	\label{fig:eigenvalues}
\end{figure}

At the special scaling dimensions \(\rho=1/2\) and \(\rho=1\), corresponding
respectively to \(Z=R_Q/2\) and \(Z=R_Q\), the vertex operator and the secular coupling \(A_m^\dagger+A_m\) admit an exact
fermionic representation, although of a different nature at the two points.
The \(\rho=1/2\) case is the free-fermion or Toulouse point
\cite{FendleySaleurWarner1994,AmeduriKonikLeClair1995,Konik1995,LeClairLesageSaleur1996},
whereas the \(\rho=1\) case is associated with a current-algebra representation
\cite{CallanKlebanovLudwigMaldacena1994,PolchinskiThorlacius1994}.
In the following, we adopt a pedestrian derivation, in the spirit of~\cite{vonDelftSchoeller1998}, keeping the strip-mode operators explicit
and avoiding any conformal-field-theory construction. The two refermionizable impedances stand out as special points where the spectrum of \(A_m^\dagger+A_m\) collapses onto a small number of highly degenerate eigenvalues as shown in Fig.~\ref{fig:eigenvalues}. At \(Z=R_Q/2\), only two eigenvalues remain, suggesting that a single fermionic orbital couples to the Cooper-pair tunneling degree of freedom. At \(Z=R_Q\), the spectrum forms a discrete ladder whose eigenvalues are given by the zeros of Hermite polynomials. This structure reflects the underlying bosonic nature of \(A_m+A_m^\dagger\), since the eigenvalues of a harmonic-oscillator quadrature in a truncated Hilbert space are also given by the zeros of Hermite polynomials. 

At resonance, the degenerate Floquet manifolds in Fig.~\ref{fig:fockstateladder} form a ladder with an energy spacing
$\hbar\omega_0$, smaller than the Floquet periodicity $\hbar\omega_J=m\hbar\omega_0$ and are not equivalent at first sight. For example, the spectrum in Fig.~\ref{fig:eigenvalues} is computed in the manifold that contains the vacuum. However, within the secular approximation, the spectrum, or more precisely the spectral support, of the Josephson coupling is identical in every manifold. The reason is that adding a
single photon in a very high mode, chosen so that its frequency equals
$\omega_0$ modulo $\omega_J$, moves a state from one manifold to the next while
leaving the junction essentially unaffected. This holds whenever the Josephson coupling is a
relevant perturbation, which is the case when $Z<R_Q$. Every dressed state of one manifold therefore
has approximate copies in all the others, and the secular quasienergy spectrum shown in Fig.~\ref{fig:eigenvalues}
repeats with the period $\hbar\omega_0$, rather than the Floquet period, except for the fine structures due to truncation. At $Z=R_Q/2$ and $Z=R_Q$ the same conclusion follows directly from the fermionic and bosonic algebras of the resonant operators. 

\subsection[Toulouse point]{Toulouse point $Z=R_Q/2$}
\label{sec:toulouse_floquet}
\begin{figure}
	\centering
	\includegraphics{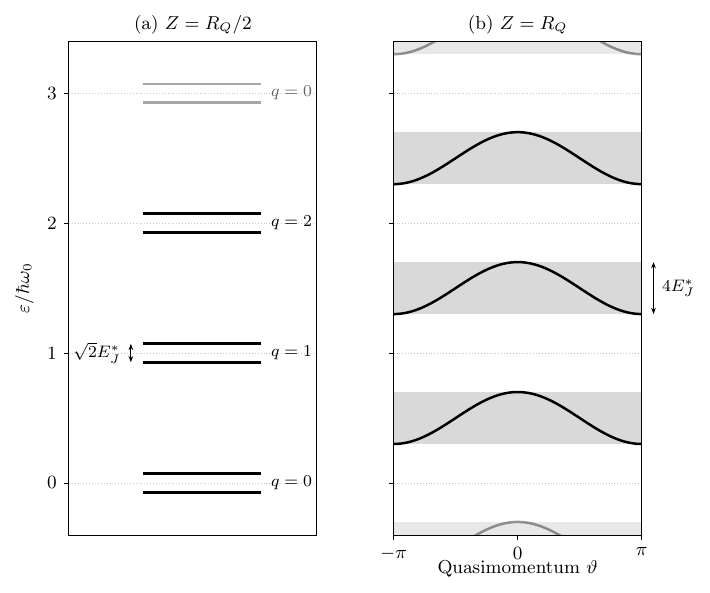}
	\caption{Quasienergy spectrum at resonance, $\omega_J=3\omega_0$, within the
		secular approximation ($E_J^\ast/(\hbar\omega_0)=0.1$).
		(a) At $Z=R_Q/2$, each resonant manifold $q$ splits into a doublet
		$q\hbar\omega_0\pm E_J^\ast/\sqrt2$. The doublets repeat every $\hbar\omega_0$,
		and the Floquet replica of $q=0$ appears at $3\hbar\omega_0$. 
		(b) At $Z=R_Q$, the one-body Floquet spectrum consists of minibands of width
		$4E_J^\ast$, which are eigenenergies of uniform tight-binding chains. The miniband centers lie at half-integer multiples of $\hbar\omega_0$.
		Note that panel (a) shows many-body
		quasienergies, while panel (b) shows only one-body quasienergies.}
\label{fig:quasienergy_spectrum}
\end{figure}

At \(Z=R_Q/2\), the vertex operator \(V\) has scaling dimension \(1/2\)
and can be represented in terms of a single fermionic field \(\psi(x)\).
With the field normalization left implicit, the Josephson term has the
schematic form
\begin{equation}
	\mathcal{H}_J = iE_J^\ast K \left[T\psi(0) + T^\dagger\psi^\dagger(0) \right],
\end{equation}
where \(K\) is a Majorana Klein factor. Since the fermionized form is used here only as a guide, we do not need the precise definitions of the fermion field and of the Klein factor. The fermionized form suggests that \(A_n\) is the Fourier mode of a single fermionic field, and therefore satisfies canonical anticommutation relations up to normalization. We show in Appendix~\ref{app:toulouse} that these operators indeed satisfy
\begin{equation}
	\{A_m,A_n^\dagger\}=\frac12\delta_{mn},\qquad m,n>0,
\end{equation}
as well as 
\begin{equation}
	\{A_m^\dagger,A_n^\dagger\} = \{A_m,A_n\} =0 \ {\rm for} \ m,n>0.
\end{equation}
These relations can be derived directly in the bosonic representation. The set of rescaled operators $c_n=\sqrt{2}\,A_n$, with $n>0$, obeys the canonical fermionic anticommutation relations. The Floquet Hamiltonian is then
\begin{equation}
	\begin{aligned}
	\mathcal H_F =
	\sum_{n>0}\hbar n\omega_0\,c_n^\dagger c_n
	-\hbar\omega_J\hat N & -\frac{E_J^\ast}{2}(T+T^\dagger)\mathcal P
	\\
	& +
	\frac{E_J^\ast}{\sqrt2}
	\sum_{n>0}
	\left[  T c_n^\dagger + T^\dagger c_n +  (-1)^n (  T c_n + T^\dagger c_n^\dagger) \right],
	\end{aligned}
\label{eq:HFloquetToulouse}
\end{equation}
where fermionic orbital indices \(n\) run over all positive integers. The operator
\(\mathcal P\) is the parity of the number of fermionic excitations; it
comes from the \(A_0\) term of the Josephson Hamiltonian,
\begin{equation}
	\mathcal P = V_0 = (-1)^{\sum_{n>0} c_n^\dagger c_n}.
\end{equation}
The two contributions inside the Josephson coupling term have a simple Floquet interpretation. The terms
$Tc_n^\dagger+T^\dagger c_n$ describe processes in which the transfer of a Cooper pair is accompanied by the creation or annihilation of a fermionic excitation, whereas
$Tc_n+T^\dagger c_n^\dagger$ are their counter-rotating counterparts. The secular approximation retains only the resonant term and \eqref{eq:HFloquetToulouse} simplifies to
\begin{equation}
	\mathcal H_{\rm sec}
	=
	\sum_{n>0}
	\hbar n\omega_0\,c_n^\dagger c_n
	-\hbar\omega_J\hat N
	+
	\frac{E_J^\ast}{\sqrt{2}}
	\left(
	Tc_m^\dagger
	+
	T^\dagger c_m
	\right).
	\label{eq:MajoranaSecularHamiltonian}
\end{equation}
All fermionic modes with \(n\neq m\) are spectators in this approximation.
Since the resonant mode is fermionic, its occupation can only be zero or one.
For any fixed configuration of the spectator modes, the Josephson term
therefore couples only the two states
\begin{equation}
	|\lambda,0;N\rangle
	\qquad\text{and}\qquad
	|\lambda,1;N+1\rangle,
\end{equation}
where \(\lambda\) denotes the occupations of all modes \(n\neq m\). Denoting by
\(E_\lambda\) the spectator energy, the secular Floquet Hamiltonian restricted to
this two-dimensional subspace is
\begin{equation}
	\mathcal H_{\lambda,N}
	=
	E_\lambda-\hbar\omega_JN
	+
	\begin{pmatrix}
		0 & E_J^\ast/\sqrt{2} \\
		E_J^\ast/\sqrt{2} & -\hbar\delta
	\end{pmatrix}.
	\label{eq:MajoranaSecularDoublet}
\end{equation}
Every spectator configuration thus produces the same avoided crossing, with
the two quasienergies
\begin{equation}
	\varepsilon_{\lambda,\pm}
	=
	E_\lambda-\hbar\omega_JN
	-\frac{\hbar\delta}{2}
	\pm
	\frac{1}{2}
	\sqrt{
		(\hbar\delta)^2
		+
		2(E_J^\ast)^2
	}.
	\label{eq:MajoranaSecularSpectrum}
\end{equation}
At exact resonance \((\delta=0)\), the quasienergy spectrum separates into
\(m\) families of doublets with centers separated by \(\hbar\omega_0\) (see Fig.~\ref{fig:quasienergy_spectrum}), each with
the splitting \(\sqrt2E_J^\ast=\hbar\Omega_R\).

\subsection[Bosonic point]{Bosonic point $Z=R_Q$}
\label{sec:bosonic_floquet}
At \(Z=R_Q\), the vertex operator has scaling dimension \(1\), and its
fermionic representation requires two fields, \(\psi_1(x)\) and
\(\psi_2(x)\). With their normalization left implicit, the Josephson term
has the schematic form
\begin{equation}
		\mathcal{H}_J = 2iE_J^\ast \Big[ T\psi_1^\dagger(0)\psi_2(0)
		- T^\dagger \psi_2^\dagger(0)\psi_1(0) \Big].
\end{equation}
The bilinear \(\psi_1^\dagger(0)\psi_2(0)\) transfers weight
between the two fermion fields at the boundary. This motivates an
\(SU(2)\) current-algebra representation, with transverse components
\(J_+(0)=\psi_1^\dagger(0)\psi_2(0)\) and
\(J_-(0)=\psi_2^\dagger(0)\psi_1(0)\), and a longitudinal component
\(J_z(0)=\frac12[\psi_1^\dagger(0)\psi_1(0)
-\psi_2^\dagger(0)\psi_2(0)]\). As shown in Appendix~\ref{app:bosonic}, the \(A_m\) operators
indeed obey the corresponding current-mode algebra,
\begin{equation}
	[A_m,A_n^\dagger]
	=
	m\delta_{mn}
	-
	\mathcal J_{n-m},
	\qquad m,n>0.
\end{equation}
The current-mode operator \( \mathcal J_\ell\) is defined for \(\ell\in\mathbb Z\) as follows.
For even \(\ell\), \(\mathcal J_\ell=0\), and for odd \(\ell\) we obtain \(\mathcal J_\ell=2\sqrt \ell \,a_\ell^\dagger\) for positive \(\ell\) and
\(\mathcal J_{\ell}=2\sqrt{-\ell}\,a_{-\ell}\) for negative \(\ell\). Consequently, within either the
positive even or the positive odd sector separately, the rescaled operators
\(A_m/\sqrt m\) obey canonical bosonic commutation relations. The Floquet Hamiltonian reads
\begin{equation}
	\mathcal H_F
	=
	\hbar\omega_0
	\sum_{\substack{r>0\\ r\ {\rm odd}}}
	A_r^\dagger A_r
	-
	\hbar\omega_J \hat N
	+
	E_J^\ast
	\sum_{n\in\mathbb Z}
	\left(
	T A_n^\dagger
	+
	T^\dagger A_n
	\right). \label{eq:HFloquetRQ}
\end{equation}
The free Hamiltonian involves only the positive odd \(A_r\) modes, whereas the Josephson term contains all Fourier components. For odd \(r\), the free evolution \(A_r(t)=e^{-ir\omega_0t}A_r\) follows directly from the diagonal form of \(H_0\). The even components likewise evolve as \(A_m(t)=e^{-im\omega_0t}A_m\), through their nontrivial commutation relations with the odd modes. Although the full algebra contains nontrivial couplings between the two parity sectors, its underlying structure remains essentially bosonic. The two terms $T A_n^\dagger+T^\dagger A_n$ describe processes in which Cooper-pair
transfer creates or annihilates a mostly bosonic excitation. As we show below, \(A_n^\dagger\) creates a coherent superposition of particle--hole excitations of a filled Fermi sea, each carrying energy \(n\hbar\omega_0\).

In the secular approximation, we retain only the resonant $n=m$ term in the Josephson coupling term. A Cooper-pair transfer therefore excites a collective oscillator with annihilation operator
\begin{equation}
	B_m=\frac{A_m}{\sqrt m},
	\qquad [B_m,B_m^\dagger]=1 .
	\label{eq:defBm}
\end{equation}
Each forward transfer adds one quantum to this oscillator and increases the transferred-pair number \(\hat N\) by one. The coherent secular dynamics thus conserves \(\hat N-B_m^\dagger B_m\). For a fixed value of this difference, the oscillator occupation determines the Floquet index. Eliminating this index
gives the secular Floquet Hamiltonian, equivalent to the collective-oscillator Hamiltonian in the rotating frame, 
\begin{equation}
	H_{\rm sec}
	=E_\lambda-\hbar\delta B_m^\dagger B_m
	+\sqrt m\,E_J^\ast(B_m^\dagger+B_m),
	\label{eq:HRWAdisplaced}
\end{equation}
where \(E_\lambda\) includes the spectator energy and the chosen Floquet offset.
For \(\delta\ne0\), the drive displaces the oscillator by \(\sqrt m\,E_J^\ast/(\hbar\delta)\). The Hamiltonian can be diagonalized with the exact quasienergies
\[
\varepsilon_{\lambda,n}
=E_\lambda-n\hbar\delta+\frac{m(E_J^\ast)^2}{\hbar\delta},
\qquad n=0,1,2,\ldots .
\]
At exact resonance, the Hamiltonian is proportional to the quadrature \(B_m+B_m^\dagger\), whose spectrum is continuous and unbounded. Truncating the oscillator to a finite number of Fock states gives discrete quadrature eigenvalues proportional to the zeros of a Hermite polynomial. This explains the discrete spectrum at \(Z=R_Q\) in Fig.~\ref{fig:eigenvalues}.

The current algebra underlying the \(A_m\) operators suggests expressing \(A_m^\dagger\) as a sum of fermion bilinears that create particle–hole excitations acting on a filled Fermi sea~\cite{PolchinskiThorlacius1994}. The calculation is detailed in Appendix~\ref{app:bosonic}.  In this new representation, the free Hamiltonian and the resonant tunneling operator $A_m^\dagger$ are
\begin{align}
	H_{0}&=\hbar\omega_0\sum_{j\in\mathbb Z+1/2}
	j\,{:}d_j^\dagger d_j{:},
	\label{eq:H0RefermionRWA}\\
	A_m^\dagger&=(-1)^{m+1}\sum_{j\in\mathbb Z+1/2}
	\eta_j d_j^\dagger d_{j-m},
	\qquad \eta_j=(-1)^{j+1/2},
	\label{eq:AmRefermionRWA}
\end{align}
where normal ordering is with respect to the filled sea. The fermion space is extended with respect to the original problem and Eq.~\eqref{eq:H0RefermionRWA} equals the free strip Hamiltonian 
only on the physical sector defined in the Appendix. 
Each term in \(A_m^\dagger\) promotes one fermion by \(m\) levels.  Acting on the filled Fermi sea, the tunneling operator therefore creates a coherent superposition of particle--hole excitations of energy \(m\hbar\omega_0\).

To describe the dynamics generated by this quadratic Hamiltonian, we consider the associated one-particle Floquet problem. Its basis states \(\lvert j,N\rangle_d\) are labeled by a fermionic orbital \(j\) and a Floquet index \(N\). A forward transfer connects \(\lvert j,N\rangle_d\) to \(\lvert j+m,N+1\rangle_d\), conserving \(q=j-mN\). The problem therefore separates into independent chains labeled by \(q\), whose sites are the states \(\lvert q+mN,N\rangle_d\), with one site for each \(N\). The diagonal energy at site \(N\) is
\[
\hbar\omega_0(q+mN)-\hbar\omega_JN
=\hbar\omega_0q-\hbar\delta N.
\]
The hopping matrix element from \(N\) to \(N+1\) is \(E_J^\ast s_{q,N}\), where \(s_{q,N}=(-1)^{m+1}\eta_{q+m(N+1)}\). We absorb these signs into the phases of the chain states by writing
\begin{equation}
	\lvert q,N\rangle_c=g_{q,N}\lvert q+mN,N\rangle_d,
	\qquad g_{q,N+1}=s_{q,N}g_{q,N},
	\label{eq:gaugeTransformationRQ}
\end{equation}
yielding a tight-binding Hamiltonian with uniform hopping,
\begin{equation}
	h_q=\sum_N(\hbar\omega_0q-\hbar\delta N)c_{q,N}^\dagger c_{q,N}
	+E_J^\ast\sum_N
	\bigl(c_{q,N+1}^\dagger c_{q,N}+\mathrm{H.c.}\bigr).
	\label{eq:RWAChain}
\end{equation}
The sectors \(q\) and \(q+m\) differ by a Floquet replica, leaving \(m\) inequivalent chains. Detuning produces a uniform energy step \(-\hbar\delta\) per transferred pair, giving a Wannier--Stark ladder on each chain. At resonance this step vanishes. A fermion can then propagate coherently along the chain, with quasimomentum \(\vartheta\) and one-particle dispersion
\begin{equation}
	\epsilon_q(\vartheta)=\hbar\omega_0q+2E_J^\ast\cos\vartheta .
	\label{eq:RWAminiband}
\end{equation}
Unlike at the Toulouse point, where the single active fermionic mode directly determines the splitting of the many-body Floquet doublets, the one-particle minibands obtained here (see Fig.~\ref{fig:quasienergy_spectrum}) do not directly describe the many-body spectrum, which is continuous and unbounded. Nevertheless, the one-particle minibands determine photon-emission frequencies and have a direct physical meaning.

\section[Secular master equation]{Secular master equation}
\label{sec:master}

The steady state in the strip results from the balance between the Josephson drive term  \(E_J^\ast(A_m+A_m^\dagger)\) and dissipation due to photon loss.
Losses arise from the measurement line connected at \(x=l\) and also
from intrinsic losses such as dielectric loss. We describe both baths
by a zero-temperature Lindblad master equation with one jump operator
per strip mode. The master equation for the density operator \(\varrho\) is
\begin{equation}
	\dot\varrho
	=
	-\frac{i}{\hbar}[H_{\rm sec},\varrho]
	+
	\sum_{\substack{r>0\\ r\ {\rm odd}}}\kappa_r\,\mathcal D[a_r]\varrho,
	\qquad
	\mathcal D[a]\varrho=a\varrho a^\dagger-\frac12\{a^\dagger a,\varrho\}.
	\label{eq:masterEquation}
\end{equation}
This description assumes $\kappa_r\ll\omega_0$, so that the strip modes decay independently, and that the spectral density of the output line is approximately constant over the scale $E_J^\ast/\hbar$, allowing us to use the bare mode operators as jump operators. For the circuit in Fig.~\ref{fig:circuit}, a low-impedance
termination \(Z_m\ll Z\) produces a mode-independent amplitude reflection
coefficient \(-1+2Z_m/Z\). The same fraction of energy is lost on each round trip, yielding a mode-independent coupling loss rate \(\kappa_c=(4/\pi)(Z_m/Z)\omega_0\). Other mechanisms, such as dielectric loss, produce loss rates proportional to the mode frequency and therefore correspond to a constant intrinsic quality factor. Denoting their contribution to the fundamental mode by \(\kappa_i\), the total loss rate of mode \(r\) is \(\kappa_r=\kappa_c+r\kappa_i\). When intrinsic losses dominate, the loss rate simplifies to \(\kappa_r=r\kappa_i\) and all modes share a common quality factor
\(Q=\omega_0/\kappa_i\). We refer to this limit as the constant-quality-factor limit.

\subsection{General steady-state properties}
The following general properties hold throughout the regime of validity of the secular approximation at any impedance and any finite detuning.
\begin{itemize}
	\item \emph{Current and energy conservation.} The Floquet index counts transferred Cooper pairs, and
	\(d\langle\hat N\rangle/dt\) gives the Cooper-pair rate
	\begin{equation}
		\Gamma_{\rm cp}
		=
		\frac{iE_J^\ast}{\hbar}\left\langle A_m-A_m^\dagger\right\rangle,
		\qquad
		I_{\rm cp}=2e\,\Gamma_{\rm cp},
		\label{eq:currentOperator}
	\end{equation}
	with the expectation value taken in the steady state of
	Eqs.~\eqref{eq:Hrot} and~\eqref{eq:masterEquation}. The
	energy balance between dissipation and power injected by the voltage source follows from
	\[
	\frac{d}{dt}\langle H_0\rangle
	=m\hbar\omega_0\Gamma_{\rm cp}
	-\sum_{\substack{r>0\\r\ {\rm odd}}}
	\hbar r\omega_0\kappa_r\langle a_r^\dagger a_r\rangle.
	\]
	Hence stationarity gives the energy balance
	\begin{equation}
		m\Gamma_{\rm cp}
		=\sum_{\substack{r>0\\r\ {\rm odd}}}
		r\kappa_r\langle a_r^\dagger a_r\rangle.
		\label{eq:energyBalance}
	\end{equation}
At finite detuning, this remains an exact balance for the bare strip energy \(H_0\), while full energy conservation includes the Josephson-coupling contribution \(\hbar\delta\Gamma_{\rm cp}\) to the dissipated power, giving a total \(IV=\hbar\omega_J\Gamma_{\rm cp}\).
	\item \emph{Coherent and incoherent emission.}
	We restrict attention to steady states that preserve the period of the Josephson drive, so the coherent mean field contains only integer harmonics of \(\omega_J\). This excludes subharmonic responses, such as those associated with parametric resonance, which could arise in the present system at much larger \(E_J^\ast\) in the semiclassical regime.
	Within the secular approximation, the mean field of mode \(r\) oscillates at \(r\omega_J/m\), so it can be nonzero only when \(r\) is a multiple of \(m\). Since photon-mode indices are odd, coherent response vanishes for even \(m\). Incoherent emission is not subject to this phase-locking condition and can occur around any of the odd-mode resonances \(r\omega_0\).
	\item \emph{Validity.} The secular approximation requires
	\(E_J^\ast/\hbar,|\delta|,\kappa_r\ll\omega_0\) for retained modes.
	Beyond-secular processes produce energy shifts of order
	\(E_J^{\ast2}/(\hbar\omega_0)\).
	Neglecting these shifts in a particular spectrum or coherence also
	requires \(E_J^{\ast2}/(\hbar\omega_0)\ll\hbar\gamma_{\rm line}\),
	where \(\gamma_{\rm line}\) is the relevant dissipative linewidth.
	This stronger constraint concerns the accuracy of line positions and fine structure at linewidth resolution, rather than the validity of the secular expansion itself.
\end{itemize}

\subsection[Weak drive limit]{Weak drive limit and $P(E)$ theory}
\label{sec:weak}

In the weak-drive limit, we can assume that the strip remains close to the photon vacuum and the steady state can be calculated as a small perturbation above vacuum to the second order in \(E_J^\ast\). The resulting Cooper-pair tunneling rate and photon emission correspond to those obtained by \(P(E)\) theory~\cite{IngoldNazarov1992,IngoldGrabertEberhardt1994}.

We briefly summarize the predictions of this model. Near the \(m\)th resonance, a single Cooper-pair transfer from the vacuum excites the manifold at energy \(m\hbar\omega_0\), spanned by the multimode Fock states \(\ket{f}=\ket{n_1^{(f)},n_3^{(f)},\ldots}\) satisfying \(\sum_r r n_r^{(f)}=m\). The no-jump evolution of \(A_m^\dagger\ket{0}\) determines the stationary coherences between the vacuum and this manifold. In the rotating frame, these are
\begin{equation}
	c_f\equiv\bra{f}\varrho_{\rm ss}\ket{0}
	\simeq
	\frac{E_J^\ast}{\hbar}
	\frac{\bra{f}A_m^\dagger\ket{0}}{\delta+i\Gamma_f/2},
	\qquad
	\Gamma_f=\sum_r\kappa_r n_r^{(f)},
	\label{eq:weakDriveAmplitude}
\end{equation}
where \(\Gamma_f\) is the total loss rate from \(\ket{f}\).

The first photon emission takes the system out of the resonant manifold. In the steady state, the rate of these first emissions balances the Cooper-pair tunneling rate, giving
\begin{equation}
	\Gamma_{\rm cp}
	=
	\sum_f\Gamma_f|c_f|^2
	=
	\frac{2E_J^{\ast2}}{\hbar^2}
	\sum_f w_f
	\frac{\Gamma_f/2}{\delta^2+\Gamma_f^2/4},
	\qquad
	w_f=\left|\bra{f}A_m^\dagger\ket{0}\right|^2 .
	\label{eq:weakDrive}
\end{equation}
The weights sum to \(W_m=\sum_f w_f=|f_{\rho,m}|/4\),
where \(f_{\rho,m}\) are the coefficients in the
Taylor expansion of \(F_\rho(x)\), which enters the vacuum
vertex correlator (see Appendix~\ref{app:spectral}).
Equation~\eqref{eq:weakDrive} is the lossy-mode form of the \(P(E)\) result. In the limit of vanishing linewidths, the Lorentzians reduce to the usual delta functions.
The Fock-state weights $w_f$ can be explicitly computed at any impedance as
\begin{equation}
w_f=\frac14\prod_s
\frac{(\beta_s^2)^{n_s^{(f)}}}{n_s^{(f)}!},
\qquad
\beta_s^2=\frac{4\rho}{s}. \label{eq:wf}
\end{equation}
Given the occupations of the $\ket{f}$ states, we obtain the strip-mode occupations from the entire loss cascade. Every cascade starting from \(\ket f\) releases \(n_r^{(f)}\) photons from mode \(r\), irrespective of their emission order. Since the injection rate associated with \(\ket f\) is \(\Gamma_f|c_f|^2\), the stationary photon flux from each mode satisfies
\begin{equation}
	\kappa_r\langle a_r^\dagger a_r\rangle
	=
	\sum_f n_r^{(f)}\Gamma_f|c_f|^2 .
	\label{eq:weakDriveModeFlux}
\end{equation}
This expression holds for arbitrary mode loss rates \(\kappa_r\) and includes losses through both the measurement port and intrinsic dissipation. It obeys the energy balance \(\sum_r r\kappa_r\langle a_r^\dagger a_r\rangle = m\Gamma_{\rm cp}\).

The cascade also determines the population of each intermediate Fock state. Writing the probability of occupation as \(p_{\boldsymbol n}=\bra{\boldsymbol n}\varrho_{\rm ss}\ket{\boldsymbol n}\), one starts from \(p_f\simeq |c_f|^2\) in the resonant manifold and proceeds in order of decreasing energy. For each nonvacuum state below that manifold, the stationary balance gives
\begin{equation}
	\left( \sum_r\kappa_r n_r \right) p_{\boldsymbol n}
	=\sum_r \kappa_r(n_r+1)\,p_{\boldsymbol n+\boldsymbol e_r},
	\label{eq:weakDriveCascade}
\end{equation}
where \(\boldsymbol e_r\) adds one photon to mode \(r\), and populations above the directly excited manifold vanish to this order.

\paragraph{Constant-quality-factor limit.} For a constant quality factor, the loss cascade gives a closed expression for the occupation of each strip mode. All states in the manifold have the same linewidth \(\Gamma_f=\kappa_m\) and Eq.~\eqref{eq:weakDriveModeFlux} gives the photon yield for mode $r$
\[
Y_r = 
\frac{\kappa_r\langle a_r^\dagger a_r\rangle}{\Gamma_{\rm cp}}
=
\frac{\sum_f n_r^{(f)}\Gamma_f|c_f|^2}
{\sum_f \Gamma_f|c_f|^2}
=
\frac{1}{W_m}\sum_f n_r^{(f)}w_f .
\]
Using Eq.~\eqref{eq:wf}, one sees that counting photons in mode \(r\) reduces the spectral-weight sum to that for the remaining energy \((m-r)\hbar\omega_0\), giving the yield
\begin{equation}
	Y_r=\frac{4\rho}{r}\frac{W_{m-r}}{W_m},
	\qquad
	W_0=\frac14,\qquad W_k=0\ \text{for }k<0 .
	\label{eq:weakDrivePhotonYield}
\end{equation}
Only odd modes \(r\leq m\) are populated at this order, with steady-state occupations \(\langle a_r^\dagger a_r\rangle=\Gamma_{\rm cp}Y_r/\kappa_r\). At the two special impedances, the following results then emerge.
\begin{itemize}
	\item \emph{Toulouse point, \(Z=R_Q/2\).}
	The spectral weight is \(W_m=1/2\) for every \(m\geq1\), so the resonant tunneling rate and the tunnel current decrease with voltage as $1/V\propto1/m$,
	\begin{equation}
		\Gamma_{\rm cp}=\frac{2 E_J^{\ast 2}}{m \hbar^2 \kappa_i}.\label{eq:Gamma_cp_1_P(E)}
	\end{equation}
	 The photon yield is \(Y_r=2/r\) for each odd \(r<m\), with \(Y_m=1/m\) when \(m\) is odd. Thus all odd modes below \(m\) carry equal dissipated power at resonance, although the photon flux is largest in the fundamental mode. For every \(m>1\), two fundamental photons are lost per transferred Cooper pair on average.
	
	\item \emph{Bosonic point, \(Z=R_Q\).}
	Here, the spectral weight \(W_m=m\) compensates the growth of \(\kappa_m\), so the resonant tunneling rate is independent of voltage, 
	\begin{equation}
		\Gamma_{\rm cp}=\frac{4 E_J^{\ast 2}}{\hbar^2 \kappa_i}.\label{eq:Gamma_cp_2_P(E)}
	\end{equation}
	The photon yield is \(Y_r=4(m-r)/(mr)\) for each odd \(r<m\), with \(Y_m=1/m^2\) when \(m\) is odd. The dissipated power in the modes below \(m\) is therefore proportional to \(m-r\) at resonance and is largest in the fundamental mode.
\end{itemize}

\paragraph{Dephasing.} In order to account for decoherence between states that differ in the number of transferred Cooper pairs, we introduce a dephasing rate \(\Gamma_\phi\) between Floquet states with different $N$ (see Sec.~\ref{sec:RQ_dephased}). Dephasing increases the coherence decay rate in
Eq.~\eqref{eq:weakDriveAmplitude} from \(\Gamma_f/2\) to \(\gamma_f\).
The resulting tunneling rate is
\begin{equation}
	\Gamma_{\rm cp}
	=
	\frac{2E_J^{\ast2}}{\hbar^2}
	\sum_f w_f
	\frac{\gamma_f}{\delta^2+\gamma_f^2},
	\qquad
	\gamma_f=\Gamma_\phi+\frac{\Gamma_f}{2}.
	\label{eq:weakDriveDephased}
\end{equation}
Each term in Eq.~\eqref{eq:weakDriveDephased} gives the injection rate
into \(\ket f\). Dividing this rate by \(\Gamma_f\) gives the
steady-state population of that state. When dephasing dominates photon
loss, all contributions share the same half-width \(\Gamma_\phi\), so
the relative injection rates are set by the weights \(w_f\). The
resulting photon yields are given by Eq.~\eqref{eq:weakDrivePhotonYield}
independently of the mode loss rates.

\paragraph*{}These results assume that the strip remains close to the vacuum throughout the tunneling and relaxation cycle. The population of the directly excited manifold must remain much smaller than unity. We now turn to the description of the strongly driven steady state, where this population becomes appreciable, while retaining the same resonant tunneling processes as in \(P(E)\) theory.

\section[Toulouse point steady state]{Secular steady state at the Toulouse point $Z=R_Q/2$}
\label{sec:secular_toulouse}

At the Toulouse point, the secular Hamiltonian in the rotating frame \eqref{eq:Hrot} becomes
\begin{equation}
	H_{\rm sec}
	=
	-\frac{\hbar\delta}{m}\sum_{n>0}n\,c_n^\dagger c_n
	+
	\frac{E_J^\ast}{\sqrt2}\left(c_m+c_m^\dagger\right).
	\label{eq:HrotToulouse}
\end{equation}
Each Cooper-pair transfer
fills orbital $m$, and the next transfer requires it to be emptied by photon
emission.  The emission processes
follow from the fermionic photon operators in
Eq.~\eqref{eq:PhotonOperatorCanonicalFermions} of Appendix~\ref{app:toulouse}.
Each \(a_r\) can move a fermion down by \(r\) orbitals, annihilate a pair
whose indices sum to \(r\), or annihilate a single fermion in orbital \(r\):
\begin{equation}
	a_r
	=
	\sqrt{\frac{2}{r}}
	\sum_{n\geq1}
	c_n^\dagger c_{n+r}
	+
	\sqrt{\frac{2}{r}}
	\sum_{n=1}^{(r-1)/2}
	(-1)^n c_nc_{r-n}
	-
	\frac{\mathcal P c_r}{\sqrt r}.
\end{equation} 
A fermion in the resonant orbital \(m\) can leave it by hopping to an
empty lower orbital \(p<m\), by pair annihilation with an occupied
orbital \(p>0\), or by direct annihilation when \(m\) is odd.
Only spectator orbitals of parity opposite to \(m\) participate in the
hop and pair channels. For a definite spectator configuration \(\lambda\),
the photon-jump rate for emptying an initially occupied orbital \(m\) is
\begin{equation}
	\gamma_m(\lambda)=
	\sum_{\substack{0<p<m\\p+m\ {\rm odd}}}
	\frac{2\kappa_{m-p}}{m-p}(1-n_p)
	+\sum_{\substack{p>0\\p+m\ {\rm odd}}}
	\frac{2\kappa_{m+p}}{m+p}n_p
	+\frac{\kappa_m}{m}\delta_{m\ {\rm odd}}.
	\label{eq:gammaM}
\end{equation}
Both sums run over orbitals \(p\) of parity opposite to \(m\) and
\(n_p\in\{0,1\}\) are the spectator occupations in \(\lambda\).
For \(p<m\), occupying an opposite-parity orbital blocks hopping into it but opens a pair-annihilation channel.

At the fundamental resonance \(m=1\), no spectator can be created: orbital 1
can only be emptied by the term \(\mathcal Pc_1\) of \(a_1\), at rate
\(\kappa_1\). The dynamics is then exactly that of a driven two-level system,
and
\begin{equation}
	\Gamma_{\rm cp}
	=
	\kappa_m\,
	\frac{E_J^{\ast2}/2}{\hbar^2\delta^2+\hbar^2\kappa_m^2/4+E_J^{\ast2}} \qquad (m=1).
	\label{eq:currentM1}
\end{equation}
In terms of the original modes, the two states are \(\ket{0}\) and the
one-photon state \(\ket{1}\) of the fundamental mode. The transition to
\(\ket{2}\) is forbidden because the Franck--Condon matrix element
\(\bra{2}:e^{i\varphi(0)}:\ket{1}\propto 2-4\rho\) vanishes at \(\rho=1/2\). 
The associated single-mode Franck--Condon blockade mechanism was studied
theoretically~\cite{Gramich2013,Dambach2015} and underlies the photon antibunching
observed in a high-impedance resonator~\cite{Rolland2019}.

At higher resonances ($m>1$), the two-level structure remains exact for the excitation path: only orbital $m$ is pumped by the drive. The decay channels become more complicated, but saturation of the resonant transition remains.

\subsection{Constant-quality-factor limit: optical Bloch equations for the resonant orbital}
\label{sec:QF}

In this limit, the non-Hermitian no-jump term is proportional to \(H_0\), hence
diagonal in the fermionic occupation basis. In Eq.~\eqref{eq:HrotToulouse}
it amounts to
\[
\delta\longrightarrow\delta+i\,m\kappa_i/2.
\]
Apart from the drive on \(m\), the no-jump generator preserves orbital
occupations, and each photon jump moves fermions downward or annihilates
them. Starting from the vacuum, the dynamics remains in orbitals \(1,\ldots,m\).
Within this subspace, the hop and pair rates in Eq.~\eqref{eq:gammaM}
become equal,
\begin{equation}
	\frac{2\kappa_{m-p}}{m-p}=\frac{2\kappa_{m+p}}{m+p}=2\kappa_i,
\end{equation}
and we obtain
\begin{equation}
	\gamma_m(\lambda)=m\kappa_i=\kappa_m.
\end{equation}
The equality of the emptying rates for all occupation configurations 
and the restriction of the dynamics to the lowest $m$ orbitals allow us to derive the closed equations (see Appendix~\ref{app:BlochToulouseConstantQF})
\begin{align}
	\frac{d}{dt}\langle c_m\rangle
	&=
	\left(i\delta-\frac{\kappa_m}{2}\right)\langle c_m\rangle
	+\frac{iE_J^\ast}{\sqrt2\hbar}\left(2\langle n_m\rangle-1\right),\label{eq:Bloch_c_m}
	\\
	\frac{d}{dt}\langle n_m\rangle 
	&=
	\frac{iE_J^\ast}{\sqrt2\hbar}\left\langle c_m-c_m^\dagger\right\rangle
	-\kappa_m\langle n_m\rangle . \label{eq:Bloch_n_m}
\end{align}
These are optical Bloch equations for the resonant coherence and
occupation, with Rabi frequency \(\Omega_R=\sqrt 2 E_J^\ast/\hbar\), detuning \(\delta\),
and decay rate \(\kappa_m\). The dynamics of the other orbitals requires numerical simulations.

 The first term in the equation for $\langle n_m\rangle$ describes the population injected into the resonant orbital by Cooper-pair tunneling. According to Eq.~\eqref{eq:currentOperator}, this term is precisely the net Cooper-pair transfer rate $\Gamma_{\rm cp}$. In the steady state, this injection is balanced by photon-induced decay of the orbital population at rate $\kappa_m\langle n_m\rangle$. We therefore obtain $\Gamma_{\rm cp}=\kappa_m\langle n_m\rangle$, yielding
\begin{equation}
	\Gamma_{\rm cp}
	=
	\kappa_m\,
	\frac{E_J^{\ast2}/2}{\hbar^2\delta^2+\hbar^2\kappa_m^2/4+E_J^{\ast2}}.
	\label{eq:QFcurrentToulouse}
\end{equation}
This closed form holds at every resonance and extends the $m=1$ result of Eq.~\eqref{eq:currentM1}. Only the resonant orbital enters the current. The other orbitals are fed by the cascade and remain correlated with it, but they never act back on \(\langle c_m\rangle\) and \(\langle n_m\rangle\). 
Their occupations do not obey closed rate
equations: a photon jump can map one configuration onto a coherent
superposition of degenerate configurations. We therefore compute the full
stationary density operator on this subspace by solving the sparse linear
problem $\mathcal L\varrho_{\rm ss}=0$ of
Eq.~\eqref{eq:ToulouseSteadyStateNumerics} (Appendix~\ref{app:BlochToulouseConstantQF}), which remains tractable up to $m\simeq10$. The lowest mode population can however be derived analytically and computed at any $m$.

\section[Bosonic point steady state]{Secular steady state at the bosonic point \(Z=R_Q\)}
\label{sec:secular_RQ}

At \(Z=R_Q\), successive Cooper-pair tunneling events are not Pauli blocked. In the chain picture, $A_m^\dagger$ is a sum of single-fermion hops by $m$ levels. After one hop, other fermions can still be promoted, so successive transfers are allowed.
In steady state, each net transfer supplies \(m\hbar\omega_0\), and photon emission removes the same energy on average. Finding the steady state requires following the loss channels and the subsequent evolution of the populated states, which is more intricate than at the Toulouse point.  For odd \(r>0\), the photon annihilation operator lowers a fermion by \(r\) levels with the same amplitude at every site (Appendix~\ref{app:bosonic}),
\begin{equation}
	a_r=\frac{1}{\sqrt r}\sum_j d_j^\dagger d_{j+r}.
	\label{eq:arRefermionRWA}
\end{equation}
This move changes the chain label from \(q\) to \(q-r\). Since \(r\) is odd, the Josephson hopping signs on these two chains satisfy \(s_{q-r,N}=-s_{q,N}\). The gauge choice~\eqref{eq:gaugeTransformationRQ} therefore gives the photon matrix element a factor \((-1)^N\), up to a phase independent of \(N\). In the miniband basis this factor shifts the quasimomentum by \(\pi\),
\begin{equation}
	a_r=\frac{1}{\sqrt r}\sum_{q,\vartheta}
	c_{q-r,\vartheta+\pi}^\dagger c_{q,\vartheta}.
	\label{eq:arMiniband}
\end{equation}
Here \(q\) runs over all half-integer chain labels, and the relative
phases of the chain states are chosen to remove the constant
phase factors. The squared matrix element of each allowed one-particle transition is \(1/r\).
At the Toulouse point, a constant quality factor yields closed equations for the driven orbital. At \(Z=R_Q\), however, photon loss couples different fermionic chains, and their evolution must be treated jointly.

\subsection{Constant-quality-factor and Gaussian approximation}
\label{sec:RQ_gaussian}

\begin{figure}
 	\centering
 	\includegraphics[width=0.9\textwidth]{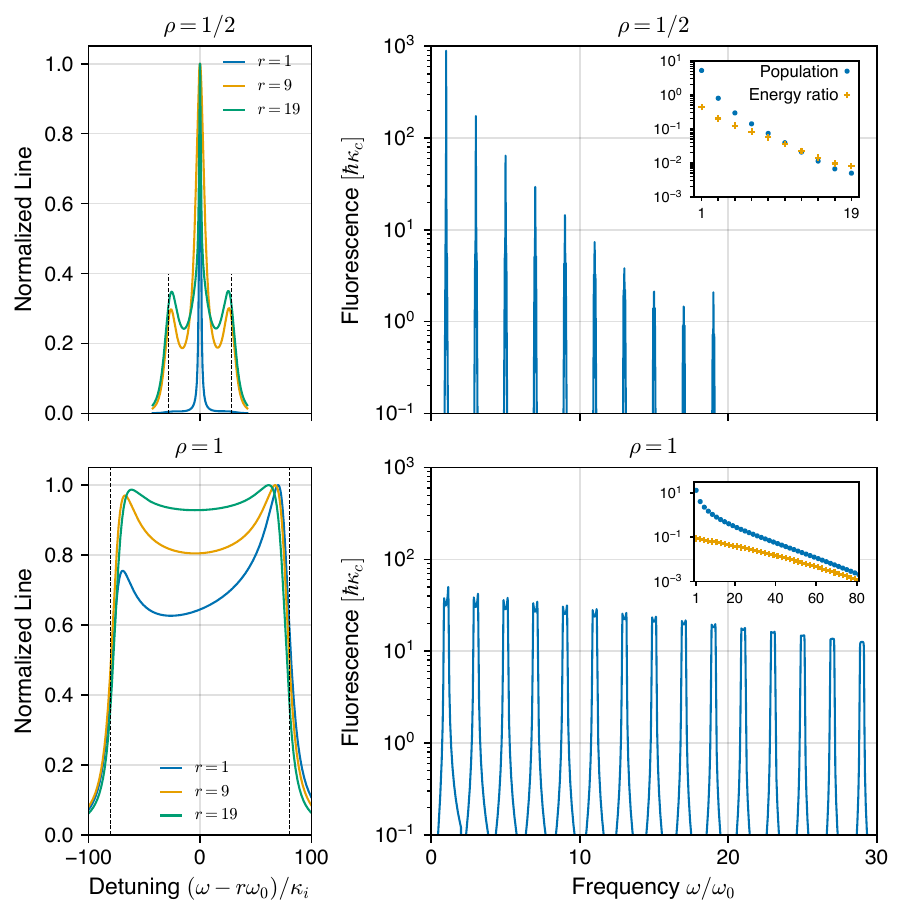}
	\caption{Fluorescence spectrum at the \(m=10\) resonance for \(\rho=1/2\)
		(upper row) and \(\rho=1\) (lower row). Fluorescence lines appear at the
		frequencies \(r\omega_0\) of the free modes, \(r=1,3,5,\ldots\).
		The left column shows the lineshapes of three selected lines,
		each normalized to its own maximum. Each line is a Mollow-like
		triplet at \(\rho=1/2\) and an emission band at \(\rho=1\).
		The dashed lines indicate $\pm \sqrt{2} E_J^\ast/\hbar$ at $\rho=1/2$ and $\pm 4E_J^\ast/\hbar$ at $\rho=1$.
		The right column shows the fluorescence spectrum, assuming a very small and frequency independent coupling rate $\kappa_c$. The relative line intensity at a given $r$ is proportional to the fraction of the total strip energy stored in the mode (orange cross in the insets). The blue dots show the mode population. 
		At \(\rho=1/2\), no mode with \(r\geq 2m\) is populated, whereas at \(\rho=1\) the emission extends well beyond.
		All simulations use the secular approximation at exact resonance, with a
		constant quality factor \(Q=500\) and \(E_J^\ast/(\hbar\kappa_i)=20\).
		Frequency shifts  beyond the secular approximation of order \(E_J^{\ast 2}/(\hbar^2 \omega_0)\), which may not be small compared to $\kappa_i$,  are neglected.}
 	\label{fig:fluo}
 \end{figure}

To obtain quantitative estimates of the steady state, we consider a minimal Gaussian model in which we retain all fermionic orbital occupations and coherences in the one-body density matrix
\(C_{jj'}=\langle d_{j'}^\dagger d_j\rangle\) (Appendix~\ref{app:RQ_gaussian}). The quadratic Hamiltonian gives a closed equation for \(C\), but photon loss couples its evolution
to four-fermion averages. We assume that the fermionic state remains
Gaussian and evaluate these averages at each time by Wick factorization.
This retains Pauli blocking and exchange while neglecting the connected
four-fermion correlations generated by loss.
The resulting master equation can be solved numerically. The model applies to any loss rates $\kappa_r$, but we present results only for the constant quality factor case. The appendix explains how to extract the tunneling current and strip-mode populations from its stationary solution that we show in figures~\ref{fig:fluo} and~\ref{fig:bosonic_current}.

\paragraph{Strongly driven regime} In the weakly driven limit, $P(E)$ theory predicts a resonant tunneling rate independent of $m$ scaling as $E_J^{\ast 2}$. For a constant quality factor, we can estimate strongly driven effects from the dynamics of the fermionic chains beyond $P(E)$. A quasiparticle keeps climbing the miniband at each tunneling event, until it emits. It moves as a Bloch wave with quasimomentum $\vartheta$ and group velocity $(2E_J^\ast/\hbar) \sin \vartheta$ sites per unit time. Photon emission moves the quasiparticle down by \(r\) levels. Its squared matrix element is \(1/r\) [Eq.~\eqref{eq:arMiniband}], so each available final state contributes a rate \(\kappa_r/r=\kappa_i\). A particle \(j\) levels above the Fermi surface has of order \(j\) possible downward moves, subject to the occupation of the final levels. Its estimated emission rate therefore grows as \(\kappa_i j\).

During an upward excursion at a speed of order \(m E_J^\ast/\hbar\) in level index, the cumulative emission probability becomes appreciable when \(\int_0^{j_\ast}\kappa_i j\,\hbar\,\mathrm{d}j/(m E_J^\ast)\) is of order unity. This sets the characteristic width of the excited region around the Fermi surface,
\begin{equation}
	j_\ast\sim\sqrt{\frac{mE_J^\ast}{\hbar\kappa_i}} .
	\label{eq:climbHeightRQ}
\end{equation}
If this region contains of order \(j_\ast\) mobile particles and holes, each contributing a pair-transfer rate of order \(E_J^\ast/\hbar\) and storing an energy of order \(j_\ast\hbar\omega_0\), the resulting strong-drive estimates are
\begin{equation}
	\Gamma_{\rm cp}\approx \sqrt m\,
	\frac{(E_J^\ast/\hbar)^{3/2}}{\sqrt{\kappa_i}},
	\qquad
	\frac{\langle H_0\rangle}{\hbar\omega_0}
	=\sum_r r\langle a_r^\dagger a_r\rangle
	\approx \frac{mE_J^\ast}{\hbar\kappa_i}.
	\label{eq:strongDriveRQ}
\end{equation}
These estimates are consistent with the energy balance \(m\Gamma_{\rm cp}=\kappa_i\sum_r r^2\langle a_r^\dagger a_r\rangle\) when the energy-weighted mean mode index is of order \(j_\ast\).
The excited band therefore spreads to a height of about $j_\ast$ levels above and below the Fermi level.

 \section{Fluorescence spectrum}
 
 \subsection{Mollow triplets at the Toulouse point}
 \label{sec:toulouse_triplet}
 
 Fluorescence arises from transitions between the dressed 
 states of Eq.~\eqref{eq:MajoranaSecularDoublet} shown in Fig.~\ref{fig:quasienergy_spectrum}(a). The emission
 frequencies are given by quasienergy differences, shifted by
 integer multiples of \(\omega_J\). The photon-operator matrix
 elements determine which transitions are allowed.
 At resonance, the allowed transitions in mode \(r\)
 connect doublets whose centers differ by \(r\hbar\omega_0\).
 Transitions between the same branches produce the central line,
 while transitions between opposite branches produce sidebands
 separated from it by the doublet splitting \(\Omega_R\).
The fluorescence spectrum is thus a series of Mollow triplets~\cite{Mollow1969} at frequencies
 \begin{equation}
 	\omega=r\omega_0,\qquad \omega=r\omega_0\pm\Omega_R,\qquad r=1,3,\ldots
 	\label{eq:MajoranaMollowFrequencies}
 \end{equation}
 Figure~\ref{fig:fluo} shows the simulated triplet lineshapes, which are obtained through the quantum regression theorem applied to the stationary state (Appendix~\ref{app:BlochToulouseConstantQF}).

\paragraph{Constant-quality-factor and saturation limit.} Fluorescence is concentrated around the modes with \(r<2m\), when orbitals above \(m\) are weakly populated. Indeed, photon emission annihilates at most two fermions, whose orbital indices sum to at most \(2m-1\) if neither exceeds \(m\). For a constant quality factor, orbitals above \(m\) remain exactly empty and only bosonic modes \(r\le2m-1\) are occupied, which explains the sharp cutoff in Fig.~\ref{fig:fluo}. The highest active mode, $r=2m-1$, then shows perfect zero-delay antibunching at any drive. Every
emission into this mode empties the driven orbital $m$, $a_{2m-1}\propto c_{m-1}c_m$, and no other emission path exists. Because of Pauli blocking, a second photon can only be emitted after the drive has refilled orbital $m$. The lower modes are not necessarily antibunched because they can also be emitted through transitions among the spectator orbitals.
 
In the saturation regime at resonance, orbital $m$ is half occupied, and a
constant quality factor fixes its emptying rate at $m\kappa_i$ independently
of the spectator occupations, so that $\Gamma_{\rm cp}=m\kappa_i/2$
[Eq.~\eqref{eq:QFcurrentToulouse}]. The analytical formula in Eq.~\eqref{eq:fundamentalFiniteDrivePopulation} for the lowest mode occupation 
 gives $\langle a_1^\dagger a_1\rangle=m/2$ . The loss rate in the fundamental mode,
$\kappa_1\langle a_1^\dagger a_1\rangle=m\kappa_i/2$, therefore equals
$\Gamma_{\rm cp}$: on average, one fundamental-mode photon is lost per
transferred Cooper pair, whereas the weak-drive regime of $P(E)$ theory gives
two per pair for $m>1$. Numerical solutions of the master equation show that the energy stored per mode decreases with the mode index (inset of Fig.~\ref{fig:fluo}). The fluorescence is thus stronger around the fundamental mode. 

\paragraph{Coherent emission} Coherent emission only appears when $m$ is odd, in which case only \(\langle a_m\rangle\) is allowed to be nonzero. On the invariant subspace, its evolution equation is
\begin{equation}
	\frac{d}{dt}\langle a_m\rangle
	=\left(i\delta-\frac{\kappa_m}{2}\right)\langle a_m\rangle
	+\frac{iE_J^\ast}{\hbar\sqrt{2m}}\langle\mathcal P\rangle
	\qquad(m\ {\rm odd}).
	\label{eq:QFcoherentMean}
\end{equation}
The steady-state value follows directly; it depends on the average parity
\(\langle\mathcal P\rangle\), which is not given by the closed equations for the driven orbital $m$. At weak drive \(\langle\mathcal P\rangle \approx 1\), and, when $E_J^\ast$ increases, \(\langle\mathcal P\rangle\) tends to zero as the transition saturates and the coherent emission vanishes. 

 \subsection{Fluorescence bands at the bosonic point}
 
 At $\rho=1$, photon emission moves a fermion from an occupied orbital to an empty one.
 All other occupations remain unchanged, so the many-body transition energy
 is the difference between the initial and final one-particle quasienergies, which was not the case at the Toulouse point.
 At resonance, the laboratory emission frequency is thus
 \begin{equation}
 	\omega_{\rm em}
 	=\frac{\epsilon_q(\vartheta)-\epsilon_{q-r}(\vartheta+\pi)}{\hbar}
 	=r\omega_0+\frac{4E_J^\ast}{\hbar}\cos\vartheta .
 	\label{eq:FluorescenceFrequencyRQ}
 \end{equation}
 The allowed transitions in mode \(r\) therefore span a band of total width \(8E_J^\ast/\hbar\) centered at \(r\omega_0\). Its edges come from the miniband dispersion (see Fig.~\ref{fig:quasienergy_spectrum}). The stationary particle and hole occupations determine where the emission weight lies within the band, and dissipation broadens the individual transitions. The lineshapes shown in Fig.~\ref{fig:fluo} are obtained from the stationary Gaussian state of Eq.~\eqref{eq:GaussianClosureRQ} by propagating the two-time fermion contractions with the stationary effective one-body Hamiltonian, an additional factorization detailed in Appendix~\ref{app:RQ_gaussian}.

  \paragraph{Constant-quality-factor limit.} Emission into mode $r$ moves a quasiparticle down by $r$ levels. As shown above, within the constant-quality-factor approximation, every allowed odd jump occurs at the same rate $\kappa_i$, regardless of its size. An excited particle can drop into any empty level below it, including holes deep below the Fermi level. Particle--hole separations up to a few \(j_\ast\) on each side of the Fermi level thus produce photons with $r$ of several \(j_\ast\), with no penalty from the matrix elements. As a result, one fluorescence photon can carry the energy of several pairs and the emission spectrum extends to modes with $r \gg m$ (see Fig.~\ref{fig:fluo}). 
  
In the strongly driven regime, we have shown above that the energy stored in the modes grows as 
\begin{equation}
\frac{\langle H_0\rangle}{\hbar\omega_0}
=\sum_r r\langle a_r^\dagger a_r\rangle
\approx \frac{mE_J^\ast}{\hbar\kappa_i}.
\end{equation}
At strong drive, numerical solutions of the Gaussian kinetic model show that the energy per mode decreases more slowly with mode index than at the Toulouse point. The fundamental-mode population is approximately
\begin{equation}
	\langle a_1^\dagger a_1 \rangle \approx 0.92 \sqrt{\frac{m E_J^\ast}{\hbar \kappa_i}}. \label{eq:RQ_pop}
\end{equation}
 
 \section{Cooper-pair mean current and second-order correlation function}

 \begin{figure}
 	\centering
 	\includegraphics[width=0.9\textwidth]{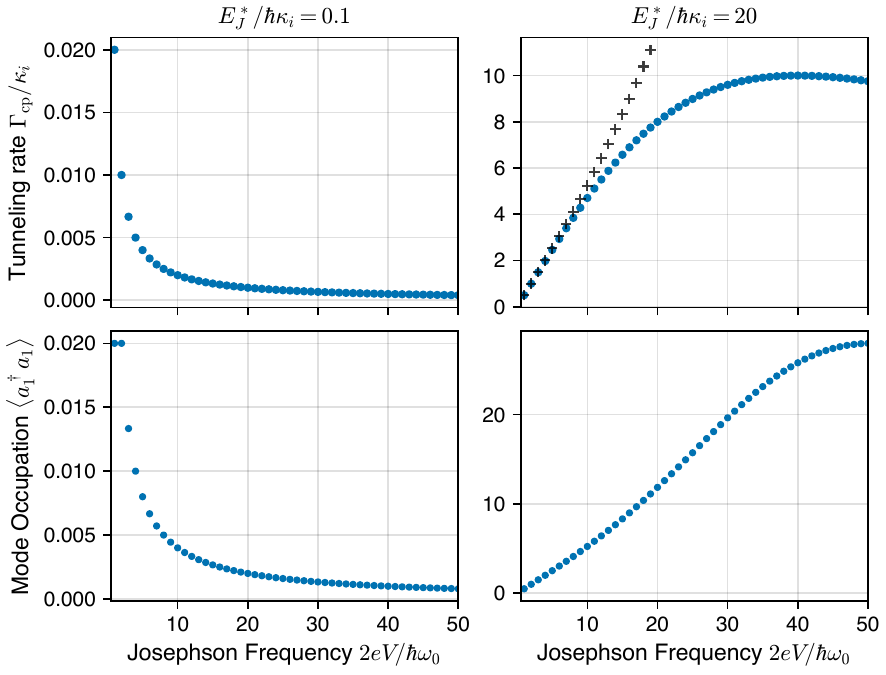}
 	\caption{{\bf Toulouse point $Z=R_Q/2$.} Normalized Cooper-pair tunneling rate \(\Gamma_{\rm cp}/\kappa_i\) (upper row) and mean fundamental-mode occupation \(\langle a_1^\dagger a_1\rangle\) (lower row) versus voltage, for two values of the Josephson coupling \(E_J^\ast\) in the weakly driven (left) and strongly driven regime (right). We assume a constant quality factor $Q=500$. 
	In the strongly driven regime, resonances are saturated at low voltage, and the peak heights increase linearly with \(m\). The tunneling rate at resonance reaches a maximum when $m\approx 2E_J^\ast/(\hbar \kappa_i)$  before decreasing as the higher resonances enter the weak-drive regime.
	The fundamental-mode occupation follows a similar trend, but its maximum occurs at a higher voltage.
	Gray crosses in the upper row reproduce the occupation data from the lower row, illustrating the relation \(\Gamma_{\rm cp}/\kappa_i\simeq\langle a_1^\dagger a_1\rangle\) at the saturated resonances.}
 	\label{fig:toulouse_current}
 \end{figure}
 
 \begin{figure}
 	\centering
 	\includegraphics[width=0.9\textwidth]{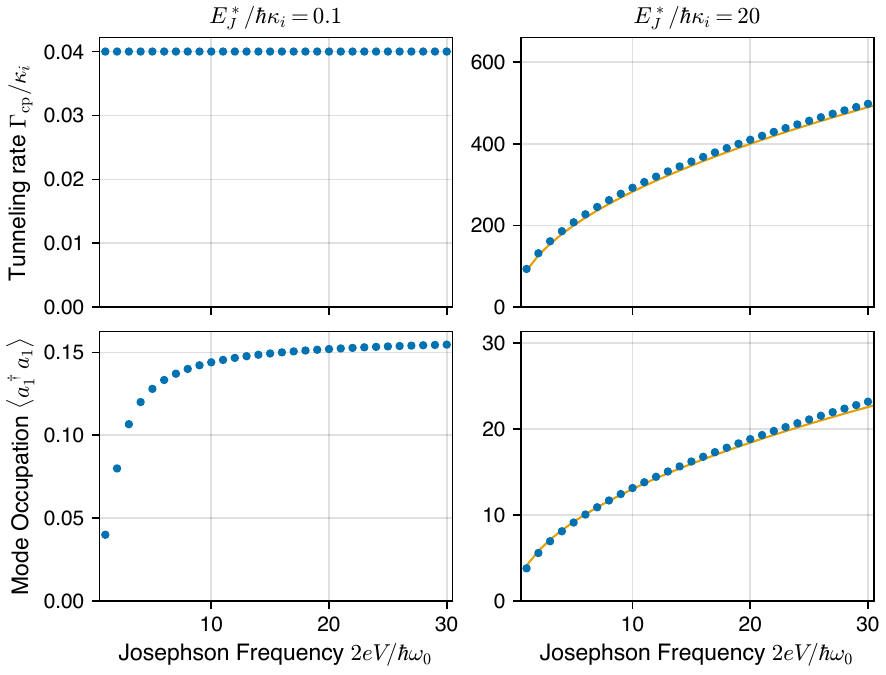}
 	\caption{{\bf Bosonic point $Z=R_Q$.} Normalized Cooper-pair tunneling rate \(\Gamma_{\rm cp}/\kappa_i\) (upper row) and mean fundamental-mode occupation \(\langle a_1^\dagger a_1\rangle\) (lower row) versus voltage, for two values of the Josephson coupling \(E_J^\ast\) in the weakly driven (left) and strongly driven regime (right). We assume a constant quality factor $Q=500$. We use $P(E)$ theory for the weakly driven regime and our Gaussian kinetic model in the case of strong driving. The solid line in the strongly driven regime shows the prediction of Eq.~\eqref{eq:strongDriveRQ} and Eq.~\eqref{eq:RQ_pop} with a $\sqrt{V}$ dependence.}
 	\label{fig:bosonic_current}
 \end{figure}
 
 \subsection{Antibunching at the Toulouse point}
 \label{sec:antibunching}
 At the Toulouse point, the resonant transfer operator is fermionic and
 \begin{equation}
 	(A_m^\dagger)^2=0.
 	\label{eq:nilpotent}
 \end{equation}
 A forward Cooper-pair transfer, generated by \(TA_m^\dagger\),
 fills the resonant orbital and blocks an immediate second forward
 transfer. Conditioning on a forward transfer applies \(A_m^\dagger\)
 to the state, so two consecutive forward transfers probe
 \((A_m^\dagger)^2\). The normalized zero-delay coincidence,
 defined through successive short charge-counting intervals, is therefore
 \begin{equation}
 	g^{(2)}_{{\rm cp},++}(0)
 	=\frac{\langle A_m^2(A_m^\dagger)^2\rangle}
 	{\langle A_mA_m^\dagger\rangle^2}=0.
 	\label{eq:forwardCPcoincidence}
 \end{equation}
 The relation between charge transfer and orbital occupation is also
 expressed by the exact Hamiltonian conservation law
 \begin{equation}
 	[\mathcal H_{\rm sec},\hat N-n_m]=0,
 	\qquad n_m=c_m^\dagger c_m.
 	\label{eq:chargeOrbitalConservation}
 \end{equation}
The antibunching of two successive forward tunneling events is independent of the exact loss model and is a strong feature of the secular model. The finite-delay recovery time depends on photon loss. Below we obtain a simple expression in the incoherent limit.
 
\paragraph{Tunneling current in the constant-quality-factor limit.} Within the constant-quality-factor approximation, the Cooper-pair
tunneling rate of Eq.~\eqref{eq:QFcurrentToulouse} leads to the tunneling current at resonance
\begin{equation}
 	I_{\rm cp}
 	=
 \frac{\pi V}{Q R_Q}\,
 \frac{1}{1+\left(\dfrac{eV}{QE_J^\ast}\right)^2}.
\end{equation}
The crossover from weak drive to saturation occurs at
$E_J^\ast = m\hbar\kappa_i/2$. At fixed voltage, the current initially increases as $E_J^{\ast2}$ and saturates at $em\kappa_i =\pi V/(Q R_Q)$.
At fixed coupling $E_J^\ast$, resonances are saturated at low voltages
(unless $E_J^\ast\ll\hbar\kappa_i/2$), and the current at resonance
increases linearly with voltage as $\pi V/(Q R_Q)$. It reaches a maximum
$eE_J^\ast/\hbar$ at $V^\ast \approx QE_J^\ast/e$.
At higher voltages, the weakly driven regime is recovered and the current
decreases as $2QE_J^{\ast2}/(\hbar V)$ (see Fig.~\ref{fig:toulouse_current}). The figure~\ref{fig:toulouse_current} also shows the lowest mode occupation, which follows the same trend as the current.

 \subsection{Bunching at the bosonic point}
 At the bosonic point, subsequent Cooper-pair tunneling events are not Pauli blocked but, on the contrary, stimulated. The drive term \(B_m+B_m^\dagger\) in \eqref{eq:HRWAdisplaced} is bosonic, and the amplitude for adding a collective excitation to a state already containing \(n\) quanta is \(\sqrt{m(n+1)}\). The second-order correlation function at zero delay defined in Eq.~\eqref{eq:forwardCPcoincidence} therefore becomes
 \begin{equation}
 	\begin{aligned}
 		g^{(2)}_{{\rm cp},++}(0)
 		&=\frac{\langle(n_B+1)(n_B+2)\rangle}{(\langle n_B \rangle +1)^2}\\
 		&=1+\frac{\operatorname{Var}(n_B)+ \langle n_B \rangle +1}{(\langle n_B \rangle +1)^2}>1.
 	\end{aligned}
 	\label{eq:bosonicForwardCoincidence}
 \end{equation}
 where \(n_B= B_m^\dagger B_m\) and we use the identity \(B_m^2(B_m^\dagger)^2=(n_B+1)(n_B+2)\).
 Forward transfers are therefore bunched at every resonance \(m\) for any state with finite first and second moments, independently of our Gaussian assumption. In the vacuum limit, we recover \(g^{(2)}_{{\rm cp},++}(0)\to2\).

\paragraph{Constant-quality-factor limit.} The rate estimate of Eq.~\eqref{eq:strongDriveRQ} suggests that, at fixed voltage, the tunneling current crosses over from quadratic growth \(E_J^{\ast2}\) in
the weak-drive regime to growth with an exponent of \(3/2\) at strong drive. Because
the collective matrix element grows as \(\sqrt m\), the cross-over voltage is set by
\(m\gg(2E_J^\ast/\hbar\kappa_i)^2\), much larger than at the Toulouse point. We therefore expect a current at resonance given by
\begin{equation}
I_{\rm cp}(V)
\approx
\frac{\pi}{R_Q}\,
\frac{E_J^\ast}{\hbar\omega_0}\,
\sqrt{\frac{2Q E_J^\ast V}{e}}, \label{eq:Icp_bosonic}
\end{equation}
over a large voltage range when $E_J^\ast/\hbar \gg \kappa_i$ as verified in figure~\ref{fig:bosonic_current}. 

\section{Incoherent Cooper-pair transfer}
\label{sec:RQ_dephased}

Coherences between neighboring Floquet sectors \(N\) and \(N+1\) can also decay through processes unrelated to photon loss. In a dc-biased circuit, the dominant mechanism is low-frequency voltage noise of the bias circuit. We describe this dephasing by adding \(2\Gamma_\phi\mathcal D[\hat N]\varrho\) to the master equation, so that \(\bra{N}\varrho\ket{N'}\) decays at the rate \(\Gamma_\phi(N-N')^2\). This form assumes that the noise is white on the time scales \(\Gamma_\phi^{-1}\), \(\kappa_r^{-1}\), and \(\hbar/E_J^\ast\). Quasi-static fluctuations, such as \(1/f\) noise or drift, instead average the steady state over the detuning \(\delta\).

\subsection{Toulouse point}
Within the constant-quality-factor approximation, the only effect of this dephasing is to replace \(\kappa_m/2\) by \(\Gamma_2=\kappa_m/2+\Gamma_\phi\) in Eq.~\eqref{eq:Bloch_c_m}. Equation~\eqref{eq:Bloch_n_m} and the invariant-subspace argument are unchanged, and the steady state gives
\begin{equation}
	\Gamma_{\rm cp}=\frac{\kappa_m\Gamma_J}{2(\kappa_m+\Gamma_J)},
	\qquad
	\Gamma_J=\frac{2E_J^{\ast2}}{\hbar^2}\,\frac{\Gamma_2}{\Gamma_2^2+\delta^2},
	\label{eq:dephasedCurrentToulouse}
\end{equation}
which reduces to Eq.~\eqref{eq:QFcurrentToulouse} for \(\Gamma_\phi=0\).

In the regime \(E_J^\ast/\hbar,\kappa_m\ll\Gamma_\phi\ll\omega_0\), the coherence adiabatically follows the occupation, and the optical Bloch equations reduce to
\(d\langle n_m\rangle/dt=\tfrac{\Gamma_J}{2}(1-\langle n_m\rangle)-\tfrac{\Gamma_J}{2}\langle n_m\rangle-\kappa_m\langle n_m\rangle\), with \(\Gamma_J\simeq2E_J^{\ast2}/(\hbar^2\Gamma_\phi)\) near resonance. Forward transfers fill orbital \(m\) and backward transfers empty it, both at the rate \(\Gamma_J/2\); these rates are equal because the noise is classical. The photon cascade empties the orbital at the rate \(\kappa_m\). Orbital \(m\) thus behaves as a single resonant level in sequential tunneling, coupled to a half-filled reservoir, the dephased junction, and to an empty one, the photon cascade.
The picture of Sec.~\ref{sec:antibunching} is otherwise unchanged:
\begin{itemize}
	\item Since \((A_m^\dagger)^2=0\), forward transfers remain antibunched, and the rate still saturates at \(\kappa_m/2\).
	\item In this limit, the long-time Fano factor of the net transferred-pair number, including backward transfers, is \(F=\lim_{t\to\infty}\operatorname{Var}[\Delta N(t)]/\langle\Delta N(t)\rangle=1-\kappa_m\Gamma_J/(\kappa_m+\Gamma_J)^2\geq3/4\). It is sub-Poissonian, and backward transfers through the same orbital prevent it from dropping below \(3/4\). Appendix~\ref{app:toulouse_counting} derives this result from the two-state charge-counting generator.
\end{itemize}
The same rate equation gives the recovery of the antibunching dip in the second-order correlation function,
\begin{equation}
	g^{(2)}_{{\rm cp},++}(\tau)
	=1-e^{-(\kappa_m+\Gamma_J)\tau}.
	\label{eq:dephasedForwardRecovery}
\end{equation}
In the fluorescence spectrum, the Mollow sidebands, which are signatures of coherent dressing, are washed out at strong \(\Gamma_\phi\).

\subsection{Bosonic point}
We suppose that \(\Gamma_\phi\) exceeds both the loss rates and the stimulated transfer rate \((E_J^\ast/\hbar) \allowbreak \sqrt{m(n_B+1)}\), while remaining small compared with \(\omega_0\). The coherences between successive Floquet indices can then be eliminated adiabatically, and the Josephson term becomes a pair of incoherent jumps with equal rates,
\begin{equation}
	\mathcal L_J\varrho=\Gamma_J\left(\mathcal D[TA_m^\dagger]+\mathcal D[T^\dagger A_m]\right)\varrho,
	\qquad
	\Gamma_J=\frac{2E_J^{\ast2}}{\hbar^2}\,\frac{\Gamma_\phi}{\Gamma_\phi^2+\delta^2}.
	\label{eq:dephasedJumpsRQ}
\end{equation}
Two results hold exactly within this master equation.
\begin{itemize}
	\item \emph{Cooper-pair rate.} The Cooper-pair rate becomes \(\Gamma_{\rm cp}=\Gamma_J\langle A_mA_m^\dagger-A_m^\dagger A_m\rangle=m\Gamma_J\). In the fermion ladder of Eq.~\eqref{eq:AmRefermionRWA}, this is the counting identity \(\sum_j(\langle n_{j-m}\rangle-\langle n_j\rangle)=m\), with \(n_j=d_j^\dagger d_j\): whatever the occupations, upward moves by \(m\) levels outnumber downward moves by exactly \(m\). The rate depends neither on the strip state nor on the loss rates. The resonance is a Lorentzian of half-width \(\Gamma_\phi\) and height \(2mE_J^{\ast2}/(\hbar^2\Gamma_\phi)\), and the strong-drive behavior of Eq.~\eqref{eq:strongDriveRQ} is absent. This predicts a tunneling current at resonance, which increases linearly with $V$ as
	\begin{equation}
		I_{\rm cp}=\frac{4\pi}{R_Q}\,\frac{(E_J^\ast/\hbar)^2}{\omega_0\Gamma_\phi}\,V .
	\end{equation}
	\item \emph{Fluorescence.} Incoherent Cooper-pair transfers cause diffusion along each fermion chain
	and wash out the fluorescence bands associated with coherent minibands.
	The commutation relation \([a_1,A_m^\dagger]=2A_{m-1}^\dagger\) gives a lower bound on the steady-state population of the fundamental mode,
	\begin{equation}
		\langle a_1^\dagger a_1\rangle
		\geq \frac{4(m-1)\Gamma_J}{\kappa_1+8\Gamma_J},
		\qquad m>1 .
	\end{equation}
	At fixed $\Gamma_J$ and $\kappa_1$, this lower bound grows linearly with
	the resonance index. In the weak drive limit, the population is
	$4(m-1)\Gamma_J/\kappa_1$ to leading order, in agreement with the
	weak-drive yield of Eq.~\eqref{eq:weakDrivePhotonYield}. For
	$\Gamma_J\gg\kappa_1$ within the incoherent regime, the bound approaches
	$(m-1)/2$ photons.
\end{itemize}

\section{Beyond the secular approximation}
\label{sec:beyond}

The secular regime discussed so far requires \(E_J^\ast \ll \hbar\omega_0\), but Josephson couplings above this threshold are experimentally relevant, especially at low impedance. Finding the many-body steady state without the secular approximation is, however, difficult and beyond the scope of this paper. Here we take a first step in this direction by computing the one-particle Floquet eigenstates and their quasienergies.

At both special impedances, the coupling term is quadratic in the fermionic representation, and the one-body evolution generated by the full time-dependent Hamiltonian can be solved exactly, now retaining all components of the vertex operator and not only the resonant one:
\begin{equation}
	H(t)
	=
	H_0
	-\frac{E_J^\ast}{2}
	\left[
	e^{-i\omega_Jt}\,V+e^{i\omega_Jt}\,V^\dagger
	\right],
	\qquad
	V=\sum_{n\in\mathbb Z}V_n .
	\label{eq:Hfull}
\end{equation}
Using \(V_n^\dagger=(-1)^nV_{-n}\), the Josephson term
can be written as
\(-\frac{E_J^\ast}{2}\sum_n\lambda_n(t)V_n\), with
\begin{equation}
	\lambda_n(t)=e^{-i\omega_Jt}+(-1)^ne^{i\omega_Jt}.
\end{equation}

\subsection{Toulouse point: Bogoliubov--de Gennes Floquet problem}
\label{sec:beyond_toulouse}

\begin{figure}
	\centering
	\includegraphics[width=0.9\textwidth]{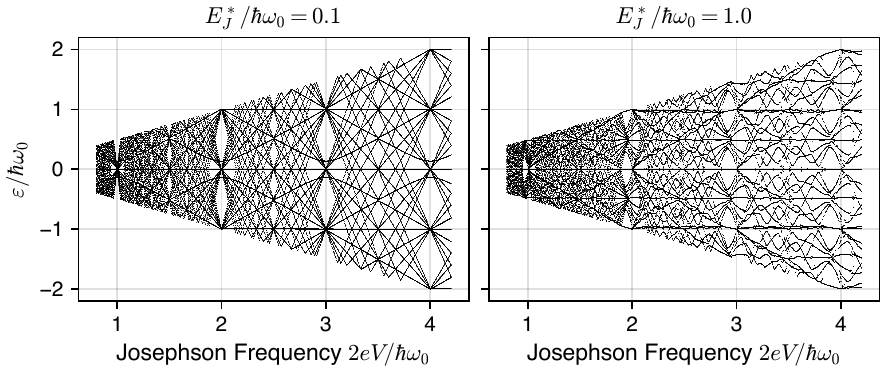}
\caption{{\bf Toulouse point $Z=R_Q/2$.} One-body BdG Floquet spectrum as a function of voltage for two Josephson couplings $E_J^\ast$. The first Floquet zone extends from \(-\hbar\omega_J/2\) to \(+\hbar\omega_J/2\). At weak coupling near a resonance, the two particle--hole branches describe a single active fermionic mode; its positive excitation energy is the splitting of the physical doublet discussed in Sec.~\ref{sec:toulouse_floquet}.}
	\label{fig:toulouse_bdg_energies}
\end{figure}

At the Toulouse point, the operators \(c_n=\sqrt2A_n\) introduced in
Sec.~\ref{sec:toulouse_floquet} provide a fermionic representation of the
bosonic strip states. Their free energies are \(n\hbar\omega_0\), with
\(n>0\). The full Josephson coupling is linear in these operators and in
\(V_0=\mathcal P\). Introducing an auxiliary Majorana \(\eta\) and representing
\(\mathcal P\) by a second Majorana \(\gamma_0\) makes the Hamiltonian quadratic~\cite{Konechny2019,AmeduriKonikLeClair1995}:
\begin{equation}
	\begin{split}
		H(t)={}&\sum_{n>0}\hbar n\omega_0\,c_n^\dagger c_n
		-iE_J^\ast\cos(\omega_Jt)\,\eta\gamma_0\\
		&+\frac{iE_J^\ast}{\sqrt2}\sum_{n>0}\lambda_n(t)\,
		\eta\left[c_n^\dagger+(-1)^nc_n\right].
	\end{split}
	\label{eq:HextToulouse}
\end{equation}
Here \(\eta^2=\gamma_0^2=1\); the two Majoranas anticommute with each other
and with all \(c_n\). The construction embeds the physical problem in a
doubled Hilbert space containing two equivalent Floquet spectra. The
physical sector has total enlarged-system fermion parity \(+1\);
the quasiparticle occupations must obey
Eq.~\eqref{eq:toulousePhysicalOccupations}, derived in
Appendix~\ref{app:toulouse_sector}. We use the same notation \(c_n\) for the
embedded operators.

The Heisenberg equations are now linear. Diagonalizing the evolution over one drive period defines Bogoliubov quasiparticle operators \(\Gamma_\alpha\), which combine particle and hole components through their common coupling to \(\eta\). The corresponding one-body quasienergies occur in pairs \(\pm\varepsilon_\alpha\). Since the junction couples to the strip only through the field amplitudes at the boundary, the strip degrees of freedom can be eliminated using the following two resolvent sums:
\begin{align}
	g_0(z)&=\sum_{n\in\mathbb Z}\frac{1}{z-n\hbar\omega_0}
	=\frac{\pi}{\hbar\omega_0}\cot\!\left(\frac{\pi z}{\hbar\omega_0}\right),\\
	g_1(z)&=\sum_{n\in\mathbb Z}\frac{(-1)^n}{z-n\hbar\omega_0}
	=\frac{\pi}{\hbar\omega_0}\csc\!\left(\frac{\pi z}{\hbar\omega_0}\right).
	\label{eq:toulouseBoundarySums}
\end{align}
These functions retain the
discrete spectrum of the finite strip while summing its ideal harmonic
ladder to arbitrarily high energies; a physical high-frequency cutoff can
be included in the sums using symmetric limits.

Let \(\mu_k\) be the Fourier amplitudes of the auxiliary Majorana $\eta$,
with time dependence \(e^{-i\varepsilon t/\hbar}\sum_k \mu_ke^{-ik\omega_Jt}\),
and write \(\Omega_k=\varepsilon+k\hbar\omega_J\). Eliminating the strip gives
\begin{equation}
	\Omega_k \, \mu_k=E_J^{\ast2}\sum_{s=\pm1}
	\left[g_0(\Omega_{k+s})\mu_k+g_1(\Omega_{k+s})\mu_{k+2s}\right].
	\label{eq:BdGquasienergy}
\end{equation}
Each term describes propagation from the auxiliary Majorana into the strip and back. The two coupling steps can involve opposite drive harmonics, leaving the harmonic index unchanged, or the same harmonic, changing it by two. The corresponding coupling phases give the resolvent sums \(g_0\) and \(g_1\), respectively. The even and odd harmonic chains therefore separate,
and the resulting tridiagonal equation can be solved by continued fractions or by locating the quasienergies at which an eigenvalue of the tridiagonal matrix crosses zero (Appendix~\ref{app:toulouse_green}). Figure~\ref{fig:toulouse_bdg_energies} shows the quasienergy spectrum versus voltage for two coupling strengths. At weak coupling (left panel), the resonant branches correspond to a single active fermionic mode, while the remaining states follow the uncoupled energy branches. The BdG pair \(\pm\varepsilon_b\), with \(\varepsilon_b=\sqrt{(\hbar\delta)^2+2E_J^{\ast2}}\), represents the particle and hole components of this mode, whereas the physical many-body doublet has levels \(\pm\varepsilon_b/2\) relative to its center [Eq.~\eqref{eq:MajoranaSecularSpectrum}].

\paragraph{Exact spectrum at resonance.}

\begin{figure}
	\centering
	\includegraphics{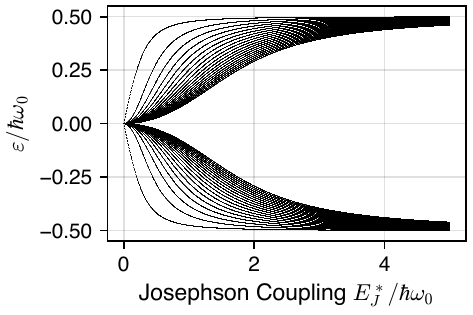}
	\caption{\textbf{Toulouse point $Z=R_Q/2$.} One-body BdG roots of Eq.~\eqref{eq:toulouseIntegerSpectrum}, folded into the Floquet zone at \(m=1\). The two branches linear in $E_J^\ast$ at weak coupling are the particle--hole pair of a single active fermionic mode, whose excitation energy sets the physical doublet splitting. At strong coupling, the roots approach \(\pm\hbar\omega_0/2\), the equivalent edges of this Floquet zone.}
	\label{fig:toulouse_bdg_EJ}
\end{figure}

For \(\omega_J=m\omega_0\), with \(m\ge1\), Eq.~\eqref{eq:BdGquasienergy} can be solved exactly. The quasienergies are given by the roots of the following equation (Appendix~\ref{app:toulouse_integer}):
\begin{equation}
	x=\mathcal G\cot(\pi x),\qquad
	\mathcal G=2\pi\left(\frac{E_J^\ast}{\hbar\omega_0}\right)^2,
	\qquad
	\varepsilon=\hbar\omega_0x\pmod{\hbar\omega_J}.
	\label{eq:toulouseIntegerSpectrum}
\end{equation}
This equation determines the regular roots away from integer poles. 
For nonzero coupling, there is one root in each interval \((r,r+1)\),
\(r\in\mathbb Z\). The two roots nearest zero recover
\(\hbar\omega_0x\simeq\pm\sqrt2E_J^\ast\) at weak coupling. The unfolded roots are
independent of \(m\), while their folding gives the Floquet quasienergies that
depend on \(m\). Figure~\ref{fig:toulouse_bdg_EJ} shows these roots folded at \(m=1\) as a function of $E_J^\ast$.

\paragraph{Strong coupling and emergent fermions.}
At strong coupling, the unfolded roots approach
\((r+1/2)\hbar\omega_0\), with quasienergies defined modulo \(\hbar\omega_J\). This spectrum corresponds to that of a free Majorana channel with antiperiodic boundary conditions. This is due to strong hybridization with the auxiliary Majorana, which produces a boundary
scattering phase of \(\pi\) (Appendix~\ref{app:toulouse_strong}).
The effective Floquet Hamiltonian of the system is then given by a sum of free fermion modes with frequencies \((n+1/2)\omega_0\) with $n=0,1,\ldots$. 

\subsection{Bosonic point: a chiral fermion with a driven point contact}
\label{sec:beyond_RQ}

\begin{figure}
	\centering
	\includegraphics[width=0.9\textwidth]{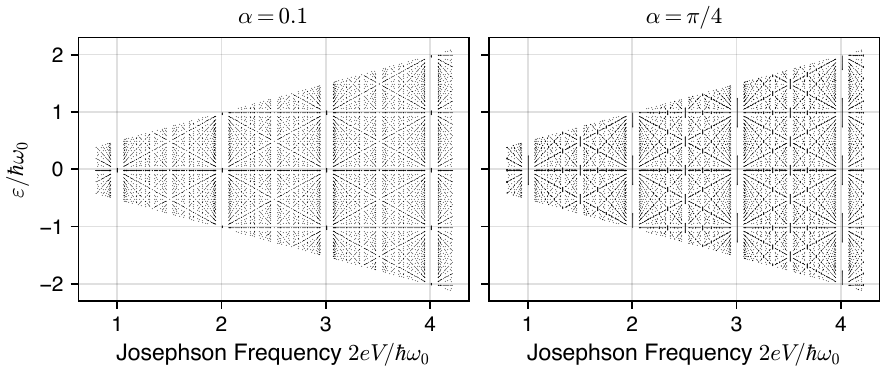}
	\caption{\textbf{Bosonic point $Z=R_Q$.} One-body Floquet spectrum versus voltage for two values of the contact angle $\alpha$, which controls the coupling strength. The quasienergies are computed with the passage matrices of Appendix~\ref{app:RQ_rational} at every rational bias \(\omega_J/\omega_0=p/q\) with coprime \(p,q\) and \(q\leq16\), and are folded into the first Floquet zone. They are measured from \(\hbar\omega_0/2\), so that the free levels lie at integer multiples of \(\hbar\omega_0\). Minibands form at each rational bias; their widths scale as \(\alpha^q\) at weak coupling. At integer values of $\omega_J/\omega_0$, the gaps close at $\alpha=\pi/2$, which corresponds to the maximum angle reached in the strong-coupling limit when the junction capacitance \(C_J\) sets the cutoff.}
	\label{fig:bosonic_bdg_energies}
\end{figure}

At \(Z=R_Q\), the Josephson vertex transfers fermions between the levels of the ladder introduced above and detailed in Appendix~\ref{app:bosonic}. The full Hamiltonian conserves fermion number, and the Josephson term creates particle--hole excitations when it acts on the filled Fermi sea. In position space, the same dynamics describes a chiral fermion circulating around a ring, with the Josephson junction coupling two points half a turn apart. Appendix~\ref{app:RQ_contact} derives this representation and the corresponding Floquet evolution.

The one-body Floquet modes have a simple interpretation in the limit of a narrow wave packet. Its two components start half a turn apart at \(t=0\) and circulate at the same speed. Each time they reach the junction, an \(SU(2)\) rotation mixes their amplitudes, with a rotation axis set by the Josephson phase at arrival. When \(\omega_J/\omega_0=p/q\), the two components return to their starting positions after \(q\) revolutions, or \(p\) drive periods. If their initial amplitudes form an eigenvector of the product of rotations, the packet returns to itself up to an overall phase. Varying its starting position shifts the arrival times relative to the drive, and the resulting eigenphases trace out the quasienergy minibands.

The scattering angle \(\alpha\) controls the mixing at a single passage. At integer resonance, after removing the common free-propagation phase, the eigenvalues of the return map over one revolution are \(e^{\pm i\beta}\), with \(\sin(\beta/2)=\sin\alpha\,\sin\theta\) for a suitable choice of origin for the arrival phase \(\theta\). The parameters \(\alpha\), \(\omega_0\), and \(\omega_J\) fully specify the ideal Floquet problem. Relating \(\alpha\) to the physical Josephson energy \(E_J\) requires a prescription for the high-frequency cutoff.

Here, we assume that the junction capacitance \(C_J\) provides the high-frequency cutoff. As \(E_J/E_C\) increases, with \(E_C=e^2/(2C_J)\), the low-energy contact angle evolves from \(\alpha=0\) toward \(\alpha=\pi/2\). These limits correspond to an open boundary and a phase-pinned boundary, respectively, with the latter reached as \(E_J/E_C\to\infty\) \cite{LukyanovWerner2007,Paris2025}.
At an integer resonance, \(\omega_J=m\omega_0\), the sequence of drive phases repeats after each revolution. The quasienergy minibands are centered on the free ladder \((k+\tfrac12)\hbar\omega_0\), with offsets
\begin{equation}
	\varepsilon(\theta)=\pm\frac{\hbar\omega_0}{\pi}
	\arcsin\!\left(\sin\alpha\,\sin\theta\right),
	\label{eq:exactBandRQ}
\end{equation}
where the arrival phase \(\theta\) coincides, up to an additive constant, with the quasimomentum introduced in Sec.~\ref{sec:secular_RQ}. For \(\alpha\ll1\), the dispersion reduces to \(\varepsilon(\theta)\simeq\pm(\hbar\omega_0/\pi)\alpha\sin\theta\).
As \(\alpha\) increases, the dispersion departs from a sinusoid and the full miniband width grows as \(2\alpha\hbar\omega_0/\pi\). At \(\alpha=\pi/2\), adjacent minibands touch and the dispersion becomes piecewise linear.

At a noninteger rational bias, \(\omega_J/\omega_0=p/q\), the miniband dispersion and half-width depend only on $q$ and are given by
\begin{equation}
	\sin\!\left(\frac{q\pi\varepsilon(\theta)}{\hbar\omega_0}\right)
	=\pm(\sin\alpha)^q\sin\theta,
	\qquad
	W_{q}=\frac{\hbar\omega_0}{\pi q}
	\arcsin\!\left(|\sin\alpha|^q\right).
	\label{eq:RQRationalBands}
\end{equation}
Folding the spectrum modulo \(\hbar\omega_J\) leaves \(p\) distinct miniband centers, as shown in Fig.~\ref{fig:bosonic_bdg_energies}.
In Appendix~\ref{app:RQ_periodicities}, we show that the unfolded one-body quasienergy spectrum is periodic in
\(\omega_J\) with period \(\omega_0\) and in \(\varepsilon\) with
period \(\hbar\omega_0\).
Quasienergies are also equivalent modulo \(\hbar\omega_J\).
At rational bias \(\omega_J/\omega_0=p/q\), with \(p\) and \(q\)
coprime, the spectrum repeats with spacing \(\hbar\omega_0/q\),
giving \(p\) repetitions within one Floquet zone.

At weak coupling, the half-width reduces to \(W_{q}\simeq \frac{\hbar\omega_0}{\pi q}\left(\frac{2\pi E_J^\ast}{\hbar\omega_0}\right)^q\). The \(q\)th-order scaling reflects the transfer of \(q\) Cooper pairs, which supplies the resonant energy \(q\hbar\omega_J=p\hbar\omega_0\). The widths therefore decrease exponentially with the denominator \(q\).

\subsection{Many-body Floquet states}
\label{sec:manybody_floquet}

At both impedances, the quadratic fermion construction must be restricted
to the physical sector before counting many-body Floquet states.
At the Toulouse point, the Bogoliubov occupations, including the zero
mode, have fixed total parity as specified in
Eq.~\eqref{eq:toulousePhysicalOccupations}; they cannot all be chosen
independently. At \(Z=R_Q\), the physical states have zero charge and
are annihilated by every positive even density mode,
Eq.~\eqref{eq:RQPhysicalSector}. Imposing this condition can require
linear combinations within degenerate spaces of Slater determinants.
The two projections are detailed in
Appendices~\ref{app:toulouse_sector} and~\ref{app:RQ_physical_sector}.
In the selected sectors, many-body quasienergies are sums of the allowed
one-body quasienergies, modulo
\(\hbar\omega_J\). The uniform part of these sums diverges with the cutoff, logarithmically at \(Z=R_Q/2\) and linearly at \(Z=R_Q\).
This is consistent with the marginal character of the boundary cosine at \(\rho=1\).

\section{Conclusion}
\label{sec:conclusion}

Our main results concern the nature of the excitations created by
a secular component of the Josephson drive. Their fermionic character
at \(\rho=1/2\) and bosonic character at \(\rho=1\) lead to predictions
for antibunching and bunching of forward Cooper-pair transfers, as well as the qualitative structure of the fluorescence spectrum. These predictions are robust and independent
of the details of dissipation and decoherence.

Quantitative predictions for the tunneling current and fluorescence
line intensities are more sensitive to the loss model. We have shown
that a constant quality factor, possibly combined with strong dephasing between
different Floquet indices, simplifies the description and facilitates
quantitative predictions for the steady state. These assumptions,
particularly that of a constant quality factor, may nevertheless
be difficult to justify in realistic experiments.

We now conclude by assessing the regimes which are within reach with a standard Josephson junction made of Al. We assume that the high-frequency cutoff is set by the junction capacitance
\(C_J\), as is reasonable for many realizations of high-inductance
transmission lines. A first requirement is that sufficiently many strip
modes lie below the cutoff frequency \(1/(ZC_J)\). Their number is
approximately
\begin{equation}
	N_m \sim \frac{1}{2\rho C_J\omega_0R_Q}.
\end{equation}
A small Josephson junction typically has \(C_J\sim1\,{\rm fF}\), corresponding
to a charging energy \(E_C/h\sim20\,{\rm GHz}\). A mode spacing of
\(1\,{\rm GHz}\) or less then places more than ten modes below the cutoff.
In the following, we take \(\omega_0=2\pi\times500\,{\rm MHz}\), which ensures
\(N_m>10\) at both impedances. Thermal occupation of the fundamental mode
can be suppressed at sufficiently low millikelvin temperatures.

The secular condition itself can readily be satisfied. The renormalized
Josephson coupling is estimated as~\cite{LukyanovWerner2007}
\begin{equation}
	E_J^\ast \simeq E_J
	\left(
	\frac{e^{-\gamma}}{2}
	\frac{\omega_0}{\omega_c}
	\right)^\rho,
	\qquad
	\omega_c=\frac{1}{ZC_J},
\end{equation}
where \(\gamma\) is the Euler constant. At the two special
impedances, this gives
\begin{align}
	\frac{E_J^\ast}{\hbar\omega_0}
	&\simeq 0.33\,\frac{E_J}{\sqrt{E_C\hbar\omega_0}},
	\qquad \rho=\frac12,\\
	\frac{E_J^\ast}{\hbar\omega_0}
	&\simeq 0.22\,\frac{E_J}{E_C},
	\qquad \rho=1.
\end{align}
For \(C_J\sim1\,{\rm fF}\) and
\(\omega_0=2\pi\times500\,{\rm MHz}\), the requirement
\(E_J^\ast\ll\hbar\omega_0\) becomes
\(E_J/h\ll9\,{\rm GHz}\) at \(\rho=1/2\) and
\(E_J/h\ll90\,{\rm GHz}\) at \(\rho=1\).
Normal-state resistances between \(20\,{\rm k}\Omega\) and
\(1\,{\rm M}\Omega\), accessible using multiple oxidation steps, correspond
to bare Josephson couplings \(E_J/h\) ranging from approximately
\(150\,{\rm MHz}\) to \(7\,{\rm GHz}\).
The secular condition is therefore satisfied throughout this range at
\(\rho=1\). It is also readily accessible at \(\rho=1/2\), for example
with a double-oxidation barrier yielding a normal-state resistance above
\(200\,{\rm k}\Omega\). The more stringent condition
\[
\frac{E_J^{\ast2}}{\hbar\omega_0}<\hbar\kappa_i,
\]
which keeps nonsecular energy shifts below the intrinsic linewidth, can
also be reached with highly resistive junctions and quality factors of a
few hundred.

The constant-quality-factor assumption is less generally applicable and
may hold only over a limited range of modes. It also entails an
experimental trade-off: intrinsic losses must dominate the coupling to
the measurement port, reducing the fraction of emitted radiation that can
be collected. For a measurement-line impedance \(Z_m\simeq10\,\Omega\),
subsequently tapered to \(50\,\Omega\), the coupling-limited quality factor
is \(\omega_0/\kappa_c\sim500\) at \(\rho=1\).
A typical situation is then that the quality factor of the lowest modes is limited by the coupling and becomes mode independent for higher modes. 

Finally dephasing can be estimated from the measured linewidths in previous observations of inelastic Cooper pair tunneling, the lowest reported rates are about \(\Gamma_\phi/2\pi\approx1\)--\(3\,{\rm MHz}\)~\cite{Martel2025}. This is typically in the same range as the loss rates and the junction is neither in the strongly dephased regime nor in the negligible dephasing regime. 

We therefore expect that the main predictions can be tested with current
experimental techniques, through measurements of current noise and
microwave fluorescence spectra. On the theoretical side, the same
refermionization approach could be extended to inelastic quasiparticle
tunneling, offering the prospect of exact calculations beyond the
Born--Markov approximation. Another direction is to investigate the
\(\rho=2\) point discussed in~\cite{MorelMora2021}.

\paragraph{Acknowledgements}
We thank Gianluca Aiello, Marco Aprili, Denis Basko, Julien Basset, Isabelle Bouchoule, Samuel Cailleaux, Jérôme Dubail, Serge Florens, Julien Gabelli, and Nicolas Roch for fruitful discussions. We acknowledge the use of artificial intelligence tools to assist with some calculations and the writing of this manuscript.

\appendix

\section[Vertex--vertex correlator]{Spectral weights of the vertex--vertex correlator}
\label{app:spectral}

The normal ordering of two vertex operators introduces the function
\begin{equation}
	V(u)V(v)
	=
	F_\rho\!\left(\frac{v}{u}\right)
	:V(u)V(v):,
\end{equation}
with
\begin{equation}
	F_\rho(x)
	=
	\left(\frac{1-x}{1+x}\right)^{2\rho}
	=
	\sum_{n=0}^{\infty}f_{\rho,n}x^n,
	\qquad |x|<1.
	\label{eq:FrhoExpansion}
\end{equation}
The coefficients $f_{\rho,n}$ determine the relative spectral weight
associated with the different ways in which a given energy can be
distributed among the strip modes. The same function appears in
$P(E)$ theory, when one calculates how the energy released by a tunneling Cooper pair is
distributed among the modes of the electromagnetic environment.

The large-$n$ behavior follows directly from the dominant singularity
of $F_\rho(x)$ at $x=-1$. Close to this point,
\begin{equation}
	F_\rho(x)
	\sim
	2^{2\rho}(1+x)^{-2\rho},
\end{equation}
and therefore
\begin{equation}
	f_{\rho,n}
	\sim
	(-1)^n
	\frac{2^{2\rho}}{\Gamma(2\rho)}
	n^{2\rho-1}.
	\label{eq:FrhoAsymptotic}
\end{equation}
Thus the magnitude of the spectral weights is asymptotically independent
of $n$ at the Toulouse point $Z=R_Q/2$, for which $\rho=1/2$, while it
grows linearly with $n$ at the bosonic point $Z=R_Q$, for which
$\rho=1$.

At these two special points, the expansions take the closed forms:
\begin{align}
	F_{1/2}(x)
	&=
	\frac{1-x}{1+x}
	=
	1+2\sum_{n=1}^{\infty}(-1)^n x^n,
	\\
	F_1(x)
	&=
	\left(\frac{1-x}{1+x}\right)^2
	=
	1+4\sum_{n=1}^{\infty}(-1)^n n x^n.
\end{align}
As we show below, these spectral weights underlie the fermionic and
bosonic algebras obtained at $Z=R_Q/2$ and $Z=R_Q$, respectively.

The coefficients of the series expansion of $F_\rho$ also appear in $P(E)$ theory. The phase correlator central to \(P(E)\) theory is
\begin{equation}
	C(t)=\langle 0 | e^{i\varphi(t)}e^{-i\varphi(0)} |0 \rangle
	=e^{-S}F_\rho^{-1}(e^{-i\omega_0t})
	=e^{-S}\sum_{n\geq0}p_{\rho,n}e^{-in\omega_0t},
	\label{eq:physicalPEcorrelator}
\end{equation}
with the same ultraviolet regularization and convergence prescription
as the vertex contractions. Since \(1/F_\rho(x)=F_\rho(-x)\), the coefficients of the physical phase correlator satisfy \(p_{\rho,n}=(-1)^n f_{\rho,n}=|f_{\rho,n}|\). Each \(p_{\rho,n}\) sums the weights of all multimode Fock configurations with total energy \(n\hbar\omega_0\). With the normalization of the tunneling operator used in Sec.~\ref{sec:weak}, the corresponding total weight is \(W_n=p_{\rho,n}/4\), while the common vacuum factor \(e^{-S}\) is absorbed into \(E_J^{\ast2}\). For \(n\geq1\), the weights are constant at the Toulouse point, \(p_{1/2,n}=2\), and grow linearly at the bosonic point, \(p_{1,n}=4n\). The vacuum coefficient is \(p_{\rho,0}=1\) for all impedances.

\section[Algebra at the Toulouse point]{Algebra at the Toulouse point $Z=R_Q/2$}
\label{app:toulouse}

\subsection{Canonical anticommutation relations}

We first derive the algebra of the Laurent modes of the vertex operator,
\begin{equation}
	V(u)=\sum_{n\in\mathbb Z}V_nu^n .
\end{equation}
At $\rho=1/2$, normal ordering gives
\begin{align}
	\{V(u),V(v)\}
	&=
	\left[
	F_{1/2}\!\left(\frac{u}{v}\right)
	+
	F_{1/2}\!\left(\frac{v}{u}\right)
	\right]
	:V(u)V(v):
	\nonumber\\
	&=
	2\sum_{p=-\infty}^{\infty}
	(-1)^p u^p v^{-p}
	:V(u)V(v): .
\end{align}
Since only odd harmonics enter the vertex operator,
\begin{equation}
	:V(u)V(-u):=1.
\end{equation}
Hence the nonconstant part of the normal-ordered product vanishes for
$v=-u$ and may be written as
\begin{equation}
	:V(u)V(v):
	=
	1+(u+v)R(u,v),
\end{equation}
where $R(u,v)$ is an operator-valued Laurent series. Furthermore,
\begin{equation}
	(u+v)
	\sum_{p=-\infty}^{\infty}
	(-1)^p u^p v^{-p}
	=0,
\end{equation}
as follows by shifting the summation index. The normal-ordered factor
therefore drops out and
\begin{equation}
	\{V(u),V(v)\}
	=
	2\sum_{p=-\infty}^{\infty}
	(-1)^p u^p v^{-p}.
	\label{eq:majoranaVertexCAR}
\end{equation}
Comparing equal powers of $u$ and $v$ gives
\begin{equation}
		\{V_p,V_q\}
		=
		2(-1)^p\delta_{p,-q}.
	\label{eq:majoranaVertexModeCAR}
\end{equation}
For $m>0$, define the canonical fermionic operators
\begin{equation}
	c_m
	\equiv
	\sqrt{2}\,A_m
	=
	-\frac{(-1)^m}{\sqrt{2}}V_{-m},
	\quad
	c_m^\dagger
	\equiv
	\sqrt{2}\,A_m^\dagger
	=
	-\frac{1}{\sqrt{2}}V_m.
	\label{eq:canonicalMajoranaFermions}
\end{equation}
Equation~\eqref{eq:majoranaVertexModeCAR} immediately gives
\begin{equation}
	\{c_m,c_n^\dagger\}
	=
	\delta_{mn},
	\quad
	\{c_m,c_n\}
	=
	\{c_m^\dagger,c_n^\dagger\}
	=
	0.
\end{equation}
Equivalently,
\begin{equation}
	\{A_m,A_n^\dagger\}
	=
	\frac12\delta_{mn}.
\end{equation}

\subsection{Free Hamiltonian and completeness of the fermionic basis}

The free strip Hamiltonian is
\begin{equation}
	H_0
	=
	\hbar\omega_0
	\sum_{\substack{r>0\\r\ {\rm odd}}}
	r\,a_r^\dagger a_r .
	\label{eq:H0OddBosons}
\end{equation}
A simple way of expressing it in the fermionic basis follows
from the commutator of $H_0$ with the vertex operator. 
Using
\begin{equation}
	[H_0,a_r^\dagger]
	=
	\hbar\omega_0 r\,a_r^\dagger,
	\qquad
	[H_0,a_r]
	=
	-\hbar\omega_0 r\,a_r,
\end{equation}
we see that the commutator with $H_0$ acts on every term of the vertex
operator in the same way as $\hbar\omega_0u\partial_u$. Hence
\begin{equation}
	[H_0,V(u)]
	=
	\hbar\omega_0\,u\partial_uV(u).
\end{equation}
Expanding $V(u)=\sum_nV_nu^n$, we therefore obtain
\begin{equation}
	[H_0,V_n]
	=
	n\hbar\omega_0V_n .
\end{equation}
Using Eq.~\eqref{eq:canonicalMajoranaFermions},
\begin{equation}
	[H_0,c_m^\dagger]
	=
	m\hbar\omega_0c_m^\dagger,
	\quad
	[H_0,c_m]
	=
	-m\hbar\omega_0c_m .
	\label{eq:H0FermionCommutator}
\end{equation}
Moreover, acting on the bosonic vacuum,
\begin{equation}
	V(u)|0\rangle
	=
	\exp\left[
	\sum_{\substack{r>0\\r\ {\rm odd}}}
	\sqrt{\frac{2}{r}}\,
	a_r^\dagger u^r
	\right]|0\rangle ,
\end{equation}
which contains no negative powers of $u$. Hence
\begin{equation}
	V_{-m}|0\rangle=0,
	\qquad
	c_m|0\rangle=0,
	\qquad m>0.
\end{equation}
The states
\begin{equation}
	c_{m_1}^\dagger\cdots c_{m_k}^\dagger|0\rangle,
	\qquad
	m_1<\cdots<m_k,
	\label{eq:FermionicFockStates}
\end{equation}
are thus eigenstates of $H_0$ with energy
\begin{equation}
	E=\hbar\omega_0\sum_{j=1}^k m_j .
\end{equation}
The fermionic Fock states exhaust the original bosonic Hilbert space by
the standard Euler partition identity underlying the boson--fermion
correspondence: partitions into odd integers are in one-to-one
correspondence with partitions into distinct integers. Thus the
appearance of fermionic modes at every positive integer $m$ does not
introduce additional degrees of freedom. Equation~\eqref{eq:H0FermionCommutator} therefore fixes the free
Hamiltonian uniquely:
\begin{equation}
		H_0
		=
		\sum_{m=1}^{\infty}
		\hbar m\omega_0\,c_m^\dagger c_m
		=
		2\sum_{m=1}^{\infty}
		\hbar m\omega_0\,A_m^\dagger A_m.
	\label{eq:H0MajoranaAm}
\end{equation}

\subsection{Parity operator}

The zero Laurent mode has a direct interpretation in the fermionic
representation. Setting $p=q=0$ in
Eq.~\eqref{eq:majoranaVertexModeCAR} gives
\begin{equation}
	V_0^2=1,
\end{equation}
while setting one of the indices to zero gives, for $m>0$,
\begin{equation}
	\{V_0,c_m\}
	=
	\{V_0,c_m^\dagger\}
	=
	0.
\end{equation}
Thus $V_0$ anticommutes with every $c_m$ and $c_m^\dagger$.
The sign of $V_0$ on the vacuum follows directly from the expansion above:
the coefficient of $u^0$ in $V(u)|0\rangle$ is $|0\rangle$, so that
\begin{equation}
	V_0|0\rangle=|0\rangle.
\end{equation}
Anticommuting $V_0$ through a fermionic Fock state then gives
\begin{equation}
	V_0
	c_{m_1}^\dagger\cdots c_{m_k}^\dagger|0\rangle
	=
	(-1)^k
	c_{m_1}^\dagger\cdots c_{m_k}^\dagger|0\rangle .
\end{equation}
Since these states form a complete basis,
\begin{equation}
		\mathcal P
		\equiv
		V_0
		=
		(-1)^{\hat N_f},
		\qquad
		\hat N_f
		=
		\sum_{m>0}c_m^\dagger c_m.
	\label{eq:MajoranaParity}
\end{equation}

\subsection{Photon annihilation operators in the fermionic representation}

We finally express the physical photon operators $a_r$, with
$r=1,3,5,\ldots$, in the fermionic basis. From the vertex operator one
immediately finds
\begin{equation}
	[a_r,V(u)]
	=
	\sqrt{\frac{2}{r}}\,
	u^rV(u),
\end{equation}
and therefore
\begin{equation}
	[a_r,V_n]
	=
	\sqrt{\frac{2}{r}}\,
	V_{n-r}.
	\label{eq:arVertexModeCommutator}
\end{equation}
Using
\begin{equation}
	c_n
	=
	-\frac{(-1)^n}{\sqrt{2}}V_{-n},
	\qquad
	c_n^\dagger
	=
	-\frac{1}{\sqrt{2}}V_n,
	\qquad n>0,
\end{equation}
together with $V_0=\mathcal P$, Eq.~\eqref{eq:arVertexModeCommutator}
becomes
\begin{equation}
	[a_r,c_n]
	=
	-\sqrt{\frac{2}{r}}\,
	c_{n+r},
	\label{eq:arCnCommutator}
\end{equation}
and
\begin{equation}
	[a_r,c_n^\dagger]
	=
	\begin{cases}
		\displaystyle
		\sqrt{\frac{2}{r}}\,c_{n-r}^\dagger,
		& n>r,
		\\[2mm]
		\displaystyle
		-\frac{\mathcal P}{\sqrt r},
		& n=r,
		\\[2mm]
		\displaystyle
		-\sqrt{\frac{2}{r}}\,
		(-1)^n c_{r-n},
		& 0<n<r.
	\end{cases}
	\label{eq:arCnDaggerCommutator}
\end{equation}
Since the fermionic Fock states form a complete basis, these commutation
relations determine the photon operator up to an additive multiple of the
identity, which is fixed by the vacuum condition $a_r|0\rangle=0$. For odd $r$, the
resulting expression is
\begin{equation}
		a_r
		=
		\sqrt{\frac{2}{r}}
		\sum_{n\geq1}
		c_n^\dagger c_{n+r}
		+
		\sqrt{\frac{2}{r}}
		\sum_{n=1}^{(r-1)/2}
		(-1)^n c_nc_{r-n}
		-
		\frac{\mathcal P c_r}{\sqrt r}.
	\label{eq:PhotonOperatorCanonicalFermions}
\end{equation}
Indeed, using the canonical anticommutation relations, the first term
reproduces Eq.~\eqref{eq:arCnCommutator} together with the $n>r$ part of
Eq.~\eqref{eq:arCnDaggerCommutator}; the second reproduces the $n<r$
part, while the parity term gives the special contribution at $n=r$. Each term lowers the energy of $H_0$ by $r\hbar\omega_0$, as required for annihilation of a photon in the
physical mode $r$.

Every term annihilates the vacuum, so this operator also satisfies the
vacuum condition and is therefore $a_r$.

\section[Algebra at the bosonic point]{Algebra at the bosonic point $Z=R_Q$}
\label{app:bosonic}

\subsection{Vertex-mode commutation relations}

We now turn to the point $\rho=1$. At the Toulouse point $\rho=1/2$,
the anticommutator of two vertex modes is a $c$-number. At $\rho=1$,
the commutator instead contains an operator-valued contribution:
for $[A_m,A_n^\dagger]$ with $m\neq n$, this contribution is a photon operator
of the original strip modes, nonzero only when $m-n$ is odd.
The canonical relation $[A_m,A_m^\dagger]=m$ follows as a particular
case of this more general algebra.

We introduce the normal-ordered product
\begin{equation}
	\mathcal N(u,v)
	\equiv
	:V(u)V(v):
	=
	\sum_{p,q\in\mathbb Z}
	C_{p,q}u^pv^q .
\end{equation}
At $\rho=1$, the normal-ordering function is
\begin{equation}
	F_1(x)
	=
	\left(\frac{1-x}{1+x}\right)^2
	=
	1+4\sum_{k=1}^{\infty}(-1)^k k x^k .
\end{equation}
The commutator of two vertex operators is therefore
\begin{align}
	[V(u),V(v)]
	&=
	\left[
	F_1\!\left(\frac{v}{u}\right)
	-
	F_1\!\left(\frac{u}{v}\right)
	\right]
	\mathcal N(u,v)
	\nonumber\\
	&=
	4\sum_{k=-\infty}^{\infty}
	(-1)^k k\,u^{-k}v^k
	\mathcal N(u,v).
	\label{eq:bosonicVertexCommGenerating}
\end{align}
Comparing the coefficient of $u^pv^q$ gives
\begin{equation}
	[V_p,V_q]
	=
	4\sum_{k\in\mathbb Z}
	(-1)^k k\,
	C_{p+k,q-k}.
	\label{eq:VpVqIntermediate}
\end{equation}
Let $\ell=p+q$ and change the summation variable to $s=q-k$. We obtain
\begin{equation}
	[V_p,V_q]
	=
	4(-1)^q
	\left[
	q\sum_s(-1)^sC_{\ell-s,s}
	-
	\sum_s(-1)^s s\,C_{\ell-s,s}
	\right].
	\label{eq:VpVqTwoSums}
\end{equation}
Both sums can be evaluated directly from the normal-ordered product:
the first by setting $v=-u$, and the second by applying
$v\partial_v$ before setting $v=-u$.

Since only odd harmonics enter the vertex operator,
\begin{equation}
	\mathcal N(u,-u)
	=
	:V(u)V(-u):
	=
	1.
\end{equation}
On the other hand,
\begin{equation}
	\mathcal N(u,-u)
	=
	\sum_\ell
	u^\ell
	\sum_s(-1)^sC_{\ell-s,s},
\end{equation}
and hence
\begin{equation}
	\sum_s(-1)^sC_{\ell-s,s}
	=
	\delta_{\ell,0}.
	\label{eq:firstBosonicSum}
\end{equation}
For the second sum,
\begin{equation}
	\left.
	v\partial_v\mathcal N(u,v)
	\right|_{v=-u}
	=
	\sum_\ell
	u^\ell
	\sum_s(-1)^s s\,C_{\ell-s,s}.
	\label{eq:weightedBosonicSum}
\end{equation}
At $\rho=1$,
\begin{equation}
	\mathcal N(u,v)
	=
	\exp\left[
	2\sum_{\substack{r>0\\r\ {\rm odd}}}
	\frac{a_r^\dagger}{\sqrt r}
	\left(u^r+v^r\right)
	\right]
	\exp\left[
	-2\sum_{\substack{r>0\\r\ {\rm odd}}}
	\frac{a_r}{\sqrt r}
	\left(u^{-r}+v^{-r}\right)
	\right].
	\label{eq:normalOrderedBosonicPoint}
\end{equation}
Both exponents vanish at $v=-u$. Differentiating before setting
$v=-u$ therefore gives
\begin{equation}
	\left.
	v\partial_v\mathcal N(u,v)
	\right|_{v=-u}
	=
	-2
	\sum_{\substack{r>0\\r\ {\rm odd}}}
	\sqrt r
	\left(
	a_r^\dagger u^r
	+
	a_r u^{-r}
	\right).
	\label{eq:normalOrderedDerivative}
\end{equation}

It is convenient to introduce the longitudinal current modes
\begin{equation}
	\mathcal J_\ell
	=
	\begin{cases}
		2\sqrt{\ell}\,a_\ell^\dagger,
		& \ell>0,\quad \ell\ {\rm odd},
		\\[1mm]
		2\sqrt{-\ell}\,a_{-\ell},
		& \ell<0,\quad \ell\ {\rm odd},
		\\[1mm]
		0,
		& \ell\ {\rm even},
	\end{cases}
	\label{eq:longitudinalCurrentModes}
\end{equation}
where in particular $\mathcal J_0=0$. Equation
\eqref{eq:normalOrderedDerivative} becomes
\begin{equation}
	\left.
	v\partial_v\mathcal N(u,v)
	\right|_{v=-u}
	=
	-\sum_{\ell\in\mathbb Z}
	\mathcal J_\ell u^\ell,
\end{equation}
and comparison with Eq.~\eqref{eq:weightedBosonicSum} gives
\begin{equation}
	\sum_s(-1)^s s\,C_{\ell-s,s}
	=
	-\mathcal J_\ell.
	\label{eq:secondBosonicSum}
\end{equation}
Substituting Eqs.~\eqref{eq:firstBosonicSum} and
\eqref{eq:secondBosonicSum} into Eq.~\eqref{eq:VpVqTwoSums}, we obtain
the full vertex-mode algebra
\begin{equation}
		[V_p,V_q]
		=
		4(-1)^q
		\left[
		q\,\delta_{p,-q}
		+
		\mathcal J_{p+q}
		\right].
	\label{eq:fullBosonicVertexAlgebra}
\end{equation}
The first term is the central contribution, while the second is an
operator-valued contribution proportional to the longitudinal current.
Using
\begin{equation}
	A_m
	=
	-\frac{(-1)^m}{2}V_{-m},
	\qquad
	A_m^\dagger
	=
	-\frac{1}{2}V_m,
	\qquad m>0,
\end{equation}
we obtain, for $m,n>0$,
\begin{equation}
		[A_m,A_n^\dagger]
		=
		m\delta_{mn}
		-
		\mathcal J_{n-m}.
	\label{eq:fullAmAnDaggerAlgebra}
\end{equation}
For the same positive indices,
\begin{align}
	[A_m,A_n]
	&=
	(-1)^m\mathcal J_{-(m+n)},
	\label{eq:AmAnBosonicPoint}
	\\
	[A_m^\dagger,A_n^\dagger]
	&=
	(-1)^n\mathcal J_{m+n}.
	\label{eq:AmDaggerAnDaggerBosonicPoint}
\end{align}

The definitions of $A_m$ and $A_m^\dagger$ extend to every $m\in\mathbb Z$,
with $A_{-m}=(-1)^mA_m^\dagger$, and Eq.~\eqref{eq:fullAmAnDaggerAlgebra}
then holds for all $m,n\in\mathbb Z$. The positive-index restriction is
essential only for Eqs.~\eqref{eq:AmAnBosonicPoint}
and~\eqref{eq:AmDaggerAnDaggerBosonicPoint}: for arbitrary indices,
Eq.~\eqref{eq:fullBosonicVertexAlgebra} gives
$[A_m,A_n]=(-1)^m\left[m\,\delta_{m+n,0}+\mathcal J_{-(m+n)}\right]$,
which contains an additional central term when $m+n=0$.

The full set of $A_m$ operators therefore does not consist of mutually
independent bosonic modes. If $m,n>0$ have the same parity, however,
both $\mathcal J_{n-m}$ and $\mathcal J_{m+n}$ vanish, and
\begin{equation}
	[A_m,A_n^\dagger]
	=
	m\delta_{mn},
	\qquad
	[A_m,A_n]
	=
	0.
	\label{eq:SameParityAAlgebra}
\end{equation}
Thus, within either parity sector separately,
\begin{equation}
	b_m=\frac{A_m}{\sqrt m}
\end{equation}
obeys canonical bosonic commutation relations. The nontrivial part of
the algebra arises entirely from commutators between modes of opposite
parity.

\subsection[Twisted SU(2) current algebra]{Twisted $SU(2)_1$ current algebra}

The structure of the $A_m$ algebra becomes more transparent when written
in terms of three current components. For positive odd $r$, define the
longitudinal modes
\begin{equation}
	J_r^z
	=
	\frac{\sqrt r}{2}\,a_r,
	\qquad
	J_{-r}^z
	=
	\frac{\sqrt r}{2}\,a_r^\dagger .
	\label{eq:JzOriginalBosons}
\end{equation}
They are related to the modes introduced above by
\begin{equation}
	\mathcal J_\ell
	=
	4J_{-\ell}^z,
	\qquad
	\ell\ {\rm odd}.
	\label{eq:calJandJz}
\end{equation}
The two transverse components are constructed from the vertex modes:
\begin{equation}
	J_m^x
	=
	\frac{i}{4}V_{-m},
	\qquad
	m\ {\rm odd},
	\label{eq:JxVertex}
\end{equation}
and
\begin{equation}
	J_m^y
	=
	\frac{1}{4}V_{-m},
	\qquad
	m\ {\rm even}.
	\label{eq:JyVertex}
\end{equation}
Using
\begin{equation}
	V_n^\dagger=(-1)^nV_{-n},
\end{equation}
one immediately verifies
\begin{equation}
	(J_m^a)^\dagger=J_{-m}^a.
\end{equation}

The mode structure is therefore twisted:
\begin{equation}
	J_m^x,\ J_m^z:
	\quad m\ {\rm odd},
	\qquad
	J_m^y:
	\quad m\ {\rm even}.
	\label{eq:TwistedCurrentModing}
\end{equation}
In particular, only the $y$ component possesses a zero mode,
\begin{equation}
	J_0^y=\frac14V_0.
	\label{eq:JyZeroMode}
\end{equation}

For positive indices, Eqs.~\eqref{eq:JxVertex} and
\eqref{eq:JyVertex} may also be written directly in terms of the
$A_m$ operators:
\begin{align}
	J_m^x
	&=
	\frac{i}{2}A_m,
	&
	J_{-m}^x
	&=
	-\frac{i}{2}A_m^\dagger,
	&& m>0,\quad m\ {\rm odd},
	\label{eq:JxA}
	\\
	J_m^y
	&=
	-\frac{1}{2}A_m,
	&
	J_{-m}^y
	&=
	-\frac{1}{2}A_m^\dagger,
	&& m>0,\quad m\ {\rm even}.
	\label{eq:JyA}
\end{align}
Thus the odd and even $A_m$ modes represent two different transverse
components of the same current algebra.

We now verify that these components obey the current algebra stated in
Eq.~\eqref{eq:TwistedSU2CurrentAlgebra} below. For example, if $m$ and $n$ are
odd, Eq.~\eqref{eq:fullBosonicVertexAlgebra} gives
\begin{align}
	[J_m^x,J_n^x]
	&=
	-\frac{1}{16}
	[V_{-m},V_{-n}]
	\nonumber\\
	&=
	\frac{m}{4}\delta_{m,-n},
	\label{eq:JxJxAlgebra}
\end{align}
since $\mathcal J_{-(m+n)}=0$ for $m+n$ even. The same calculation for
the even $y$ modes gives
\begin{equation}
	[J_m^y,J_n^y]
	=
	\frac{m}{4}\delta_{m,-n},
\end{equation}
while the original photon commutation relations give
\begin{equation}
	[J_m^z,J_n^z]
	=
	\frac{m}{4}\delta_{m,-n}.
\end{equation}
For a mixed commutator, let $m$ be odd and $n$ even. Then
\begin{align}
	[J_m^x,J_n^y]
	&=
	\frac{i}{16}
	[V_{-m},V_{-n}]
	\nonumber\\
	&=
	\frac{i}{4}
	\mathcal J_{-(m+n)}
	\nonumber\\
	&=
	iJ_{m+n}^z.
	\label{eq:JxJyAlgebra}
\end{align}
The remaining mixed relations follow from
\begin{equation}
	[a_r,V_n]
	=
	\frac{2}{\sqrt r}\,
	V_{n-r},
	\qquad
	r>0,\quad r\ {\rm odd},
	\label{eq:arVnBosonicPoint}
\end{equation}
and its Hermitian conjugate. For instance,
\begin{align}
	[J_r^z,J_m^x]
	&=
	\frac{\sqrt r}{2}
	\frac{i}{4}
	[a_r,V_{-m}]
	\nonumber\\
	&=
	\frac{i}{4}
	V_{-(m+r)}
	=
	iJ_{m+r}^y.
\end{align}
Similarly, for even $m$ and positive odd $n$,
\begin{equation}
	[J_m^y,J_n^z]
	=
	\frac{\sqrt n}{8}[V_{-m},a_n]
	=
	-\frac14 V_{-(m+n)}
	=
	iJ_{m+n}^x.
\end{equation}
Hermitian conjugation extends these two relations to negative
longitudinal-mode indices.

Thus, with the allowed mode parities understood, the three components
obey
\begin{equation}
		[J_m^a,J_n^b]
		=
		i\epsilon^{abc}J_{m+n}^c
		+
		\frac{m}{4}
		\delta^{ab}\delta_{m,-n}.
	\label{eq:TwistedSU2CurrentAlgebra}
\end{equation}
The original photon operators are contained directly in the full
$A_m$ algebra. For $r>0$ odd, setting $n=m+r$ in
Eq.~\eqref{eq:fullAmAnDaggerAlgebra} gives, for any $m>0$,
\begin{equation}
	[A_m,A_{m+r}^\dagger]
	=
	-2\sqrt r\,a_r^\dagger.
\end{equation}
Hence
\begin{equation}
		a_r^\dagger
		=
		-\frac{1}{2\sqrt r}
		[A_m,A_{m+r}^\dagger],
		\quad
		a_r
		=
		-\frac{1}{2\sqrt r}
		[A_{m+r},A_m^\dagger].
	\label{eq:arFromA}
\end{equation}
Since $r$ is odd, the two $A$ operators necessarily belong to opposite
parity sectors. The mixed commutators between the odd and even sectors
therefore reconstruct the original longitudinal current.

\subsection{Free Hamiltonian from a current rotation}

The zero mode $J_0^y$ generates rotations between the $x$ and $z$
components of the current. Setting one index to zero in
Eq.~\eqref{eq:TwistedSU2CurrentAlgebra} gives
\begin{equation}
	[J_0^y,J_r^z]
	=
	iJ_r^x,
	\qquad
	[J_0^y,J_r^x]
	=
	-iJ_r^z.
	\label{eq:JyZeroModeRotationAlgebra}
\end{equation}
Thus, defining
\begin{equation}
	U(\theta)
	=
	\exp\left(-i\theta J_0^y\right),
\end{equation}
we obtain
\begin{align}
	U(\theta)J_r^zU^\dagger(\theta)
	&=
	J_r^z\cos\theta
	+
	J_r^x\sin\theta,
	\\
	U(\theta)J_r^xU^\dagger(\theta)
	&=
	J_r^x\cos\theta
	-
	J_r^z\sin\theta.
\end{align}
In particular,
\begin{equation}
	U\left(\frac{\pi}{2}\right)
	J_r^z
	U^\dagger\left(\frac{\pi}{2}\right)
	=
	J_r^x.
	\label{eq:CurrentRotationZtoX}
\end{equation}

The rotation leaves the free Hamiltonian invariant. As at the Toulouse
point, the explicit form of the vertex operator gives
\begin{equation}
	[H_0,V(u)]
	=
	\hbar\omega_0\,u\partial_uV(u),
\end{equation}
and therefore
\begin{equation}
	[H_0,V_n]
	=
	n\hbar\omega_0V_n.
	\label{eq:H0VertexModesBosonicPoint}
\end{equation}
Since $J_0^y=V_0/4$,
\begin{equation}
	[H_0,J_0^y]=0,
\end{equation}
and hence
\begin{equation}
	U(\theta)H_0U^\dagger(\theta)=H_0.
	\label{eq:H0RotationInvariant}
\end{equation}
For positive odd $r$,
\begin{equation}
	J_r^z
	=
	\frac{\sqrt r}{2}a_r,
	\qquad
	J_r^x
	=
	\frac{i}{2}A_r.
\end{equation}
Equation~\eqref{eq:CurrentRotationZtoX} therefore directly implies
\begin{equation}
		\frac{A_r}{\sqrt r}
		=
		-i\,
		U\left(\frac{\pi}{2}\right)
		a_r
		U^\dagger\left(\frac{\pi}{2}\right),
	\quad
	r=1,3,5,\ldots .
	\label{eq:RotatedBosonicModes}
\end{equation}
Thus the canonical operators
\begin{equation}
	b_r
	\equiv
	\frac{A_r}{\sqrt r},
	\qquad
	r=1,3,5,\ldots,
\end{equation}
are simply the original photon modes rotated in current space, up to an
irrelevant phase.

The original free Hamiltonian is
\begin{equation}
	H_0
	=
	\sum_{\substack{r>0\\r\ {\rm odd}}}
	\hbar r\omega_0\,a_r^\dagger a_r.
\end{equation}
Using Eq.~\eqref{eq:H0RotationInvariant} and the transformation
\eqref{eq:RotatedBosonicModes}, we immediately obtain
\begin{equation}
		H_0
		=
		\sum_{\substack{r>0\\r\ {\rm odd}}}
		\hbar r\omega_0\,b_r^\dagger b_r
		=
		\hbar\omega_0
		\sum_{\substack{r>0\\r\ {\rm odd}}}
		A_r^\dagger A_r.
	\label{eq:H0BosonicAm}
\end{equation}

The odd $A_r$ operators therefore provide a complete canonical bosonic
representation of the original transmission line: they are related to
the physical photon modes by a unitary rotation generated by the
Hermitian zero mode $J_0^y=V_0/4$. The even $A_m$ operators are not
additional independent degrees of freedom; they represent the second
transverse component of the same twisted current algebra, while mixed
odd--even commutators reconstruct the original longitudinal current.

\subsection{Fermionic representation of the vertex modes}

A simultaneous representation of all vertex modes is obtained by introducing
canonical fermions $d_j$, indexed by half-integers
$j\in\mathbb Z+1/2$, with
\begin{equation}
	\{d_j,d_{j'}^\dagger\}=\delta_{jj'},
	\qquad
	\{d_j,d_{j'}\}=0.
\end{equation}
The reference state is the filled Fermi sea
\begin{equation}
	d_j^\dagger d_j|\mathrm{FS}\rangle
	=
	\begin{cases}
		|\mathrm{FS}\rangle,&j<0,\\
		0,&j>0,
	\end{cases}
\end{equation}
and normal ordering is defined with respect to this state. The corresponding
free fermion-ladder Hamiltonian is
\begin{equation}
	H_{0,f}
	=
	\hbar\omega_0
	\sum_{j\in\mathbb Z+1/2}
	j:d_j^\dagger d_j: .
	\label{eq:H0SingleFermionLadder}
\end{equation}
The full ladder Fock space is larger than the strip Hilbert space.
The identification of $H_{0,f}$ with $H_0$ requires the physical-sector
restriction derived in Appendix~\ref{app:RQ_physical_sector} below.

On this same ladder, define, for every $m>0$,
\begin{equation}
	A_m^\dagger
	=
	(-1)^{m+1}
	\sum_{j\in\mathbb Z+1/2}
	\eta_j\,
	d_j^\dagger d_{j-m},
	\qquad
	\eta_j=(-1)^{j+1/2},
	\label{eq:AmSingleFermionRepresentation}
\end{equation}
with $A_m=(A_m^\dagger)^\dagger$. The relative phases in
Eq.~\eqref{eq:AmSingleFermionRepresentation} match those of the vertex
modes and must be kept when comparing different $m$. In particular,
$\eta_{j+\ell}=(-1)^\ell\eta_j$ gives
$A_m=-\sum_j\eta_j d_j^\dagger d_{j+m}$.

At the one-particle level, $A_m^\dagger$ simply moves a fermion
upward by $m$ levels,
\begin{equation}
	A_m^\dagger:
	\qquad
	|j-m\rangle
	\longrightarrow
	(-1)^{m+1}\eta_j\,|j\rangle .
	\label{eq:AmSingleParticleTranslation}
\end{equation}
Thus $A_m^\dagger$ is the second-quantized form of an $m$-step raising
operator. Equation~\eqref{eq:H0SingleFermionLadder} immediately gives
\begin{equation}
	[H_{0,f},A_m^\dagger]
	=
	m\hbar\omega_0 A_m^\dagger ,
\end{equation}
as required for the original vertex mode.

The bosonic commutation relation follows from the filled Fermi sea.
Using the canonical anticommutation relations,
\begin{equation}
	[A_m,A_m^\dagger]
	=
	\sum_j
	\left(
	d_{j-m}^\dagger d_{j-m}
	-
	d_j^\dagger d_j
	\right).
\end{equation}
When the occupation operators are written relative to the Fermi sea, the
normal-ordered contributions cancel after shifting the summation index,
while the vacuum contribution counts the $m$ half-integer levels
$j=1/2,\ldots,m-1/2$ that cross the Fermi surface. Hence
\begin{equation}
	[A_m,A_m^\dagger]=m .
	\label{eq:AmBosonicFromFermions}
\end{equation}
Moreover,
\begin{equation}
	A_m|\mathrm{FS}\rangle=0,
\end{equation}
since each term of $A_m$ moves a fermion from level $j$ down to level
$j-m$: for $j>0$ the initial level $j$ is empty, while for $j<0$ the
final level $j-m$ is already occupied. Consequently,
\begin{equation}
	b_m=\frac{A_m}{\sqrt m}
\end{equation}
is a canonical bosonic annihilation operator on the Fock space generated
from $|\mathrm{FS}\rangle$.

To check that all these operators realize the same vertex algebra, let
$E_{jk}=:d_j^\dagger d_k:$ and $n_j^{\rm FS}=\Theta(-j)$. The bilinear
commutator is
\begin{equation}
\begin{split}
 [E_{ij},E_{kl}]={}&\delta_{jk}E_{il}-\delta_{il}E_{kj}\\
 &+\delta_{jk}\delta_{il}(n_i^{\rm FS}-n_j^{\rm FS}).
\end{split}
\end{equation}
Writing $B_\ell=\sum_j E_{j,j-\ell}$, the staggering factors then give,
for arbitrary $m,n>0$,
\begin{align}
 [A_m,A_n^\dagger]
 &=m\delta_{mn}+\bigl[(-1)^{m+n}-1\bigr]B_{n-m},\nonumber\\
 [A_m,A_n]
 &=\bigl[(-1)^m-(-1)^n\bigr]B_{-(m+n)},\nonumber\\
 [A_m^\dagger,A_n^\dagger]
 &=\bigl[(-1)^n-(-1)^m\bigr]B_{m+n}.
 \label{eq:AmMixedFermionAlgebra}
\end{align}
The central term in the first line is the Fermi-surface contribution
evaluated above. These three relations reproduce
Eqs.~\eqref{eq:fullAmAnDaggerAlgebra}--\eqref{eq:AmDaggerAnDaggerBosonicPoint}
with the fermionic longitudinal modes
\begin{equation}
 \mathcal J_\ell=\bigl[1-(-1)^\ell\bigr]B_\ell.
 \label{eq:longitudinalCurrentFermions}
\end{equation}
Thus same-parity modes commute apart from the diagonal central term,
whereas opposite-parity modes generate the longitudinal current. Its
identification with the original photon operators is established below.

The zero vertex mode is
\begin{equation}
 V_0=1+2\sum_j\eta_j:d_j^\dagger d_j:,
 \qquad V_0|\mathrm{FS}\rangle=|\mathrm{FS}\rangle.
 \label{eq:V0BosonicFermions}
\end{equation}
The constant is fixed by the original normal-ordered vertex,
$V_0|0\rangle=|0\rangle$. The same bilinear calculation gives
\begin{equation}
 [V_0,A_m^\dagger]=2\mathcal J_m,
 \qquad [V_0,A_m]=-2\mathcal J_{-m},
\end{equation}
which supplies the zero-index cases of
Eq.~\eqref{eq:fullBosonicVertexAlgebra}. Together with
$V_m=-2A_m^\dagger$ and $V_{-m}=-2(-1)^m A_m$, this realizes all
vertex modes on one ladder, as required by the non-secular Hamiltonian
in Sec.~\ref{sec:beyond_RQ}.

\subsection{Photon annihilation operators in the fermionic representation}

The fermionic form of the physical photon operators follows by combining
Eq.~\eqref{eq:arFromA} with
Eq.~\eqref{eq:AmSingleFermionRepresentation}. For any $m>0$ and odd
$r>0$, the latter gives
\begin{align}
 [A_{m+r},A_m^\dagger]
 &=\sum_j\bigl(\eta_j\eta_{j+r}-\eta_j^2\bigr)
 d_j^\dagger d_{j+r}\nonumber\\
 &=-2\sum_j d_j^\dagger d_{j+r}.
 \label{eq:photonFromMixedFermions}
\end{align}
Here $\eta_j\eta_{j+r}=(-1)^r=-1$, and there is no central term
because the net shift is nonzero. All dependence on $m$ cancels.
Equation~\eqref{eq:arFromA} and Hermitian conjugation therefore give
\begin{equation}
	a_r
	=
	\frac{1}{\sqrt r}
	\sum_{j\in\mathbb Z+1/2}
	d_j^\dagger d_{j+r},
	\qquad
	a_r^\dagger
	=
	\frac{1}{\sqrt r}
	\sum_j
	d_{j+r}^\dagger d_j .
	\label{eq:arSingleFermionRepresentation}
\end{equation}
Thus $a_r$ moves a fermion downward by $r$ levels, while $a_r^\dagger$ moves it upward by $r$. In particular,
\begin{equation}
	[H_{0,f},a_r]
	=
	-r\hbar\omega_0 a_r,
	\qquad
	[H_{0,f},a_r^\dagger]
	=
	r\hbar\omega_0 a_r^\dagger,
\end{equation}
as required for a photon of frequency $r\omega_0$.

The canonical photon algebra again follows from the filled Fermi sea. Using the fermionic anticommutation relations and normal ordering with respect to $|\mathrm{FS}\rangle$, one obtains
\begin{equation}
	[a_r,a_s^\dagger]
	=
	\delta_{rs},
	\qquad
	[a_r,a_s]
	=
	0.
\end{equation}
For $r=s$, the central term counts the $r$ fermionic levels crossing the
Fermi surface; the normalization factors $1/\sqrt r$ reduce it to unity.
This checks the normalization of the photon operators already identified
by Eq.~\eqref{eq:arFromA}. Their vacuum module is the physical strip
space, characterized explicitly in the following subsection.

The representations of $A_m^\dagger$ and $a_r^\dagger$ have a useful common interpretation. Both are second-quantized translations on the fermionic energy lattice,
\begin{equation}
	A_m^\dagger
	=
	(-1)^{m+1}
	\sum_j
	\eta_j d_j^\dagger d_{j-m},
	\qquad
	a_r^\dagger
	=
	\frac{1}{\sqrt r}
	\sum_j
	d_{j+r}^\dagger d_j ,
\end{equation}
but the Josephson vertex mode carries the alternating phase $\eta_j$, whereas the physical photon operator does not. This representation is used in Sec.~\ref{sec:secular_RQ} to evaluate photon-emission matrix elements, and in Sec.~\ref{sec:beyond_RQ} to solve the one-body Floquet problem beyond the secular approximation.

\subsection{Physical strip sector of the fermion ladder}
\label{app:RQ_physical_sector}
The fermion ladder contains more states than the physical strip.
Its odd density modes are the strip photon modes, while its even
density modes are auxiliary. The physical strip states are obtained
by creating odd density excitations on the filled Fermi sea.
Equivalently, they have zero fermion charge and no even density
excitations. Below we show that the Josephson drive and photon loss
preserve this subspace.

Define the charge and unstaggered density modes, normal ordered relative
to the filled sea, by
\begin{equation}
 Q=\sum_j:d_j^\dagger d_j:,
 \qquad \rho_s=\sum_j:d_j^\dagger d_{j+s}:,
 \qquad s\in\mathbb Z,\quad \rho_0=Q.
 \label{eq:RQDensityModes}
\end{equation}
The bilinear algebra gives
\begin{equation}
 [\rho_s,\rho_t]=s\delta_{s,-t},\qquad
 [Q,\rho_s]=0,\qquad
 [H_{0,f},\rho_s]=-s\hbar\omega_0\rho_s.
 \label{eq:RQDensityAlgebra}
\end{equation}
In particular, $\rho_s|\mathrm{FS}\rangle=0$ for $s>0$, and
$a_r=\rho_r/\sqrt r$ for positive odd $r$.
Translations through an even number of levels leave the staggering
unchanged, $\eta_{j+2k}=\eta_j$. Their commutators with vertex modes
therefore vanish; the possible central term is an alternating sum over
$2|k|$ levels and also vanishes. Thus
\begin{equation}
 [Q,\mathcal O]=[\rho_{2k},\mathcal O]=0,
 \qquad
 \mathcal O\in\{A_n,A_n^\dagger,a_r,a_r^\dagger,V_0\},
 \label{eq:RQAuxiliaryCommutant}
\end{equation}
for $n\in\mathbb Z$, positive odd $r$, and nonzero integer $k$.
The charge and even density oscillators are auxiliary degrees of freedom.

On the finite-excitation domain of the full ladder, the free Hamiltonian
has the density-mode decomposition
\begin{align}
 H_{0,f}
 &=\hbar\omega_0\left[\frac{Q^2}{2}
       +\sum_{s>0}\rho_s^\dagger\rho_s\right]\nonumber\\
 &=H_0[a_r]+\hbar\omega_0\left[\frac{Q^2}{2}
       +\sum_{k>0}\rho_{2k}^\dagger\rho_{2k}\right].
 \label{eq:RQHamiltonianDecomposition}
\end{align}
Here $H_0[a_r]=\hbar\omega_0\sum_{r>0,\,r\ {\rm odd}}r a_r^\dagger a_r$.
One way to fix the normalization and constant is to use
Eq.~\eqref{eq:RQDensityAlgebra}: both sides have the same commutators
with all density modes, while the charge-$Q$ filled sea has energy
$\hbar\omega_0Q^2/2$. Density creation operators generate its
fixed-charge Fock module, so this determines the operator identity.

The physical sector is precisely
\begin{equation}
 \mathcal H_{\rm phys}^{(R_Q)}
 =\{\,|\psi\rangle:\ Q|\psi\rangle=0,\quad
       \rho_{2k}|\psi\rangle=0\ \text{for every }k>0\,\}.
 \label{eq:RQPhysicalSector}
\end{equation}
It is generated from $|\mathrm{FS}\rangle$ by the odd $a_r^\dagger$ alone,
and Eq.~\eqref{eq:RQHamiltonianDecomposition} gives $H_{0,f}=H_0$ on
this sector. For example, the $Q=0$ sector has two orthonormal states
at energy $2\hbar\omega_0$,
\begin{equation}
 \frac{\rho_{-1}^2}{\sqrt2}|\mathrm{FS}\rangle,
 \qquad \frac{\rho_{-2}}{\sqrt2}|\mathrm{FS}\rangle.
\end{equation}
Only the first is physical: it is the two-photon state of mode $1$;
$\rho_2$ maps the second to $\sqrt2|\mathrm{FS}\rangle$.
Equation~\eqref{eq:RQAuxiliaryCommutant} and the free evolution in
Eq.~\eqref{eq:RQDensityAlgebra} show that the physical sector is invariant
under the full driven Hamiltonian, its secular approximation, and the
photon-loss jump operators and their adjoints. No even density excitation
or nonzero charge can be generated from a physical initial state by the
exact dynamics.

\section[Toulouse point with a constant quality factor]{Steady state and fluorescence at the Toulouse point}
\label{app:BlochToulouseConstantQF}

This appendix collects the equations used to compute the steady state and the fluorescence spectrum at \(Z=R_Q/2\). We consider the secular master equation~\eqref{eq:masterEquation} with the Hamiltonian~\eqref{eq:HrotToulouse} and the photon operators of Eq.~\eqref{eq:PhotonOperatorCanonicalFermions}. Sums over \(r\) run over positive odd integers and the quality factor is constant, \(\kappa_r=r\kappa_i\).

\subsection{Invariant subspace}
\label{app:toulouseInvariantSubspace}

For a constant quality factor,
\begin{equation}
	\sum_r\kappa_r a_r^\dagger a_r=\kappa_i\sum_{n>0}n\,c_n^\dagger c_n ,
	\label{eq:QFNoJump}
\end{equation}
since both sides equal \(\kappa_iH_0/(\hbar\omega_0)\). Apart from the drive on orbital \(m\), the no-jump evolution is therefore diagonal in the fermionic occupation basis. Every term of Eq.~\eqref{eq:PhotonOperatorCanonicalFermions} moves fermions down or annihilates them. Starting from the vacuum, the state thus remains supported on the \(2^m\) configurations of the orbitals \(1,\ldots,m\). On this subspace, \(a_r\) is obtained by keeping the terms of Eq.~\eqref{eq:PhotonOperatorCanonicalFermions} that involve only orbitals \(n\leq m\), and only the modes \(r\leq2m-1\) emit. The drive changes \(H_0\) by \(\pm m\hbar\omega_0\), and each photon jump lowers it by \(r\hbar\omega_0\). The density operator therefore only couples configurations whose energies differ by multiples of \(m\hbar\omega_0\).

\subsection{Resonant orbital}
\label{app:toulouseResonant}

Place orbital \(m\) first in the fermionic ordering. In its empty/occupied basis, the photon operators take the block form
\begin{equation}
	a_r=\begin{pmatrix}U_r&B_r\\0&U_r\end{pmatrix}.
	\label{eq:arBlockForm}
\end{equation}
Here \(U_r\) collects the terms of \(a_r\) that do not involve orbital \(m\), and \(B_r\) the terms that empty it:
\begin{equation}
	B_r=\begin{cases}
		\sqrt{2/r}\,c_{m-r}^\dagger, & r<m,\\
		-\mathcal P_s/\sqrt m, & r=m,\\
		-(-1)^m\sqrt{2/r}\,c_{r-m}, & m<r<2m,
	\end{cases}
\end{equation}
where \(\mathcal P_s\) is the parity of the spectator orbitals \(1,\ldots,m-1\). Inserting Eq.~\eqref{eq:arBlockForm} into Eq.~\eqref{eq:QFNoJump} gives
\begin{equation}
	\sum_r\kappa_rU_r^\dagger U_r=\kappa_i\sum_{n<m}n\,c_n^\dagger c_n,\qquad
	\sum_r\kappa_rU_r^\dagger B_r=0,\qquad
	\sum_r\kappa_rB_r^\dagger B_r=m\kappa_i.
	\label{eq:QFBlockIdentities}
\end{equation}
The last two identities state that orbital \(m\) is emptied at the rate \(\kappa_m=m\kappa_i\) for every spectator configuration, without interference with the spectator jumps. On the invariant subspace, the dissipator therefore acts on \(c_m\) and \(n_m=c_m^\dagger c_m\) as on a single decaying level,
\begin{equation}
	\sum_r\kappa_r\mathcal D_r^\dagger(c_m)=-\frac{\kappa_m}{2}c_m,\qquad
	\sum_r\kappa_r\mathcal D_r^\dagger(n_m)=-\kappa_m n_m,
\end{equation}
where \(\mathcal D_r^\dagger(X)=a_r^\dagger Xa_r-\frac12\{a_r^\dagger a_r,X\}\) is the adjoint dissipator. Together with the commutators of the Hamiltonian~\eqref{eq:HrotToulouse} with \(c_m\) and \(n_m\), these relations give the optical Bloch equations~\eqref{eq:Bloch_c_m} and~\eqref{eq:Bloch_n_m}.

\subsection{Steady state and mode populations}
\label{app:toulouseSteadyState}

The orbitals below \(m\) are fed by the photon cascade. For \(m\geq3\), however, their occupations do not obey a closed rate equation. A photon jump can take one configuration into a superposition of degenerate configurations, for instance
\begin{equation}
	a_1\ket{\{1,3\}}=\sqrt2\ket{\{1,2\}}+\ket{\{3\}}
\end{equation}
at \(m=3\), where \(\ket{\{\ldots\}}\) lists the occupied orbitals with creation operators in ascending order. The mode populations involve these coherences. We therefore compute the full stationary density operator on the invariant subspace,
\begin{equation}
	\mathcal L\varrho_{\rm ss}\equiv
	-\frac{i}{\hbar}\left[H_{\rm sec},\varrho_{\rm ss}\right]
	+\sum_{r\leq2m-1}\kappa_r\mathcal D[a_r]\varrho_{\rm ss}=0,
	\qquad {\rm Tr}\,\varrho_{\rm ss}=1,
	\label{eq:ToulouseSteadyStateNumerics}
\end{equation}
with \(H_{\rm sec}\) given by Eq.~\eqref{eq:HrotToulouse}. This is a sparse linear problem. Its block structure in the energy reduces the number of unknowns from \(4^m\) to about \(4^m/m\), i.e., about \(10^5\) at \(m=10\), and we find a unique stationary state. The mode populations and orbital occupations are
\begin{equation}
	\langle a_r^\dagger a_r\rangle={\rm Tr}\left(a_r^\dagger a_r\,\varrho_{\rm ss}\right),
	\qquad
	f_l={\rm Tr}\left(c_l^\dagger c_l\,\varrho_{\rm ss}\right).
	\label{eq:ToulouseModePopulations}
\end{equation}
The Cooper-pair rate \(\Gamma_{\rm cp}=\kappa_mf_m\), Eq.~\eqref{eq:QFcurrentToulouse}, and the energy balance~\eqref{eq:energyBalance} provide checks of the numerical solution.

\subsection{Fluorescence spectrum}
\label{app:toulouseSpectrum}

In the laboratory frame, the photon operators of the rotating frame acquire the phase \(e^{-ir\omega_Jt/m}\). We measure frequencies in mode \(r\) from \(r\omega_J/m\), \(\Delta=\omega-r\omega_J/m\). The quantum regression theorem gives the incoherent spectrum of mode \(r\),
\begin{equation}
	\mathcal S_r(\Delta)
	=2\,{\rm Re}\,{\rm Tr}\left[\delta a_r^\dagger\,(i\Delta-\mathcal L)^{-1}\left(\delta a_r\,\varrho_{\rm ss}\right)\right],
	\qquad \delta a_r=a_r-\langle a_r\rangle ,
	\label{eq:ToulouseRegressionSpectrum}
\end{equation}
normalized as \(\int\mathcal S_r\,d\Delta/2\pi=\langle a_r^\dagger a_r\rangle-|\langle a_r\rangle|^2\). The mean field \(\langle a_r\rangle\) is nonzero only for \(r=m\) with \(m\) odd. It adds a coherent peak \(2\pi|\langle a_r\rangle|^2\delta(\Delta)\). The spectral density of the emitted power is
\begin{equation}
	p(\omega)=\frac{\hbar\omega}{2\pi}\sum_r\kappa_{\rm out}(r)\,\mathcal S_r(\omega-r\omega_J/m),
	\label{eq:ToulousePowerSpectrum}
\end{equation}
with \(\kappa_{\rm out}(r)=\kappa_r\) for the total dissipated power and \(\kappa_{\rm out}(r)=\kappa_c\) for the power collected at the measurement port.

\subsection{Fundamental-mode population at arbitrary drive}
\label{app:toulouseFundamental}

The direct solution of Eq.~\eqref{eq:ToulouseSteadyStateNumerics} becomes impractical beyond \(m\simeq10\). At resonance, the fundamental-mode population nevertheless has a closed form for every \(m\ge2\),
\begin{equation}
	\langle a_1^\dagger a_1\rangle
	=\frac{ms}{1+s}\,
	\frac{(3m-4)m^2s^2+3m(2m^2-3m+2)s+C_m}
	{2(3m-4)m^2s^2+2m(5m-2)s+C_m},
	\label{eq:fundamentalFiniteDrivePopulation}
\end{equation}
with \(s=4E_J^{\ast2}/(\hbar^2m^2\kappa_i^2)\) and \(C_m=(m-1)(m+2)(3m-2)\). It follows from a nine-operator closure of the adjoint generator: \(a_1^\dagger a_1\), together with the identity, \(n_m\), \(i(c_m-c_m^\dagger)\), and five operators generated from \(a_1^\dagger a_1\) by the drive and the dissipator, spans a space invariant under the adjoint generator. Equation~\eqref{eq:fundamentalFiniteDrivePopulation} reduces to \(ms\) at weak drive, i.e., two fundamental photons per transferred pair, as in Sec.~\ref{sec:weak}, and to \(m/2\) at saturation. For \(m=1\), the two-level result is \(\langle a_1^\dagger a_1\rangle=s/[2(1+s)]\). Figure~\ref{fig:toulouse_current} uses these expressions.

\subsection{Charge counting in the incoherent limit}
\label{app:toulouse_counting}

In the strongly dephased, constant-quality-factor regime, the resonant orbital is empty or occupied.
Write \(q=\Gamma_J/2\), \(\kappa=\kappa_m\), and \(K=\kappa+2q\).
A forward transfer fills the orbital at rate \(q\) and increments the
net transferred-pair number \(\Delta N\); a backward transfer empties it
at rate \(q\) and decrements \(\Delta N\). Photon loss empties it at
rate \(\kappa\) without changing \(N\). Define
\(p_j(\chi,t)=\sum_{\Delta N}e^{i\chi\Delta N}P(j,\Delta N;t)\),
where \(j=0,1\) is the orbital occupation. The counting parameter
\(\chi\) is conjugate to the change in the Floquet index \(N\), and
\(\boldsymbol p=(p_0,p_1)^{\mathsf T}\) obeys
\begin{equation}
 \frac{d\boldsymbol p}{dt}=L(\chi)\boldsymbol p,
 \qquad
 L(\chi)=\begin{pmatrix}
 -q & \kappa+q e^{-i\chi}\\
 q e^{i\chi} &-(\kappa+q)
 \end{pmatrix}.
 \label{eq:dephasedCountingGenerator}
\end{equation}
The eigenvalue that vanishes at \(\chi=0\) determines the long-time
cumulant-generating function,
\begin{equation}
 \lim_{t\to\infty}\frac1t\ln\langle e^{i\chi\Delta N(t)}\rangle
 =\lambda(\chi)
 =\frac{-K+\sqrt{K^2+4\kappa q(e^{i\chi}-1)}}{2},
 \label{eq:dephasedCountingEigenvalue}
\end{equation}
with the square-root branch chosen continuously from \(\chi=0\).
The first two cumulants per unit time are
\begin{equation}
 C_1=\left.\partial_{i\chi}\lambda\right|_0=\frac{\kappa q}{K},
 \qquad
 C_2=\left.\partial_{i\chi}^2\lambda\right|_0
 =C_1-\frac{2(\kappa q)^2}{K^3}.
 \label{eq:dephasedChargeCumulants}
\end{equation}
Thus \(C_1=\Gamma_{\rm cp}\) reproduces
Eq.~\eqref{eq:dephasedCurrentToulouse}, while
\begin{equation}
 F=\frac{C_2}{C_1}
 =1-\frac{2\kappa q}{K^2}
 =1-\frac{\kappa_m\Gamma_J}{(\kappa_m+\Gamma_J)^2}
 \geq\frac34.
 \label{eq:dephasedFano}
\end{equation}
The minimum occurs at \(\Gamma_J=\kappa_m\).

For the forward-transfer correlation, the stationary occupation is
\(n_{\rm ss}=q/K\), whereas immediately after a forward transfer the
orbital is filled. Its conditional occupation obeys
\(\dot n=q-Kn\), giving
\(n(\tau|+)=n_{\rm ss}+(1-n_{\rm ss})e^{-K\tau}\).
The conditional forward rate is \(q[1-n(\tau|+)]\). Dividing by the
stationary forward rate \(q(1-n_{\rm ss})\) yields
Eq.~\eqref{eq:dephasedForwardRecovery}. This correlation counts two
forward transfers, whereas \(F\) includes the signed contributions of
both transfer directions. The time dependence follows from the rate
description for \(\tau\gg\Gamma_\phi^{-1}\); the zero-delay blockade
itself follows directly from \((A_m^\dagger)^2=0\).

\section[Gaussian closure at the bosonic point]{Gaussian closure for photon loss at the bosonic point}
\label{app:RQ_gaussian}

\paragraph{Photon loss in the fermionic representation.}
A photon in an odd mode \(r\) can be emitted by any fermion with an empty
level \(r\) steps below it. The jump operator sums the amplitudes for all
these transitions. It is linear in the photon representation, where the
drive is a vertex operator. For odd \(m\), the current-space rotation of
Eq.~\eqref{eq:RotatedBosonicModes} exchanges these roles, reflecting the
two components of the twisted current in
Eq.~\eqref{eq:TwistedSU2CurrentAlgebra}. Repeated transfers can populate
arbitrarily high fermionic levels, making a direct many-body calculation
increasingly costly as the drive grows.

At resonance, emission in mode \(r\) takes a fermion from
\((q,\vartheta)\) to \((q-r,\vartheta+\pi)\) with squared matrix element
\(1/r\). If one keeps only statistically independent miniband occupations
\(f_{q,\vartheta}\), the resulting gain and loss terms are
\begin{equation}
	\left.\dot f_{q,\vartheta}\right|_{\rm loss}
	\simeq \sum_r \frac{\kappa_r}{r}
	\left[
	f_{q+r,\vartheta+\pi}(1-f_{q,\vartheta})
	- f_{q,\vartheta}(1-f_{q-r,\vartheta+\pi})
	\right].
	\label{eq:kineticRQ}
\end{equation}
Here and below, sums over \(r\) run over positive odd mode indices.
Each term requires an occupied initial level and an empty final level.

\paragraph{One-body density matrix.}
To retain the coherences generated by the drive, we evolve
\(C_{jj'}=\langle d_{j'}^\dagger d_j\rangle\).
Write \(M\) for the one-body matrix of \(A_m^\dagger\) in
Eq.~\eqref{eq:AmRefermionRWA}, and let \(J_{jj'}=j\delta_{jj'}\).
The rotating-frame one-body Hamiltonian is
\[
h=-\frac{\hbar\delta}{m}J+E_J^\ast(M+M^\dagger).
\]
For a constant quality factor, \(\kappa_r=r\kappa_i\), the no-jump loss
term is proportional to \(H_0\) in the physical subspace. Absorbing the
factor \(\sqrt r\) into the photon jump gives
\[
L_r=\sqrt r\,a_r=\sum_j d_j^\dagger d_{j+r},\qquad
(S_r)_{jj'}=\delta_{j',j+r},\qquad
\kappa_r\mathcal D[a_r]=\kappa_i\mathcal D[L_r].
\]
Here \(S_r\) is the one-body matrix of \(L_r\). Since \(L_r\) is a fermion
bilinear, its dissipator couples the equation for \(C\) to four-fermion
averages. We close the equation by applying Wick factorization at each
time, discarding the connected part of these averages.
Equation~\eqref{eq:masterEquation} then gives
\begin{equation}
	\dot C=-\frac{i}{\hbar}[h,C]
	+\frac{\kappa_i}{2}
	\left([X-X^\dagger,C]+\{P,1-C\}-\{Q,C\}\right),
	\label{eq:GaussianClosureRQ}
\end{equation}
where
\begin{equation}
	\begin{aligned}
		P&=\sum_r S_r C S_r^\dagger, &
		Q&=\sum_r S_r^\dagger(1-C)S_r,\\
		X&=\sum_r \langle L_r\rangle^\ast S_r, &
		\langle L_r\rangle&=\mathrm{Tr}(S_r C).
	\end{aligned}
	\label{eq:GaussianClosureRQterms}
\end{equation}
The gain matrix \(P\) counts particles arriving from higher levels.
The loss matrix \(Q\) counts holes available below, and \(X\) accounts
for the mean photon field.

\paragraph{Current, photon populations, and energy balance.}
The Cooper-pair tunneling rate depends only on the one-body density matrix,
\[
\Gamma_{\rm cp}=\frac{iE_J^\ast}{\hbar}
\mathrm{Tr}\bigl[(M^\dagger-M)C\bigr].
\]
This relation is exact for a given \(C\). Evaluating it with the stationary
solution of Eq.~\eqref{eq:GaussianClosureRQ} gives the Gaussian estimate.
Wick factorization gives the corresponding photon populations,
\begin{equation}
	\langle a_r^\dagger a_r\rangle_{\rm G}
	=\frac{1}{r}\left[
	\big|\mathrm{Tr}(S_r C)\big|^2
	+\mathrm{Tr}\bigl(S_r^\dagger(1-C)S_r C\bigr)
	\right].
	\label{eq:photonNumberGaussianRQ}
\end{equation}
The first term is \(\lvert\langle a_r\rangle\rvert^2\). The second is the
incoherent population, with \(C\) selecting occupied initial levels and
\(1-C\) selecting available final levels. Since
\([J,M]=mM\) and \([J,S_r]=-rS_r\), tracing the stationary equation for
\(C\) against \(J\) gives the secular energy balance,
\[
m\Gamma_{\rm cp}
=\sum_r r\kappa_r\langle a_r^\dagger a_r\rangle_{\rm G},
\]
in agreement with Eq.~\eqref{eq:energyBalance}.

The exact dynamics preserves the physical sector
\eqref{eq:RQPhysicalSector}, defined and proved invariant in
Appendix~\ref{app:RQ_physical_sector}. The Gaussian closure need not
keep the density matrix supported in this sector: a mixed
Gaussian state can acquire charge fluctuations and populate the
auxiliary even density modes. Equation~\eqref{eq:RQHamiltonianDecomposition}
shows that $H_{0,f}$ would also count their energy. We therefore evaluate
the stored strip energy from the physical odd modes as
\(\hbar\omega_0\sum_r r\langle a_r^\dagger a_r\rangle_{\rm G}\).

\paragraph{Gaussian fluorescence spectrum.}
To estimate two-time correlations, we keep the coefficients of the
Gaussian closure fixed at their stationary values. Define the effective
one-body Hamiltonian
\begin{equation}
 h_{\rm e}=h+\frac{i\hbar\kappa_i}{2}(X-X^\dagger)
             -\frac{i\hbar\kappa_i}{2}(P+Q).
 \label{eq:GaussianEffectiveHamiltonian}
\end{equation}
Equation~\eqref{eq:GaussianClosureRQ} can then be written as
\(\dot C=-i(h_{\rm e}C-Ch_{\rm e}^\dagger)/\hbar+\kappa_iP\).
With \(U(\tau)=e^{-ih_{\rm e}\tau/\hbar}\), the two-time contractions
are approximated by
\[
 \langle d_i^\dagger(\tau)d_j(0)\rangle\simeq(CU^\dagger)_{ji},
 \qquad
 \langle d_i(\tau)d_j^\dagger(0)\rangle\simeq[U(1-C)]_{ij}.
\]
Wick factorization of the two photon bilinears therefore yields
\begin{equation}
 G_r(\tau)\equiv
 \langle a_r^\dagger(\tau)a_r(0)\rangle_{\rm inc}
 \simeq\frac1r\operatorname{Tr}\left[
 S_r C U^\dagger(\tau)S_r^\dagger U(\tau)(1-C)\right].
 \label{eq:GaussianTwoTimePhoton}
\end{equation}
For \(\Delta=\omega-r\omega_J/m\), the incoherent spectrum is
\begin{equation}
 \mathcal S_r^{\rm inc}(\Delta)
 =2\operatorname{Re}\int_0^\infty d\tau\,
 e^{-i\Delta\tau}G_r(\tau),
 \qquad
 \int\frac{d\Delta}{2\pi}\mathcal S_r^{\rm inc}(\Delta)=G_r(0).
 \label{eq:GaussianFluorescenceSpectrum}
\end{equation}
If \(\lambda_a\) are the eigenvalues of \(h_{\rm e}\),
Eq.~\eqref{eq:GaussianTwoTimePhoton} is a sum of terms
\(w_{ab}e^{i(\lambda_a^*-\lambda_b)\tau/\hbar}\); their Fourier
transforms have centers \(\operatorname{Re}(\lambda_a-\lambda_b)/\hbar\)
and half-widths \(-\operatorname{Im}(\lambda_a+\lambda_b)/\hbar\).
The full spectrum adds
\(2\pi|\langle a_r\rangle|^2\delta(\Delta)\), and its integral
recovers Eq.~\eqref{eq:photonNumberGaussianRQ}.
The power spectral density is
\[
 p(\omega)=\frac{\hbar\omega}{2\pi}
 \sum_r\kappa_{\rm out}(r)\mathcal S_r(\omega-r\omega_J/m),
\]
with \(\kappa_{\rm out}(r)=\kappa_c\) for the measurement port and
\(\kappa_{\rm out}(r)=\kappa_r\) for total losses, as plotted in
Fig.~\ref{fig:fluo}. This stationary-coefficient, two-time
factorization is an additional approximation used to estimate the
fluorescence line shapes.

\section[Boundary resolvent and Floquet reduction]{Boundary resolvent and Floquet reduction\\at the Toulouse point}
\label{app:toulouse_green}

\subsection{Auxiliary Majorana and the physical sector}
\label{app:toulouse_sector}

Using the vertex algebra of Appendix~\ref{app:toulouse}, the physical
Hamiltonian is
\begin{equation}
	\begin{split}
		H(t)={}&\sum_{n>0}\hbar n\omega_0\,c_n^\dagger c_n
		-E_J^\ast\cos(\omega_Jt)\,\mathcal P\\
		&+\frac{E_J^\ast}{\sqrt2}\sum_{n>0}\lambda_n(t)
		\left[c_n^\dagger+(-1)^nc_n\right].
	\end{split}
	\label{eq:HfullToulouse}
\end{equation}
In \(\mathcal H\otimes\mathbb C^2\), define
\begin{equation}
	\tilde c_n=c_n\otimes\sigma_z,\qquad
	\gamma_0=\mathcal P\otimes\sigma_z,\qquad
	\eta=1\otimes\sigma_x.
\end{equation}
The operators \(\tilde c_n\) are canonical fermions, and \(\eta\) and
\(\gamma_0\) anticommute with them and with each other. With
\(\tilde V_n=V_n\otimes\sigma_z\), the replacement
\(V_n\mapsto i\eta\tilde V_n\) preserves all products of pairs:
\begin{equation}
	(i\eta\tilde V_p)(i\eta\tilde V_q)=V_pV_q\otimes1.
\end{equation}
It therefore leaves the free Hamiltonian unchanged and produces
Eq.~\eqref{eq:HextToulouse}. Since
\(i\eta\tilde V_n=V_n\otimes\sigma_y\), the sectors \(\sigma_y=\pm1\)
carry the original Hamiltonian with opposite signs of \(E_J^\ast\).
Changing this sign is equivalent to shifting the time origin by half a period. The two sectors
thus have identical many-body quasienergies, and retaining \(\sigma_y=+1\)
recovers the physical problem. This duplication of many-body states is
separate from particle--hole pairing in the one-body BdG spectrum.

To express the projection in terms of quasiparticles, retain $L$ strip fermions. The auxiliary fermion
$f_{\rm aux}=(\gamma_0+i\eta)/2$ anticommutes with all $\tilde c_n$ and has parity
$1-2f_{\rm aux}^\dagger f_{\rm aux}=i\eta\gamma_0$, so the total fermion parity of the enlarged system is
\begin{equation}
	\mathcal P_{\rm tot}
	=(-1)^{\sum_{n=1}^L\tilde c_n^\dagger\tilde c_n}
	(1-2f_{\rm aux}^\dagger f_{\rm aux})
	=(\mathcal P\otimes1)(\mathcal P\otimes\sigma_y)
	=1\otimes\sigma_y.
	\label{eq:toulouseTotalParity}
\end{equation}
The physical sector, $\mathcal P_{\rm tot}=+1$, contains $2^L$ of the $2^{L+1}$ states of the enlarged Fock space and is preserved by the quadratic Hamiltonian and by the embedded photon operators $a_r\otimes1$. Let $\Gamma_\alpha$ be a complete canonical set of Floquet quasiparticles, one per particle--hole pair and including zero modes when present; Floquet replicas are not additional modes. Each occupied quasiparticle reverses the parity $p_{\rm vac}=\pm1$ of their vacuum, which is fixed by the chosen Bogoliubov basis. The physical occupations therefore satisfy
\begin{equation}
	\mathcal P_{\rm tot}
	=p_{\rm vac}(-1)^{\sum_\alpha N_\alpha}=+1,
	\qquad N_\alpha=\Gamma_\alpha^\dagger\Gamma_\alpha\in\{0,1\}.
	\label{eq:toulousePhysicalOccupations}
\end{equation}
Choosing all $L+1$ occupations freely would count the physical spectrum twice.

\subsection{Linear equations and elimination of the strip}

Particles, holes, and the zero mode are collected in the signed ladder
\begin{equation}
	\chi_n=c_n\ (n>0),\qquad
	\chi_{-n}=(-1)^nc_n^\dagger\ (n>0),\qquad
	\chi_0=-\frac{\gamma_0}{\sqrt2},\qquad
	\chi_\eta=\frac{\eta}{\sqrt2},
\end{equation}
with $\{\chi_a,\chi_b^\dagger\}=\delta_{ab}$. The negative-index operators are the hole components of the BdG representation, not additional independent fermions. The Heisenberg equations of Eq.~\eqref{eq:HextToulouse} are linear and Hermitian in this normalization,
\begin{align}
	i\hbar\dot\chi_n&=n\hbar\omega_0\chi_n
	-iE_J^\ast\lambda_n(t)\chi_\eta,\\
	i\hbar\dot\chi_\eta&=iE_J^\ast\sum_{n\in\mathbb Z}
	\lambda_n^\ast(t)\chi_n.
	\label{eq:toulouseHeisenberg}
\end{align}
For a Floquet solution, let \(u_{nk}\) and \(\mu_k\) denote the components of
\(\chi_n\) and \(\chi_\eta\), respectively, multiplying
\(e^{-i\varepsilon t/\hbar-ik\omega_Jt}\). Then
\begin{align}
	(\Omega_k-n\hbar\omega_0)u_{nk}
	&=-iE_J^\ast\left[\mu_{k-1}+(-1)^n\mu_{k+1}\right],
	\label{eq:toulouseFourierEquations}\\
	\Omega_k \mu_k&=iE_J^\ast\sum_n
	\left[u_{n,k+1}+(-1)^nu_{n,k-1}\right].
\end{align}
Substituting the first equation into the second yields Eq.~\eqref{eq:BdGquasienergy}, with the sums of Eq.~\eqref{eq:toulouseBoundarySums}. The $n=0$ term, which comes from $\gamma_0$, supplies the $1/z$ term of $g_0$.

\subsection{Reduced equation and states at poles}

Away from the poles of the strip resolvents, Eq.~\eqref{eq:BdGquasienergy} is a linear system \(J(\varepsilon)\boldsymbol\mu=0\), where \(J(\varepsilon)\) is real and symmetric. Since only harmonics \(k\) and \(k\pm2\) are coupled, the even and odd harmonics form separate tridiagonal chains. The quasienergies are the values of \(\varepsilon\) at which \(J\) has a zero eigenvalue. Between successive poles the eigenvalues increase monotonically with \(\varepsilon\), so the roots are located and counted by the change in the number of negative eigenvalues, evaluated with continued fractions. A strip state whose net coupling to the auxiliary Majorana vanishes is an exact dark state at a pole energy; its auxiliary amplitudes are zero, and it must be identified from the full equations. Convergence in the number of Floquet harmonics is checked separately from the dependence on the strip cutoff.

\subsection{Exact spectral equation at integer bias}
\label{app:toulouse_integer}

At integer bias \(\omega_J=m\omega_0\), the periodicity of the cotangent
and cosecant simplifies Eq.~\eqref{eq:BdGquasienergy}. With
\(x=\varepsilon/(\hbar\omega_0)\) and \(A_j=\mu_{2j}\), the even
harmonics satisfy
\begin{equation}
	\bigl[f(x)+2jm\bigr]A_j
	=t(x)(A_{j-1}+A_{j+1}),
	\label{eq:toulouseIntegerStark}
\end{equation}
where
\begin{equation}
	f(x)=x-\mathcal G\cot(\pi x),\qquad
	t(x)=(-1)^m\frac{\mathcal G}{2}\csc(\pi x).
\end{equation}
This is a Wannier--Stark chain with uniform hopping and level spacing
\(2m\).

The Fourier transform \(A(\theta)=\sum_jA_je^{ij\theta}\) gives the
explicit solution
\begin{equation}
	A(\theta)=C\exp\!\left[
	-\frac{if(x)}{2m}\theta+i\frac{t(x)}m\sin\theta
	\right].
	\label{eq:toulouseIntegerHarmonics}
\end{equation}
Periodicity in \(\theta\) requires \(f(x)=2m\ell\), with
\(\ell\in\mathbb Z\). These solutions are roots of \(f(x)=0\) shifted
by \(2m\ell\). The odd harmonic chain supplies the shifts by odd
multiples of \(m\). Consequently, all regular quasienergies follow
from
\[
x=\mathcal G\cot(\pi x)
\]
and folding modulo \(m\), as stated in
Eq.~\eqref{eq:toulouseIntegerSpectrum}. This scalar equation includes
the full coupling between harmonics.

\subsection{Strong-coupling limit}
\label{app:toulouse_strong}

At integer bias, let \(x_r\in(r,r+1)\) be a root of
Eq.~\eqref{eq:toulouseIntegerSpectrum}. Expanding about \(r+1/2\)
gives, at fixed \(r\),
\begin{equation}
	x_r=r+\frac12-\frac{r+1/2}{\pi\mathcal G}
	+O(\mathcal G^{-2}),\qquad \mathcal G\to\infty.
	\label{eq:toulouseStrongCouplingRoots}
\end{equation}
The approach to the half-integer ladder admits a boundary-scattering
interpretation. Using
\(\cot(\pi x)=i(e^{2\pi ix}+1)/(e^{2\pi ix}-1)\), the spectral equation
can be written as
\begin{equation}
	e^{2\pi ix}s_{\rm eff}(x)=1,\qquad
	s_{\rm eff}(x)=\frac{x-i\mathcal G}{x+i\mathcal G}.
	\label{eq:toulouseScatteringQuantization}
\end{equation}
For real \(x\), \(s_{\rm eff}\) is a pure phase. The quantization
condition expresses the accumulation of the propagation and boundary
scattering phases around the unfolded strip. At low energies,
\(|x|\ll\mathcal G\), the boundary factor approaches \(-1\), producing
a scattering phase of \(\pi\). At high energies,
\(|x|\gg\mathcal G\), it approaches \(1\), recovering the integer
ladder.

\section[Driven bosonic point contact]{Driven point contact at the bosonic impedance}
\label{app:RQ_contact}

\subsection{From the fermion ladder to the contact}

At \(\rho=1\), the simultaneous representation established in
Appendix~\ref{app:bosonic} gives, for every \(n\in\mathbb Z\),
\begin{equation}
	V_n=\delta_{n0}
	+2(-1)^n\sum_{j\in\mathbb Z+1/2}\eta_j\,:d_j^\dagger d_{j-n}:,
	\label{eq:VnBilinear}
\end{equation}
with \(\eta_j=(-1)^{j+1/2}\) and normal ordering relative to the filled
sea. The unit constant in \(V_0\), fixed in Eq.~\eqref{eq:V0BosonicFermions}, only adds the c-number
\(-E_J^\ast\cos(\omega_Jt)\) to the many-body Hamiltonian and does not enter the one-body problem.
The Hamiltonian \eqref{eq:Hfull} is therefore quadratic and conserves the number of \(d\) fermions;
no auxiliary Majorana is required.

Choose \(\langle x|j\rangle=e^{ijx}/\sqrt{2\pi}\). The wavefunctions obey
\(\psi(x+2\pi)=-\psi(x)\), and the free one-body generator is
\(-i\hbar\omega_0\partial_x\). The alternating phase acts as
\(\eta_j\mapsto i\mathcal T_\pi\), where
\((\mathcal T_\pi\psi)(x)=\psi(x+\pi)\). The one-body bilinear kernel
of the symmetric partial sum \(\sum_{|n|\le N}V_n\) is
\begin{equation}
	(v^{(N)}\psi)(x)=2iD_N(x)\psi(x+\pi),\qquad
	D_N(x)=\sum_{n=-N}^N e^{inx}.
\end{equation}
Taking \(D_N(x)\to2\pi\delta_{2\pi}(x)\), with the antiperiodic
wavefunction continued around the circle, gives the formal contact operator
\begin{equation}
	\begin{split}
		h(t)\psi(x)={}&-i\hbar\omega_0\partial_x\psi(x)\\
		&-2\pi iE_J^\ast\left[
		e^{-i\omega_Jt}\delta(x)\psi(\pi)
		-e^{i\omega_Jt}\delta(x-\pi)\psi(0)\right].
	\end{split}
	\label{eq:pointContact}
\end{equation}
The two halves of the ring form the two contact channels. This is the
one-body representation of the fermion transfer in
Sec.~\ref{sec:bosonic_floquet}, with the same Josephson phase. Products of
the delta function with the wavefunction at the contact require a
regularization, which we specify below.

\subsection{Scattering angle and contact regularization}
\label{app:RQ_matching}

In the co-moving coordinate \(y=x-\omega_0t\), the contact couples only
amplitudes separated by \(\pi\). A pair labeled by \(y\in[0,\pi)\) passes
the contact twice per revolution period \(T_0=2\pi/\omega_0\). We consider an
instantaneous, energy-independent contact with the off-diagonal mixing
structure of Eq.~\eqref{eq:pointContact}. At a passage time \(t_k\), its
scattering matrix is
\begin{equation}
	U_k=\exp\!\left[-\alpha\left(
	e^{-i\omega_Jt_k}\sigma_+-e^{i\omega_Jt_k}\sigma_-\right)\right],
	\label{eq:kick}
\end{equation}
with the roles of the pair amplitudes exchanged at the other passage.
Equation~\eqref{eq:kick} specifies the boundary condition: fixing
\(\alpha\) fixes the ideal contact and its transfer probability
\(\sin^2\alpha\). The propagation and Floquet calculations below then
require no additional contact prescription. A prescription is needed
only to relate this angle to the physical Josephson coupling.
In the weak-coupling limit, our normalization gives
\begin{equation}
	\alpha\simeq\frac{2\pi E_J^\ast}{\hbar\omega_0},
	\qquad E_J^\ast\ll\hbar\omega_0.
	\label{eq:alphaEJ}
\end{equation}
This leading relation is sufficient to recover the secular results, but
it is not an all-orders relation between angle and coupling.

For the circuit of interest, the junction capacitance sets the high-frequency cutoff, \(\omega_c=1/(ZC_J)\), i.e., \(\hbar\omega_c=4E_C/\pi\) at \(Z=R_Q\), with \(E_C=e^2/(2C_J)\). In a finite strip, it also changes the mode frequencies and their boundary amplitudes, so truncating the Laurent harmonics \(V_n\) is not equivalent to including this physical cutoff.

The low-energy contact can be matched to the capacitively regularized
circuit through its static spectrum or its normalized phase mobility
\(\mu\), defined for the corresponding long-line problem. In the notation
of Paris \textit{et al.}~\cite{Paris2025}, \(\alpha=\pi g/2\), and
\begin{equation}
	\mu=\cos^2\alpha,\qquad
	\alpha=\arccos\sqrt{\mu},\qquad 0\leq\alpha\leq\frac{\pi}{2}.
	\label{eq:RQPhysicalAngle}
\end{equation}
The capacitively regularized junction interpolates from the open
boundary, \(\mu=1\) at \(E_J=0\), toward the phase-pinned boundary,
\(\mu\to0\) as \(E_J/E_C\to\infty\)
\cite{LukyanovWerner2007,Paris2025}. Thus \(\alpha\) approaches
\(\pi/2\) at strong coupling. The nonlinear crossover must be determined
from the regularized circuit; neither normal ordering alone nor
Eq.~\eqref{eq:alphaEJ} determines it. Using this static matching in the
driven problem assumes that the drive and relevant excitation energies
remain within the low-energy regime of the contact description.

\subsection{Passage matrices and spectrum at rational bias}
\label{app:RQ_rational}

Let \(\phi=\omega_Jt\). In a fixed ordering of the two wavefunction amplitudes
at points separated by half a turn (\(\pi\)) around the ring, the two passage matrices are
\begin{equation}
	K_1(\phi)=\begin{pmatrix}
		\cos\alpha&-\sin\alpha\,e^{-i\phi}\\
		\sin\alpha\,e^{i\phi}&\cos\alpha
	\end{pmatrix},\qquad
	K_2(\phi)=\begin{pmatrix}
		\cos\alpha&-\sin\alpha\,e^{i\phi}\\
		\sin\alpha\,e^{-i\phi}&\cos\alpha
	\end{pmatrix}.
\end{equation}
These matrices include the minus sign imposed by the antiperiodic boundary
condition \(\psi(x+2\pi)=-\psi(x)\) when the second component crosses
\(x=2\pi\) and re-enters the ring at \(x=0\). The pair labeled \(y\in(0,\pi)\) passes the contact at the times \(t_j=(j\pi-y)/\omega_0\), \(j=1,2,\ldots\), alternately through \(K_2\) (odd \(j\)) and \(K_1\) (even \(j\)).

At rational bias \(\omega_J/\omega_0=p/q\), with \(p,q\) coprime positive integers, the passage sequence repeats after \(q\) revolutions, or \(p\) drive periods. The ordered product of the \(2q\) passage matrices is
\begin{equation}
	U_q(y)=K_1(\phi_{2q})K_2(\phi_{2q-1})\cdots K_1(\phi_2)K_2(\phi_1),
	\qquad \phi_j=\omega_Jt_j.
\end{equation}
Including the antiperiodic free propagation, its eigenvalues satisfy
\begin{equation}
	(-1)^q\mu_\beta(y)
	=e^{-i\varepsilon_\beta qT_0/\hbar}.
	\label{eq:RQCommonPeriodPhases}
\end{equation}
This determines quasienergies modulo \(\hbar\omega_0/q\). A drive Floquet zone of width \(\hbar\omega_J\) therefore contains \(p\) repetitions.

\paragraph{Trace identity.}
Write \(c=\cos\alpha\), \(s=\sin\alpha\), and define
\[
\varphi=-\frac pq y,\qquad
\Delta=\frac{\pi p}{q},\qquad
\phi_j=\varphi+j\Delta=\omega_Jt_j.
\]
Using \(F=\begin{pmatrix}0&-1\\1&0\end{pmatrix}\), the alternating passage matrices can be expressed through
\begin{equation}
	L_j=K_1(\phi_j)F
	=\begin{pmatrix}
		-s e^{-i\phi_j}&-c\\
		c&-s e^{i\phi_j}
	\end{pmatrix},
	\qquad
	U_q=(-1)^qL_{2q}\cdots L_1.
	\label{eq:RQRationalHalfProduct}
\end{equation}
Since \(L_{j+q}=(-1)^p\sigma_z^pL_j\sigma_z^p\), the matrix
\[
B(\varphi)=i^p\sigma_z^pL_q\cdots L_1
\]
satisfies \(\det B=1\) and \(U_q=-B^2\) for every coprime pair \(p,q\). Consequently,
\[
\frac12\operatorname{Tr}U_q
=1-\frac12\bigl(\operatorname{Tr}B\bigr)^2.
\]

It remains to calculate \(\chi(\varphi)=\operatorname{Tr}B(\varphi)\). Cyclicity gives
\[
\chi(\varphi+\Delta)=(-1)^p\chi(\varphi).
\]
The trace contains only Fourier harmonics \(-q,-q+2,\ldots,q\), because off-diagonal factors must occur in pairs. Coprimality of \(p,q\) then restricts the allowed harmonics to \(\pm q\). These arise by selecting the diagonal entry of every \(L_j\), giving
\[
\chi(\varphi)
=2(-s)^q\cos\!\left(q\varphi+\frac{\pi pq}{2}\right).
\]
Substitution yields
\begin{equation}
	\frac12\operatorname{Tr}U_q(y)
	=1-2(\sin\alpha)^{2q}\sin^2\theta,
	\qquad
	\theta=py+\frac{\pi}{2}(p+q).
	\label{eq:RQRationalTrace}
\end{equation}

\paragraph{Dispersion and widths.}
Writing the eigenvalues of \(U_q\) as \(e^{\pm i\beta}\), the trace identity gives
\(\sin^2(\beta/2)=(\sin\alpha)^{2q}\sin^2\theta\). Equation~\eqref{eq:RQCommonPeriodPhases} therefore yields
\begin{equation}
	\varepsilon_{\ell,\pm}(y)
	=\hbar\omega_0\left(\frac12+\frac{\ell}{q}\right)
	\pm\frac{\hbar\omega_0}{\pi q}
	\arcsin\!\left[(\sin\alpha)^q\sin\theta\right]
	\pmod{\hbar\omega_J},
	\label{eq:RQRationalFullSpectrum}
\end{equation}
with \(\ell=0,\ldots,p-1\). Thus \(q\) determines the miniband spacing and width, while \(p\) determines the number of centers in a drive Floquet zone and the winding of the arrival phase.
For \(q=1\) and \(p=m\), Eq.~\eqref{eq:RQRationalTrace} reduces to \(\tfrac12\operatorname{Tr}U_1=1-2\sin^2\alpha\sin^2\theta\), and Eq.~\eqref{eq:RQRationalFullSpectrum} gives the integer-resonance bands of Eq.~\eqref{eq:exactBandRQ}, with the arrival phase shifted by a constant.

At weak coupling, the half-width becomes
\begin{equation}
	W_{p/q}\simeq\frac{\hbar\omega_0}{\pi q}|\alpha|^q
	\simeq\frac{\hbar\omega_0}{\pi q}
	\left|\frac{2\pi E_J^\ast}{\hbar\omega_0}\right|^q.
	\label{eq:RQRationalWeakWidth}
\end{equation}
The scaling with the \(q\)th power of the coupling corresponds to \(q\) Cooper-pair transfers, which supply the resonant energy \(q\hbar\omega_J=p\hbar\omega_0\). For fixed \(|\sin\alpha|<1\), the half-width decreases as \(|\sin\alpha|^q/q\) at large \(q\). Determining the spectrum at irrational bias requires a separate analysis.
At complete transfer, \(|\sin\alpha|=1\), the half-width reaches \(\hbar\omega_0/(2q)\), and neighboring minibands meet.

\subsection{Spectral repetition in energy and bias}
\label{app:RQ_periodicities}

Shifting the time origin by half a drive period reverses the Josephson coupling without changing the quasienergies. Translating the fermionic orbital labels by one level produces the same sign reversal together with an energy shift \(\hbar\omega_0\). Indeed, the free ladder Hamiltonian and the vertex modes in Eq.~\eqref{eq:VnBilinear} give
\begin{equation}
	\mathsf S^\dagger h(t)\mathsf S
	=h(t+T_J/2)+\hbar\omega_0 I.
	\label{eq:RQEnergyTranslation}
\end{equation}
Combining the two operations therefore relates quasienergies separated by \(\hbar\omega_0\), at any positive bias.

At rational bias, Floquet equivalence also permits shifts by \(\hbar\omega_J=(p/q)\hbar\omega_0\). Since \(p,q\) are coprime, integer combinations of these two shifts generate the finer repetition \(\hbar\omega_0/q\), accounting for the \(p\) miniband centers in a drive Floquet zone.

Repetition in bias follows from Eq.~\eqref{eq:RQRationalFullSpectrum}. The shift \(\omega_J\to\omega_J+\omega_0\) replaces \(p\) by \(p+q\), preserving the reduced denominator \(q\). At fixed \(\alpha\), the unfolded band centers and widths are therefore unchanged. This repetition concerns the spectral support; the Floquet states and dissipative steady state can retain a dependence on \(p\).

\bibliography{bosonic_floquet_ref}

\end{document}